\documentclass[twocolumn]{aastex701}

\usepackage{graphicx}
\usepackage{enumitem}
\usepackage{amsmath}
\usepackage{pifont}
\usepackage{CJK}
\usepackage{soul}
\usepackage{makecell}

\usepackage{tikz}

\newcommand{\mgas}[0]{M_{\rm gas}}
\newcommand{\mdust}[0]{M_{\rm dust}}
\newcommand{\co}[0]{\rm CO}
\newcommand{\coo}[0]{\rm C^{18}O}
\newcommand{\twelveco}[0]{\rm {}^{12}CO}
\newcommand{\thirteenco}[0]{\rm {}^{13}CO}
\newcommand{\ntwohp}[0]{\mathrm{N_2H^{+}}}
\newcommand{\sigmadust}[0]{\Sigma_{\rm dust}}
\newcommand{\sigmagas}[0]{\Sigma_{\rm gas}}

\shorttitle{DiskMINT-GARDEN}
\shortauthors{D. Deng et al.}

\graphicspath{{./}{figures/}}

\begin{document}

\title{DiskMINT-GARDEN: Self-consistent Models to Estimate Disk Masses}

\author[0000-0003-0777-7392]{Dingshan Deng} 
\affiliation{Lunar and Planetary Laboratory, the University of Arizona, Tucson, AZ 85721, USA}
\email{dingshandeng@arizona.edu}

\correspondingauthor{Dingshan Deng}
\email{dingshandeng@arizona.edu}


\author[0000-0002-3311-5918]{Uma Gorti}
\affiliation{NASA Ames Research Center, Moffett Field, CA 94035, USA}
\affiliation{Carl Sagan Center, SETI Institute, Mountain View, CA 94043, USA}
\email{ugorti@seti.org}

\author[0000-0001-7962-1683]{Ilaria Pascucci}
\affiliation{Lunar and Planetary Laboratory, the University of Arizona, Tucson, AZ 85721, USA}
\email{pascucci@arizona.edu}

\author[0000-0003-0522-5789]{Maxime Ruaud}
\affiliation{3 Boussac, Augan, France}
\email{maxime.ruaud@gmail.com}

\begin{abstract}

We present \texttt{DiskMINT-GARDEN}, a grid of self-consistent models together with a fast, open source inference tool for disk masses. The grid is built on \texttt{DiskMINT}, a tool which couples hydrostatic disk structure, continuum/line radiative transfer, and a reduced CO chemical network including freeze-out, grain-surface conversion, and isotope-selective photodissociation. \texttt{DiskMINT-GARDEN} model grid spans a large range of stellar mass ($0.1-2.0\,M_\odot$), gas disk mass ($10^{-5}-10^{-1}\,M_\star$), dust-to-gas ratio ($0.003-0.1$), and characteristic radius ($10-300\,{\rm au}$), and provides synthetic ALMA observables. We train a machine-learning regression model to infer the disk mass, dust-to-gas mass ratio, and disk size from the dust continuum and $\mathrm{C^{18}O}$ line observations. Applying \texttt{DiskMINT-GARDEN} to archival ALMA data of 34 disks, we find gas masses in good agreement with dynamical and HD-based estimates. 
Comparing our results with estimates from chemical modeling using DALI, we find that their need for large-scale elemental or CO depletion can be accounted for by grain-surface chemistry implemented in \texttt{DiskMINT}, with CO conversion to CO$_2$ being one of the main reactions.
Therefore, extant data suggest little chemical processing due to disk evolutionary processes.
\end{abstract}

\keywords{Protoplanetary disks(1300); Astrochemistry(75); Chemical abundances(224); CO line emission(262); Planet formation(1241)}

\section{Introduction}
\label{sec:intro}

The mass of protoplanetary disks, which resides predominantly in gas, is a fundamental quantity that shapes the formation and evolution of planetary systems. 
The gas disk mass sets the reservoir available for giant planet formation, influences the efficiency of planet migration, and determines the lifetime and dynamical evolution of disks \citep[e.g.,][]{Miotello_PPVII_2023, birnstiel_dust_2024}. 
Although challenging, constraining disk masses is therefore critical.

CO is the second most abundant molecule after H$_2$ and the most widely used gas tracer. 
With many rotational lines at typical disk temperatures, CO is readily observable at millimeter wavelengths.
Transitions from rare CO isotopologues such as $\coo$ are often optically thin and can trace gas deeper in the disk toward the midplane, making them promising tracers of the disk mass. 
However, early thermochemical models found that $\co$ isotopologue emission was significantly fainter than expected from models
\citep[e.g.,][]{Williams_n_Best_GasMass_2014, Miotello2014, Miotello2016, miotello_lupus_2017, Ansdell_lupus_2016, Long_Cham_2017}. 
These results were frequently characterized as evidence for substantial depletion of gas-phase CO in disks, by up to two orders of magnitude, and often interpreted as an elemental depletion of C and/or O possibly due to sequestration into planetesimals or chemical processing \citep[e.g.,][]{Bergin_n_Williams_mass_2017, Bosman_n_Banzatti_twhya_2019, Krijt_COdep_2020, Powell_COdep_2022}. 
These processes potentially make CO an interesting tracer of disk evolution, but also imply that CO-based estimates of $\mgas$ could be unreliable.

Several studies have, however, demonstrated that faint CO emission does not necessarily require large-scale elemental depletion. 
In a recent study, \citet{ricciardi_compact_2026} found $\thirteenco$ is likely to be optically thick and faint emission might just indicate compact disks.
Moreover, when disk density structure is self-consistently computed with a coupled temperature and chemistry treatment, observed CO, [CI], and H$_2$O line fluxes can be reproduced without the need for introducing any depletion factors \citep[e.g.,][]{Molyarova2017ApJ...849..130M, Ruaud2022, Powell_COdep_2022, pascucci_noCOdep_2023, deng_diskmint_2023, ruaud_cold_2024, deng_diskmint_2025, Deng_2025_AGEPRO_III_Lupus, zwicky_dancing_2025}. 
These works highlight three physical processes that are particularly important for inferring masses from CO isotopologue line emission.
\begin{enumerate}
    \item[(i)] CO freeze-out and grain-surface chemistry are essential to include as most of the disk mass is contained in the cold, high-density midplane. While it is long known that CO freezes out at low temperatures ($\sim 20$K) in the interstellar medium and in disks \citep[e.g.,][]{caselli_co_1999, tafalla_systematic_2002, jorgensen_physical_2002, qi_resolving_2008, qi_resolving_2011}, considering grain-surface reactions that convert CO ice into more stable species such as CO$_2$ results in CO-${\rm CO_2}$ ice mixture that can exist at higher temperatures ($\sim 35$K) \citep[e.g.,][]{Aikawa_COsnowline_n2hp_2015, Molyarova2017ApJ...849..130M, Ruaud2022}. Recent JWST observations of molecular clouds and disks provide growing evidence for abundant CO$_2$ ice co-spatial with CO ice, supporting this conversion pathway in both disks, \citep[e.g.,][]{sturm_jwst_2023, sturm_edge-protoplanetary_2023, bergner_jwst_2024} and in clouds \citep[][]{Smith_ice_NatAs_2025}.
    \item[(ii)] The disk vertical structure needs to be solved self-consistently to balance pressure and gravity; this physics connects the bulk of the mass at the midplane to the emitting layer at the surface \citep[e.g.,][]{Ruaud2022, deng_diskmint_2023, paneque-carreno_vertical_2025}. Most existing disk models typically iterate between chemistry and gas temperature, but fix an initially assumed, vertically isothermal density structure. Since gas-line emission mostly originates above the vertical snowline, where gas density has decreased by orders of magnitude, neglecting vertical hydrostatic equilibrium leads to inconsistent density/temperature/chemistry structures and contributes to errors in inferred disk masses. 
    \item[(iii)] Isotope-selective photodissociation affects the abundance of rare CO isotopologues in the disk surface layers, due to reduced line overlap and self-shielding of CO from UV irradiation \citep[][]{visser_photodissociation_2009}. The lowered abundance of $\coo$ relative to $\twelveco$ in UV-irradiated regions further lowers the emergent $\coo$ emission and is important to include when converting observed fluxes to disk masses \citep[e.g.,][]{Miotello2014, Miotello2016, Ruaud2022}. 
\end{enumerate}

Based on the above, we recently introduced \texttt{DiskMINT}\footnote{\url{https://github.com/DingshanDeng/DiskMINT}}, a physically motivated yet computationally efficient disk modeling framework that couples a self-consistent vertical structure with continuum and line radiative transfer and a reduced chemical network optimized for CO isotopologues. 
Applying \texttt{DiskMINT} to spatially resolved disks such as RU~Lup and IM~Lup demonstrated that CO-based gas masses can be substantially higher than earlier estimates \citep{deng_diskmint_2023, deng_diskmint_2025} and, in the case of IM~Lup, consistent with independent dynamical mass measurements \citep{lodato_dynamical_mass_2023}.
 
Independent constraints on disk mass serve as important calibrators. 
Dynamical disk masses can be inferred using high-resolution CO kinematics \citep[e.g.,][]{lodato_dynamical_mass_2023, martire_rotation_2024, longarini_exoalma_2025}. 
However, these measurements only work for massive disks ($\gtrsim 0.05\,M_\star$), where disk self-gravity produces detectable perturbations to the Keplerian rotation curve.
Another independent tracer is hydrogen deuteride (HD), whose $J=1-0$ transition at 112~$\mu$m can provide a relatively direct probe of $\mgas$ because HD is optically thin and does not freeze out. 
However,  only a small fraction of the total disk mass ($\lesssim 1\%\,\mgas$) is in regions hot enough to excite the $E_{u} \sim 128\,{\rm K}$ upper level of HD;  therefore the HD-based $\mgas$ depends sensitively on the disk temperature \citep[see discussions in][]{Trapman_HD_2017, Ruaud2022}. To date, the HD line has been detected in only three disks due to the lack of far-infrared facilities with the required sensitivity \citep{Bergin_HD_2013, McClure_HD_2016}. 

For most disks, thermochemical modeling (e.g., \texttt{DALI} \citealt{bruderer_warm_2012, bruderer_survival_2013}, \texttt{ProDiMo} \citealt{woitke_diana_2016, woitke_diana_2019}, \texttt{DiskMINT} \citealt{deng_diskmint_2023, deng_diskmint_2025}) is a necessary step in inferring disk masses from observed CO isotopologues and HD lines.
While there is some agreement on derived disk masses \citep[e.g.,][]{Deng_2025_AGEPRO_III_Lupus}, thermochemical models arrive here with and without CO depletion factors. 
Understanding whether CO is truly depleted in disks, and the magnitude of any such effect, is important beyond the determination of $\mgas$. 
Previous work has predicted large levels of CO depletion, indicative of disk evolutionary processes such as chemistry-dependent incorporation of ices that get locked into planetesimals \citep[e.g.,][]{Krijt_COdep_2020}, and/or radial and vertical transport processes \citep[e.g.,][]{booth_planet-forming_2019, furuya_different_2022} that alter C/O abundance ratios in the disk. 
The degree of CO processing affects the carbon budget available for planet formation and influences the composition of icy planetesimals and planetary atmospheres. 
Disentangling physical effects (such as disk structure and temperature gradients) from genuine chemical depletion is therefore essential for interpreting both disk masses and disk chemistry \citep[e.g.,][]{oberg_astrochemistry_2021}.

Building on the physically motivated \texttt{DiskMINT} framework, here we present \texttt{DiskMINT-GARDEN} -- the \emph{Grid of Astrochemical Radiative Disk EmissioN}.
This work extends \texttt{DiskMINT} into a comprehensive model grid and inference framework that enables rapid and reproducible disk gas-mass estimates from commonly observed quantities. 
The grid spans a large range of stellar mass, gas disk mass, dust-to-gas ratio, and disk size, and provides synthetic observables including millimeter continuum luminosities, CO isotopologue line luminosities, and disk size metrics. 
By combining these observables, \texttt{DiskMINT-GARDEN} enables efficient first-order inference of $\mgas$, which may also guide detailed modeling of individual targets. 

The structure of the paper is as follows. 
In Section~\ref{sec:method}, we describe the construction of the \texttt{DiskMINT-GARDEN} model grid and the regression framework used to map observables to disk physical parameters. 
In Section~\ref{sec:application}, we apply the framework to an archival sample of disks and compare the inferred gas masses with independent dynamical and thermochemical estimates. 
Section~\ref{sec:future} outlines future applications of the framework, including the incorporation of new tracers such as HD from upcoming facilities (e.g., \textit{PRIMA}) and extensions of the \texttt{DiskMINT} modeling framework. 
Finally, Section~\ref{sec:summary} summarizes our main results.

\section{Model and Method}
\label{sec:method}

\subsection{Model Grids}
\label{subsec:grid_parameters}

We generate a grid of disk models using \texttt{DiskMINT} following similar setups in our previous works \citep[][]{deng_diskmint_2023, deng_diskmint_2025}.
The grid spans a wide, physically motivated range of stellar and disk properties relevant to protoplanetary disks (see Table~\ref{tab:grid_props}).

We adopt six discrete stellar masses,
$M_\star = \{0.1,\,0.3,\,0.5,\,0.7,\,1.0,\,2.0\}\,M_\odot$, covering the range of stellar masses in nearby star-forming regions.
The stellar radius and luminosity are taken at an age of 1\,Myr from the stellar evolutionary models of \citet{baraffe_new_2015} for stars with $M_\star \lesssim 1.0\,M_\odot$ and of \citet{feiden_magnetic_2016} for more massive stars, see \citet{Pascucci_mass_2016} for the rationale.

For each stellar mass, we vary the total gas disk mass over five orders of magnitude with
$M_{\rm gas} / M_\star = \{10^{-5},\,10^{-4},\,10^{-3},\,10^{-2},\,10^{-1}\}$.
The dust disk mass is set via an independently varied dust-to-gas mass ratio,
$\varepsilon = M_{\rm dust}/M_{\rm gas}=\{0.100,\,0.033,\,0.010,\,0.003\}$ 
This approach allows the grid to flexibly explore departures from the canonical interstellar value ($\varepsilon=0.01$) by decoupling dust and gas mass evolution.

Accretion onto the central star is included through a stellar-mass-dependent accretion rate,
\begin{equation}
  \label{equation:mdotacc}
  \dot{M}_{\rm acc} = 10^{-8}
  \left(\frac{M_\star}{0.7\;M_\odot}\right)^2
  \,M_\odot\,\mathrm{yr}^{-1},
\end{equation}
which is motivated by empirical scaling relations in young stellar objects \citep[e.g.,][]{alcala_x-shooter_2017} and provides a consistent source of viscous heating.

We estimate the FUV (Far ultraviolet: $6\,{\rm eV} < h\nu < 13.6\,{\rm eV}$) luminosity and flux field following \citet{gorti_photoevaporation_2009, gorti_time_2009}. 
Combining the FUV component due to accretion ($ L_{\rm acc} \sim GM_\star \dot{M}_{\rm acc}/R_\star$) and chromosphere \citep[$10^{-3.3}$L$_{\star}$;][]{Valenti_IUE_2003} components gives the local field (in Habing units) of,

\begin{equation}
  \label{equation:fuv}
  \begin{aligned}
  G_0(r) =  \Bigg[\,
      &8.51\times10^{6}
      \left(\frac{M_\star}{R_\star}\right)
      \left(\frac{\dot{M}_{\rm acc}}{10^{-8}}\right)
       \\
      &+ 4.27\times10^{5}
      \left(\frac{L_\star}{L_\odot}\right)
    \Bigg]\, \left(\frac{1\,{\rm au}}{r}\right)^{2}  .
  \end{aligned}
\end{equation}

The disk surface density follows a tapered power-law profile from the self-similar solution of viscous disks \citep[][]{lynden-bell_evolution_1974},
\begin{equation}
  \label{equation:sigma}
  \Sigma(r) = \Sigma_1
  \left(\frac{r}{1\,{\rm au}}\right)^{-\gamma}
  \exp\!\left[-\left(\frac{r}{R_{\rm c}}\right)^{2-\gamma}\right],
\end{equation}
with a fixed radial slope $\gamma = 1$.
The characteristic radius $R_{\rm c}$ is varied as
$R_{\rm c}=\{10,\ 30,\ 100,\ 300\}\,{\rm au}$. 
The grid outer radius is fixed at $R_{\rm out}=1000\,{\rm au}$, sufficiently large to avoid truncation effects on the modeled observables, while the grid inner radius is set to the dust sublimation radius for each model through iterations, which depends on the stellar luminosity and thus on $M_\star$.

Dust properties are held fixed across the grid to isolate the effects of disk mass and structure.
We adopt the DIANA standard dust composition \citep[][]{woitke_diana_2016, min_multiwavelength_2016}, consisting of a mixture of pyroxene and amorphous carbon with a mass ratio of 0.87:0.13 and a porosity of 25\%.
The grain size distribution is described as $n(a) \propto a^{-p}$ and spans from $a_{\rm min}=10^{-5}$\,cm to $a_{\rm max}=0.1$\,cm with $p = 3.5$.
We solve for vertical hydrostatic equilibrium and dust settling \citep[details see][]{deng_diskmint_2025}, where the vertical dust settling assumes a fixed viscous $\alpha$ parameter of $\alpha_{\rm v}=5\times10^{-3}$ following \citet{Ruaud2022, deng_diskmint_2025}. 
Based on previous explorations \citep[][]{deng_diskmint_2023}, the inferred gas disk mass from $\coo$ line emission is not particularly sensitive to the chosen dust properties; uncertainties associated with mass inferences are discussed later.

For each model in the grid, we compute synthetic observables using \texttt{RADMC-3D} \citep[][]{Dullemond_radmc-3d_2012}. 
Here, we include
(1) the millimeter continuum luminosity $L_{\rm mm}$ at the ALMA Band 6 ($\sim 234\,{\rm GHz} \sim 1.3\,{\rm mm}$), 
(2) the $\coo$\,($2$--$1$) and $\coo$\,($3$--$2$) line luminosities $L_{\rm C^{18}O}$ using the local thermal equilibrium (LTE) mode (verified in Appendix~\ref{appendix:optical_depth} by comparing the gas densities in the $\coo$-emitting regions with the critical densities of the two transitions), 
and (3) dust size metrics derived from synthetic continuum images, such as the radius enclosing 90\% of the continuum emission, $R_{\rm dust, 90}$. 
By focusing on these specific quantities, we ensure that our modeling framework remains directly comparable to metrics most readily obtained in large ALMA disk surveys, including the existing large programs DSHARP \citep[2016.1.00484.L;][]{andrews_disk_2018}, MAPS \citep[2018.1.01055.L;][]{oberg_molecules_2021}, AGE-PRO \citep[2021.1.00128.L;][]{zhang_alma_2025}, exoALMA \citep[2021.1.01123.L;][]{teague_exoalma_2025}, and ongoing DECO (2022.1.00875.L), CHEER (2024.1.01001.L), DiskStrat (2024.1.01212.L), and DMOST (2025.1.00324.L).

\begin{deluxetable*}{lcccccccc}
  \tablecaption{Grid Parameters for the Disk Modeling Framework\label{tab:grid_props}}
  \tabletypesize{\scriptsize}
  \tablehead{
    \colhead{Parameter} & 
    \colhead{Symbol} & 
    \multicolumn{6}{c}{Values} &
    \colhead{Unit}
  }
  \startdata
  \textit{Stellar Properties\tablenotemark{a}} & & & & & & & & \\
  Stellar mass & $M_\star$ 
  & 0.1 & 0.3 & 0.5 & 0.7 & 1.0 & 2.0 
  & $M_\odot$ \\
  Stellar radius & $R_\star$ 
  & 1.00 & 1.64 & 1.87 & 2.10 & 2.42 & 3.32 
  & $R_\odot$ \\
  Stellar luminosity & $L_\star$ 
  & 0.067 & 0.332 & 0.658 & 1.091 & 1.928 & 6.385 
  & $L_\odot$ \\
  Mass accretion rate & $\dot{M}_{\rm acc}$ 
  & $2.0\times 10^{-10}$ & $1.8\times 10^{-9}$ & $5.1\times 10^{-9}$ & $1.0\times 10^{-8}$ & $2.0\times 10^{-8}$ & $8.2\times 10^{-8}$ 
  & $M_\odot\,\mathrm{yr}^{-1}$ \\
  FUV flux field & $F_{\rm FUV}(r)$ 
  & \multicolumn{6}{c}{Equation~\ref{equation:fuv}} 
  & $G_0$ \\
  $-$ at 1 au & $F_{\rm FUV}(1\;{\rm au})$ 
  & $4.6\times 10^{4}$ & $4.3\times 10^{5}$ & $1.4\times 10^{6}$ & $3.3\times 10^{6}$ & $8.0\times 10^{6}$ & $4.6\times 10^{7}$ 
  & $G_0$ \\
  \hline
  \textit{Disk Masses} & & \multicolumn{6}{c}{} & \\
  Gas disk mass & $M_{\rm gas}$ 
  & \multicolumn{6}{c}{$\{10^{-5}, 10^{-4}, 10^{-3}, 10^{-2}, 10^{-1}\} \times M_\star$} 
  & $M_\odot$ \\
  Dust-to-gas ratio & $\varepsilon$ 
  & \multicolumn{6}{c}{\{0.100, 0.033, 0.010, 0.003\}} 
  & \dots \\
  Dust disk mass & $M_{\rm dust}$ 
  & \multicolumn{6}{c}{$M_{\rm gas} \times \varepsilon$} 
  & $M_\odot$ \\
  \hline
  \textit{Disk Structure} & & \multicolumn{6}{c}{} & \\
  Surface density & $\Sigma(r)$ 
  & \multicolumn{6}{c}{Equation~\ref{equation:sigma}} 
  & $\mathrm{g}\,\mathrm{cm}^{-2}$ \\
  $-$ Slope & $\gamma$ 
  & \multicolumn{6}{c}{1} 
  & \dots \\
  $-$ Grid Inner radius & $R_{\rm in}$ 
  & \multicolumn{6}{c}{Sublimation radii} 
  & au \\
  $-$ Char. radius & $R_{\rm c}$ 
  & \multicolumn{6}{c}{\{10, 30, 100, 300\}} 
  & au \\
  $-$ Grid Outer radius & $R_{\rm out}$ 
  & \multicolumn{6}{c}{1000} 
  & au \\
  \hline
  \textit{Dust Properties} & & \multicolumn{6}{c}{} & \\
  Composition & \dots 
  & \multicolumn{6}{c}{DIANA standard dust\tablenotemark{b}} 
  & \dots \\
  Size distribution & $n(a)$ 
  & \multicolumn{6}{c}{$\propto a^{-p}$} 
  & \dots \\
  $-$ Min size & $a_\mathrm{min}$ 
  & \multicolumn{6}{c}{$1\times10^{-5}$} 
  & cm \\
  $-$ Max size & $a_\mathrm{max}$ 
  & \multicolumn{6}{c}{0.1} 
  & cm \\
  $-$ Size slope & $p$ 
  & \multicolumn{6}{c}{3.5} 
  & \dots \\
  Viscous param. & $\alpha_{\rm v}$ 
  & \multicolumn{6}{c}{$5\times10^{-3}$} 
  & \dots \\
  \enddata
  \tablenotetext{a}{Stellar parameters adopted from 1\,Myr pre-main-sequence models: \citet{baraffe_new_2015} for $M_\star \leq 1.0\,M_\odot$, and \citet{feiden_magnetic_2016} for $M_\star=2.0\,M_\odot$.}
  \tablenotetext{b}{The DIANA standard dust \citep[][]{woitke_diana_2016, min_multiwavelength_2016}, is a mixture of pyroxene and carbon (mass ratio 0.87:0.13) with 25\% porosity.}
\end{deluxetable*}

\subsection{Disk Mass Inference}
\label{subsec:infer_mass}

We construct a global regression mapping between observables and physical parameters using the model grid. We simulate the observational quantities for each grid to form an observable vector,
\begin{equation}
  \mathbf{O} \equiv (L_{\rm C^{18}O}, \, L_{\rm mm}, \, R_{90,\rm dust}).
\end{equation}
Each model grid is characterized by a physical parameter vector,
\begin{equation}
  \mathbf{\Theta} \equiv
  (M_{\rm gas},\, \varepsilon,\, R_{\rm c}).
\end{equation}

Disk properties, specifically the gas/dust surface density distribution and the dust composition and size distribution, are not varied in the grid at present and are held constant.
Stellar properties (Table~\ref{tab:grid_props}) are a function of the stellar mass $M_\star$, which is treated as a known conditioning parameter rather than an inferred quantity, since it is typically well constrained from independent stellar characterization.
This approach enables efficient inference that can provide a fast first-order estimate of the physical parameters ($\mathbf{\Theta}$) for a large dataset.

After experimenting with different approaches, we implement this mapping using a supervised machine-learning regression model based on gradient-boosted decision trees \citep[e.g.,][]{friedman_greedy_2001}, implemented with the eXtreme Gradient Boosting (\texttt{XGBoost}) library \citep{chen_xgboost_2026}.
Details of the model architecture, hyperparameter choices, validation performance, and extrapolation control are provided in Appendix~\ref{appendix:ML_details}.
The regression model is trained on the \texttt{DiskMINT-GARDEN} grid in $\log$ space.
Regularization and early stopping are applied to prevent overfitting and ensure robust interpolation across the grid.
Once trained, the model provides fast and deterministic predictions for input observables within (or near) the grid domain.
The trained regression model and inference tools are released as part of the public \texttt{DiskMINT} \texttt{v1.7.0} on \texttt{GitHub}.

\subsection{Uncertainties}
\label{subsec:uncertainties}

We estimate statistical uncertainties by propagating observational measurement errors through the regressor and including its intrinsic interpolation scatter from validation tests, and find they are typically $\lesssim 25\%$ (see Appendix~\ref{appendix:ML_details}). 

The dominant uncertainty is systematic and comes from assumptions in the underlying thermochemical modeling (e.g., disk structure, chemistry, and radiative transfer), which can reach a factor of $\lesssim 2$ \citep{Ruaud2022, deng_diskmint_2023, deng_diskmint_2025}. 
We therefore adopt a conservative factor-of-two uncertainty when interpreting inferred disk masses.

These uncertainty estimates assume that the regression model is applied within the parameter space spanned by the training grid. 
To ensure reliable application of the regression model, we perform domain checks by comparing each target’s observables to the range spanned by the training grid, and
targets lying near or beyond grid boundaries are flagged. 
Additional details on extrapolation control are described in Appendix~\ref{appendix:ML_details}.

\subsection{Grid Applicability at High Masses}
\label{subsec:grid_applicability}

The 480 models in \texttt{DiskMINT-GARDEN} span a broad range of stellar and disk parameters.
Disk mass inferences in some regions of parameter space are less robust; in particular, when gas surface densities are high enough for the disk to become gravitationally unstable and/or when most of the $\coo$ emission from the disk is optically thick.

\subsubsection{Gravitational Instability}
\label{subsubsec:gravitational_instability}

Disks with very high gas surface densities may approach the regime of gravitational instability (GI), which may result in the formation of spiral structures or fragmentation; the \texttt{DiskMINT-GARDEN} models assume axisymmetry and cannot capture these effects.
For each grid model, we estimate Toomre's stability parameter $Q$ \citep[][]{toomre_gravitational_1964} at the outer dust radius ($R_{\rm dust,90\%}$) as
\begin{equation}
  Q(R_{\rm dust,90\%}) \;=\; \frac{c_s\,\Omega}{\pi G \sigmagas},
\end{equation}
where $c_s$ is the local sound speed, $\Omega$ is the Keplerian angular frequency, and $\Sigma_{\rm gas}$ is the gas surface density evaluated at $R_{\rm dust,90\%}$.
Models with $Q < 1$ are considered gravitationally unstable and accordingly flagged as such (more details are given in Appendix~\ref{appendix:additional_model_figure}, Figure~\ref{fig:Rdust_Mgas_withdata}).

\subsubsection{Optically Thick Emission}
\label{subsubsec:optical_depth}
At high surface densities, when the mass is high, or in the inner regions, the $\coo$ line can become optically thick.
In this regime, the emergent line emission becomes more sensitive to the accuracy of the gas temperature.
We note that in the optically thick inner regions, high gas densities ensure strong gas-dust thermal coupling, so \texttt{DiskMINT}'s dust-based temperature estimates remain reliable there.

\begin{figure*}
  \centering
  \includegraphics[width=1.0\linewidth]{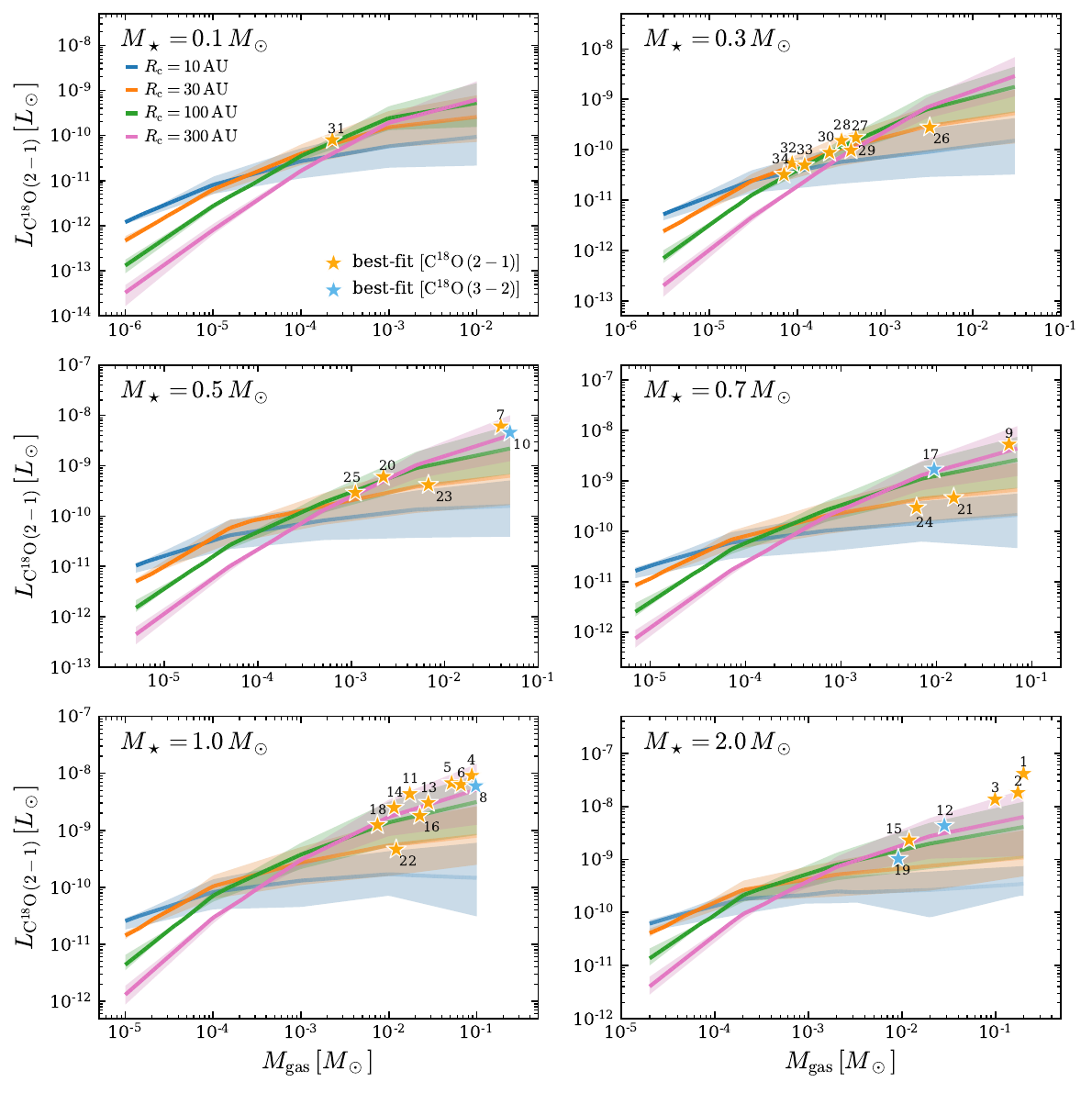}
  \caption{
    Relationship between the gas disk mass $M_{\rm gas}$ and the $\coo$\,(2-1) line luminosity, with the inferred $\coo$ line optical depth encoded in the curve transparency.
    Each panel corresponds to a different stellar mass ($M_\star=0.1$, $0.3$, $0.5$, $0.7$, $1.0$, and $2.0\,M_\odot$).
    Colored curves show grid models with different characteristic radii $R_{\rm c}$, while shaded bands indicate the variation with dust-to-gas ratio $\varepsilon$ (see Appendix~\ref{appendix:additional_model_figure} for details on the dependence of $L_{\coo}$ on $R_{\rm c}$ and $\varepsilon$).
    The transparency of each model segment indicates the line optical depth.
    More opaque segments correspond to more optically thin emission, while more transparent segments show increasingly higher optical depth (see Appendix~\ref{appendix:optical_depth}).
    Stars mark the locations of the observational sample and are labeled in order from the brightest to the faintest $L_{\rm C^{18}O}$, following Figure~\ref{fig:compare_lit}.
    Orange stars indicate sources whose $\mgas$ is inferred from the $\coo$\,(2-1) line. 
    Blue stars indicate sources whose $\mgas$ is inferred from the $\coo$\,(3-2) line when a (2-1) detection is unavailable -- for display purposes, we show their converted $\coo$\,(2-1)-equivalent luminosities, $L_{\coo\,(2-1)}^{\rm equiv} = L_{\coo\,(3-2)}/4.5$ (see Appendix~\ref{appendix:additional_model_figure}, Figure~\ref{fig:lc18o_32_mgas_withdata}, for the same plot on $L_{\rm C^{18}O\,(3-2)}$).
  }
  \label{fig:lc18o_mgas_withdata}
\end{figure*}

Figure~\ref{fig:lc18o_mgas_withdata} illustrates the relation between $L_{\rm C^{18}O}$ and $M_{\rm gas}$ in the \texttt{DiskMINT-GARDEN} grid.
The transparency of the model curves demonstrates the optical depth.
More opaque (saturated) segments correspond to optically thin emission, while more transparent (faded) segments indicate the majority of their disk masses are traced by $\coo$ with higher optical depth (see Appendix~\ref{appendix:optical_depth} for details). 

Additional model dependencies of the $L_{\rm C^{18}O}$--$M_{\rm gas}$ relation (e.g., disk size and dust-to-gas mass ratios) are discussed in Appendix~\ref{appendix:additional_model_figure}.

The analyses above define the regime where the \texttt{DiskMINT-GARDEN} inference is robust.
In the following section, we apply our framework to a compilation of protoplanetary disks with archival ALMA measurements of dust continuum and $\coo$ line emission.

\section{Application}
\label{sec:application}

We apply the \texttt{DiskMINT-GARDEN} inference framework to a sample of protoplanetary disks with published ALMA dust continuum and $\coo$ line measurements with ${\rm SNR}>3$.

Our sample consists of 34 disks: (i) 14 disks in the Lupus and Upper~Sco star-forming regions from the recent ALMA Large Program AGE-PRO \citep[][]{zhang_alma_2025, agurto-gangas_alma_2025, Deng_2025_AGEPRO_III_Lupus, vioque_alma_2025}; (ii) 14 large disks from the exoALMA Large Program \citep[][]{teague_exoalma_2025, galloway-sprietsma_exoalma_2025, trapman_exoalma_2025}, supplemented by additional archival continuum and $\coo$ measurements from the literature \citep[][]{flaherty_measuring_2020, qi_probing_2019, sturm_disentangling_2023, ribas_alma_2023, stapper_constraining_2024, loomis_unbiased_2020, qi_chemical_2015}; (iii) 6 disks drawn from DSHARP \citep[][]{andrews_disk_2018, huang_disk_2018}, MAPS \citep[][]{oberg_molecules_2021}, and other archival ALMA observations \citep[][]{paneque-carreno_directly_2023, van_der_marel_resolved_2016}.

This sample spans stellar masses $M_\star \sim 0.2$--$2.0\,M_\odot$, dust disk sizes $R_{\rm dust}\sim 10$--$500\,{\rm au}$, and ages of $\sim$1--10\,Myr.
For the analysis, we adopt ALMA Band~6 (1.3\,mm) continuum measurements together with the observed $\coo$\,(2-1) line when available; and otherwise we use the observed $\coo$\,(3-2) line. 
A small subset (five) of sources in our sample lack a detected $\coo$\,(2-1) line but have a $\coo$\,(3-2) detection. 
For them, $\mgas$ is inferred directly from the $\coo$\,(3-2) luminosity using the corresponding \texttt{DiskMINT-GARDEN} grid predictions for the (3-2) transition.

Basic target information and the \texttt{DiskMINT-GARDEN} input parameters ($L_{\coo}$, $L_{\rm mm}$, $R_{\rm dust, 90}$) are summarized in Appendix~\ref{appendix:obs_info}, Table~\ref{tab:disk_examples}.

\subsection{Inferred disk physical parameters}
\label{subsec:inferred_results}

Using the observed quantities ($L_{\coo}$, $L_{\rm mm}$, $R_{\rm dust}$) as inputs, we infer the corresponding disk physical parameters ($\mgas$, $\mdust$, and $R_{\rm c}$) for each target using the trained regression model.
The inferred parameters are summarized in Table~\ref{tab:disk_examples}, and the resulting disk masses are shown in Figure~\ref{fig:lc18o_mgas_withdata} together with the underlying model grid predictions.

For display purposes only, for the five sources with only $\coo$\,(3-2) detections, we convert $L_{\coo\,(3-2)}$ to a $\coo$\,(2-1)-equivalent luminosity, $L_{\coo\,(2-1)}^{\rm equiv} = L_{\coo\,(3-2)}/4.5$, so that all sources can be shown on a common axis in Figure~\ref{fig:lc18o_mgas_withdata}. The factor of 4.5 is derived assuming an optically thin LTE line luminosity ratio at $T_{\rm gas}\sim 35\,{\rm K}$, which is consistent with the bulk $\coo$-emitting layer and supported by the mean ratio in our thermochemical model grid. These sources are marked with blue stars in Figure~\ref{fig:lc18o_mgas_withdata}; their observed $L_{\coo\,(3-2)}$ values and the corresponding inferences based on the model $L_{\coo\,(3-2)}$ grid are shown in Appendix~\ref{appendix:additional_model_figure}, Figure~\ref{fig:lc18o_32_mgas_withdata}.

Only two disks (ID 1: HD~34282; 2: MWC~480) enter the range of models approaching gravitational instability (GI; Appendix~\ref{appendix:additional_model_figure}).
Five additional disks (ID 21: TW~Hya; 22: PDS~66; 23: Sz~71; 24: Sz~65; 26: J16221532-2511349) approach a relatively optically thick regime (Figure~\ref{fig:lc18o_mgas_withdata}), in which the $\coo$ emission starts to saturate and the line luminosity becomes less sensitive to $\mgas$. 
Even in these cases, however, the gas temperature is similar to the area-weighted dust temperature in the emitting regions \citep[see][]{Ruaud2022} and our self-consistent models infer gas masses consistent with the dynamical and HD-based estimates presented below.

\subsection{Comparison to dynamical masses}
\label{subsec:comp_dynmass}

\begin{figure}
  \centering
  \includegraphics[width=1.0\linewidth]{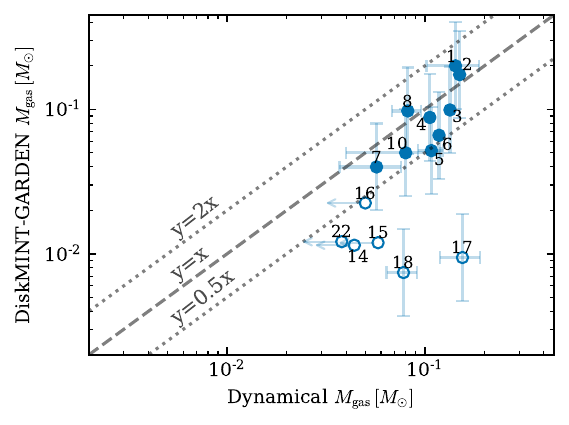}
  \caption{
      Comparison between gas disk masses inferred from \texttt{DiskMINT-GARDEN} and dynamical estimates.
      Filled circles show targets with reliable dynamical $M_{\rm gas}$, while open circles mark unreliable dynamical measurements \citep[][]{longarini_exoalma_2025}; left-pointing arrows indicate dynamical upper limits.
      Error bars denote uncertainties in both axes.
      The black dashed line shows $y=x$, and black dotted lines indicate factors of two differences.
      Numbers label sources in order of decreasing $L_{\rm C^{18}O}$, and their names are shown in Figure~\ref{fig:compare_lit} and Table~\ref{tab:disk_examples}. 
    }
  \label{fig:compare_dyn}
\end{figure}

We compare \texttt{DiskMINT-GARDEN} inferred $\mgas$ to  dynamical gas masses \citep[][]{veronesi_dynamical_2021, lodato_dynamical_mass_2023, martire_rotation_2024, longarini_exoalma_2025} in Figure~\ref{fig:compare_dyn}.
Filled circles denote disks with dynamical masses retrieved to high accuracy, while open circles denote disks with more uncertain dynamical mass estimates.

Dynamical measurements provide only upper limits (left-pointing arrows) for four disks (14: RXJ~1852.3$-$3700, 15: V~4046~Sgr, 16: AS~209, 22: PDS~66). 
Two disks (17: AA~Tau and 18: RXJ~1842.9$-$3532) have less reliable dynamical mass estimates for the following reasons (C.~Longarini, private communication). 
AA~Tau is a highly inclined disk, making it difficult to extract reliable $^{12}$CO and $^{13}$CO rotation curves and thus increasing the uncertainty in its dynamical gas-mass estimate \citep{longarini_exoalma_2025}.
The reported dynamical value implies $\mgas \sim 0.25\,M_\star$, high enough that such a disk would be gravitationally unstable.
We note that the \texttt{DiskMINT-GARDEN} inferred mass for this disk is more typical, $\sim 0.01\, M_{\odot}$.
For RXJ~1842.9$-$3532, the vertical emission surfaces of $^{12}$CO and $^{13}$CO overlap in the inner disk, complicating the recovery of the temperature structure and, in turn, the dynamical mass inference \citep{longarini_exoalma_2025, trapman_exoalma_2025}. 
Again, the disk mass inferred from \texttt{DiskMINT-GARDEN} is lower, $\sim 0.007 \, M_{\odot}$.

Overall, \texttt{DiskMINT-GARDEN} results agree with the dynamical estimates to within a factor of $\lesssim 2$ for the targets with reliable dynamical measurements (filled circles, in Fig.~\ref{fig:compare_dyn}). This level of agreement is fully consistent with the typical systematic uncertainties of order a factor of $\sim 2$ associated with both approaches \citep{lodato_dynamical_mass_2023, martire_rotation_2024, deng_diskmint_2025}.

Dynamical $\mgas$ constraints are only available for relatively massive disks ($\mgas \gtrsim 0.05\,M_\star$), which likely represent only a small fraction of the overall disk population.
For example, under the commonly assumed gas-to-dust mass ratio of 100, disks at these masses account for only $\sim 3\%$ of the population \citep[e.g.,][]{Pascucci_mass_2016, Manara2022DemographicsFormation}.
In this context, the agreement observed in Figure~\ref{fig:compare_dyn} is particularly valuable because it demonstrates that \texttt{DiskMINT-GARDEN} 
is consistent with an independent constraint and can reproduce the dynamical mass in the regime where the dynamical mass measurements are robust and where such tests are currently feasible.
Importantly, these comparisons demonstrate that \texttt{DiskMINT-GARDEN} presents a valuable method that can be used to extend $\mgas$ estimates into the lower-mass regime inhabited by most of the disk population and where dynamical constraints are not feasible. 

\subsection{Comparison to HD masses}
\label{subsec:comp_HD}

The HD\,(1-0) transition at 112\,\micron\ is often considered one of the most direct tracers of the total gas disk mass because it is optically thin and too light a molecule to freeze out in disks. 
However, HD observations require far-infrared facilities which are currently unavailable, and HD emission has been detected by \textit{Herschel} in only three (relatively massive) disks to date: GM~Aur, DM~Tau, and TW~Hya \citep[][]{Bergin_HD_2013, McClure_HD_2016}.

The HD-based masses ($M_{\rm gas,\,HD}$) for GM~Aur ($2.5\times10^{-2} - 2\times10^{-1}\,M_\odot$) and DM~Tau ($1\times10^{-2}-5\times10^{-2}\,M_\odot$) from \citet{McClure_HD_2016} agree with our inferred $\mgas$ within a factor-of-two. 
The $M_{\rm gas,\,HD}$ estimate for TW~Hya spans nearly an order of magnitude, from $6\times10^{-3}\,M_\odot$ to $6\times10^{-2}\,M_\odot$ \citep[][]{Bergin_HD_2013, Trapman_HD_2017}. 
Although this range is still consistent with our estimate, it illustrates the sensitivity of $M_{\rm gas,\,HD}$ estimates to assumptions about disk temperature and vertical structure.
The upper level energy of the HD\,(1-0) transition ($E_u/k \sim 128$\,K) is considerably higher than the temperature at the cold midplane ($T\lesssim20$\,K) where most of the gas mass lies. As a result, HD emission is sensitive to the disk thermal structure and accurate interpretation of HD emission requires thermochemical models with well-constrained vertical density and temperature profiles \citep[see discussions in e.g., ][]{Trapman_HD_2017, Ruaud2022}.

\subsection{Comparison to DALI masses}
\label{subsec:comp_dali}

For the large populations of disks with masses $\lesssim 0.05\,M_\star$, where dynamical constraints are unavailable, gas masses can currently only be estimated by thermochemical modeling. 

We next compare $\mgas$ inferred with \texttt{DiskMINT-GARDEN} to $\mgas$ estimates derived using the \texttt{DALI} thermochemical framework, which, like \texttt{DiskMINT}, also includes CO isotope chemistry and has been applied to most of the disks in our sample \citep[e.g.,][]{trapman_exoalma_2025, Trapman_AGEPRO_V_gas_masses}. 
There are, however, significant differences in the approaches between the two modeling frameworks.
\texttt{DiskMINT} models the disk structure consistently by iterating the gas density, temperature, chemistry, and dust settling simultaneously and includes grain surface chemistry as described in detail in Section~\ref{sec:intro}.

\texttt{DALI}, on the other hand, uses a parametrized, fixed initial density structure and only iterates over gas chemistry and temperature. 
Grain surface chemistry is not included, but CO is considered to freeze-out at a temperature $\sim$ 20K \citep[e.g.,][]{Trapman_AGEPRO_V_gas_masses}. 
\texttt{DALI} infers $\mgas$ by combining CO isotopologue emission with $\ntwohp$ emission. 
$\ntwohp$ abundances are obtained by post-processing \texttt{DALI} results with a smaller chemical network; however, the resulting $\ntwohp$ line emission is typically lower than observed when ISM-like elemental abundances of C and O are assumed \citep[][]{trapman_novel_2022}.
Since CO in the gas-phase is the main destroyer of $\ntwohp$, C or CO depletion factors are introduced to boost $\ntwohp$ emission and also reconcile the observed CO isotopologue emission \citep[][]{trapman_exoalma_2025, Trapman_AGEPRO_V_gas_masses}. 

\begin{figure}
    \centering
    \includegraphics[width=\linewidth]{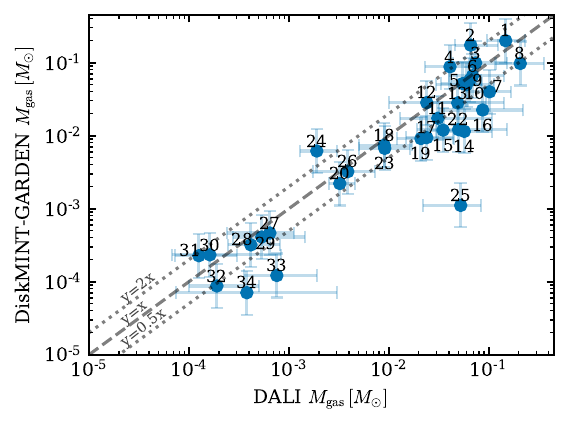}
    \caption{
    Comparison between gas disk masses inferred from \texttt{DiskMINT-GARDEN} and \texttt{DALI} estimates. 
    Notations follow Figure~\ref{fig:compare_dyn} and~\ref{fig:compare_lit}.
    }
    \label{fig:DALI-MINT}
\end{figure}

\begin{figure*}
  \centering
  \includegraphics[width=1.0\linewidth]{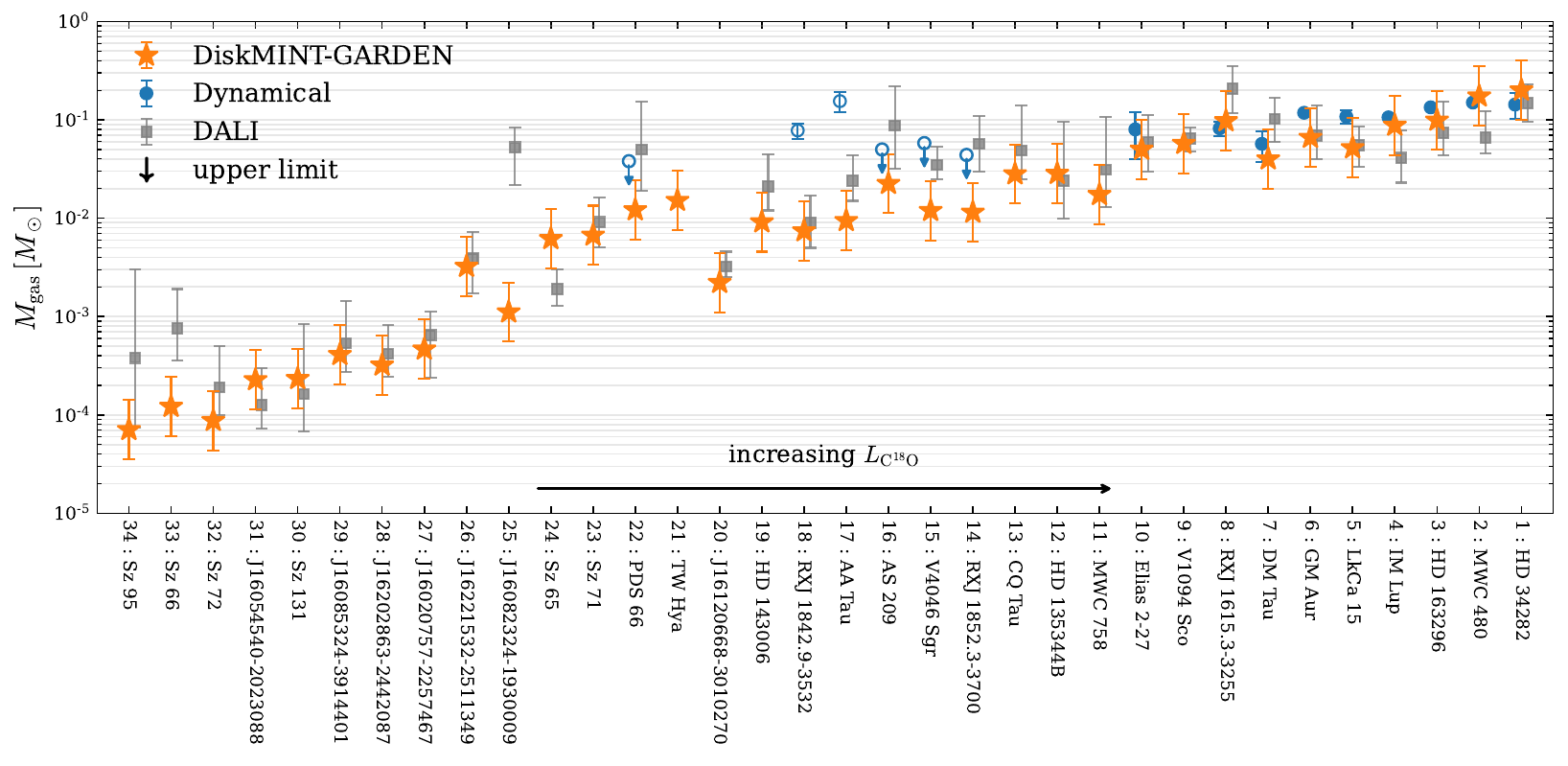}
  \caption{
    Comparison of $\mgas$ inferred with \texttt{DiskMINT-GARDEN} and independent literature measurements.
    For each disk, $\mgas$ inferred from the \texttt{DiskMINT-GARDEN} regression model are shown as orange stars, while gray squares indicate gas masses derived from DALI-based chemical modeling by \citet{Trapman_AGEPRO_V_gas_masses, trapman_exoalma_2025} and blue circles show dynamical gas mass estimates by \citet{longarini_exoalma_2025, lodato_dynamical_mass_2023, martire_rotation_2024, veronesi_dynamical_2021}.
    A conservative factor-of-two uncertainty is adopted for the inferred $\mgas$ from \texttt{DiskMINT-GARDEN}.
    Literature gas mass uncertainties reflect the quoted lower and upper bounds from the original studies.
    Hollow markers denote sources flagged as potentially unreliable estimates, and downward arrows indicate upper limits according to the original works.
    Disks are ordered along the horizontal axis by increasing observed $L_{\coo}$, and a unique ID is assigned to each disk in this work.
  }
  \label{fig:compare_lit}
\end{figure*}

Figure~\ref{fig:DALI-MINT} compares \texttt{DALI} results with ours, while Figure~\ref{fig:compare_lit} summarizes the comparison between \texttt{DALI} and \texttt{DiskMINT-GARDEN} together with the available dynamical estimates.
The gas masses from \texttt{DiskMINT-GARDEN} are typically within a factor of $\lesssim 2$ compared to \texttt{DALI} masses, comparable to the systematic uncertainties associated with disk mass measurements. 
A small subset of disks, notably ID 25: J16082324$-$1930009, 33: Sz~66, and 34: Sz~95, have larger differences, but we consider the agreement in general to be good. 

We note, however, that although the integrated gas disk masses agree, the underlying physical and chemical structures differ significantly between the two approaches. 
Since our disk models impose vertical hydrostatic equilibrium, the gas density distribution in the observable surface layers (i.e., the emitting layers) differs from the isothermal density profile adopted in \texttt{DALI} \citep[see][for a more detailed discussion]{Ruaud2022, deng_diskmint_2023}. 

\texttt{DiskMINT} and the chemical model it is based on \citep[][]{Ruaud2019} also differ in that there is no inclusion of a CO depletion factor. 
Therefore, although the derived column densities of $\coo$ are similar, in \texttt{DiskMINT} this is due to the calculated CO vertical snowline being located at higher $z$ (corresponding to $T \sim 35$\,K; \citealt{Ruaud2022}), and the overall CO abundance is not reduced ad hoc. 
In \texttt{DALI} models, by contrast, the imposed CO snowline resides at lower $z$ (corresponding to $T\sim 20$\,K) and the depletion factor derived by matching $\ntwohp$ emission results in a reduced abundance of CO  uniformly at all heights above the snowline (typically by a factor of $\sim$ 10, see \citealt{Trapman_AGEPRO_V_gas_masses}). 
According to \citet{trapman_novel_2022}, this depletion factor accounts for all processes that can result in a lower CO abundance, including errors in computing the snow line, disk chemical evolution, and planetesimal formation. 
However, since we recover disk masses similar to \texttt{DALI} without invoking any additional depletion factor, and with no other chemical sequestration besides grain surface chemistry and a more precise accounting for the vertical CO snowline (due to conversion of CO on water ice into the more tightly bound CO$_2$ molecule), we conclude that 
\texttt{DALI} models need $\ntwohp$ to improve their determinations of available gas-phase CO (see Appendix~\ref{appendix:compare_column_density}). 

More importantly, the fact that the inferred $\mgas$ differ by only a factor of two suggests that there is no significant processing of CO due to disk evolution or planetesimal formation.

We note that introducing a uniform depletion factor at all disk heights is physically distinct from a model without depletion, even if both are tuned to reproduce the same observed CO isotopologue emission.
In models with a global depletion, the reduced abundance of CO -- which is one of the strongest molecular coolants -- alters the thermal and chemical balance throughout the disk.
The thermal and density structure, molecular abundances, and the emission from all other species are therefore coupled to this choice and need to be recomputed for a consistent solution.
In \texttt{DALI}, the density structure was already parameterized and is not updated with the chemistry and temperature iterations; introduction of the depletion factors now also makes the temperature solution inconsistent and arbitrary (see recent work by \citealt{arabhavi_minds_2025} where \texttt{ProDiMo} is used to show that C and O abundance variations strongly impact disk temperatures).
In addition, such disk structures, where the true CO snowline is replaced by a freeze-out temperature and a uniform depletion of CO above, will result in very different chemistry in the emission layers, affecting the interpretation of all molecular line emission, since CO is a dominant reservoir of C and O in the disk. 
Lastly, as also pointed out by \citet{van_t_hoff_robustness_2017, anderson_new_2022, trapman_novel_2022}, $\ntwohp$ is strongly dependent on the poorly constrained ionization rate in the disk due to cosmic rays or X-rays. 

\section{Future Applications of \texttt{DiskMINT}}
\label{sec:future}

A key advantage of \texttt{DiskMINT} lies in its observational requirements.
\texttt{DiskMINT-GARDEN} infers $\mgas$ using only widely available millimeter continuum and CO isotopologue measurements, enabling rapid and homogeneous gas-mass estimates across large disk samples.
We note that acquiring $\ntwohp$ requires a dedicated ALMA setup with integration time comparable to those required to detect $\coo$ \citep[e.g.,][]{zhang_alma_2025}, effectively doubling the requested exposure time.

Comparisons in Section~\ref{sec:application} demonstrate that \texttt{DiskMINT-GARDEN} provides $\mgas$ estimates broadly consistent with existing thermochemical and dynamical techniques, while offering a less time-demanding alternative for extending $\mgas$ measurements to less massive and more representative disk populations.
Future observations will provide new opportunities to further test and extend the \texttt{DiskMINT} framework, particularly through spatially resolved observations of CO isotopologues, and complementary tracers of gas disk mass such as hydrogen deuteride (HD).

Beyond global gas-mass estimates, spatially resolved observations offer the possibility of directly constraining the radial structure of disks.
In our recent work on IM~Lup \citep[][]{deng_diskmint_2025}, we demonstrated that \texttt{DiskMINT} can simultaneously model spatially resolved continuum and CO isotopologue emission to recover both the gas and dust surface density profiles, $\sigmagas(r)$ and $\sigmadust(r)$.
This enables the derivation of the radial dust-to-gas ratio $\varepsilon(r)$, which provides direct insight into the physical processes shaping disk evolution.
In particular, radial variations in $\varepsilon$ can trace dust growth, radial drift, and local conditions favorable for planetesimal formation through mechanisms such as streaming instability and gravitational instability \citep[e.g.,][]{youdin_streaming_2005, li_thresholds_2021, birnstiel_disk_2018, birnstiel_dust_2024}.

Upcoming far-infrared facilities, including the  \textit{POEMM}\footnote{\url{https://poemm.astro.cornell.edu/}} mission and the proposed \textit{PRIMA}\footnote{\url{https://prima.ipac.caltech.edu/}} observatory \citep[][]{Glenn_PRIMA_2025}, are expected to significantly expand the number of disks with detectable HD emission.
Combining HD measurements with CO isotopologue observations will provide strong constraints on both the disk thermal structure and the gas surface density distribution.
HD and $\coo$ emission arise from similar intermediate layers between the optically thin surface and the cold midplane.
Comparing these tracers, therefore, provides a powerful diagnostic of the disk thermochemical structure.

Any discrepancies between HD- and CO-based gas-mass and surface density estimates, even if they are found to be limited based on this work, may reveal the true extent of additional physical or chemical processes, such as CO sequestration into planetesimals or chemical conversion into other carbon-bearing species \citep[e.g.,][]{Bergin_n_Williams_mass_2017, Bosman_n_Banzatti_twhya_2019}.

Future extensions of the \texttt{DiskMINT} framework will incorporate additional chemical pathways and molecular tracers, enabling simultaneous modeling of HD and multiple CO isotopologues ($\twelveco$, $\thirteenco$, and $\coo$) that probe different vertical layers of the disk. Together, these developments will allow \texttt{DiskMINT} to connect survey-scale gas-mass estimates from \texttt{DiskMINT-GARDEN} with detailed thermochemical modeling of individual systems, providing a unified framework for interpreting gas disk masses across a wide range of observational datasets.

\section{Summary}
\label{sec:summary}

We present \texttt{DiskMINT-GARDEN}, a grid-based and computationally efficient framework for inferring protoplanetary gas disk masses from widely available millimeter continuum and CO isotopologue observations.
\texttt{DiskMINT-GARDEN} builds on \texttt{DiskMINT} \citep[][]{deng_diskmint_2023, deng_diskmint_2025}, which self-consistently couples disk vertical structure, radiative transfer, and a reduced chemical network optimized for CO isotopologues.
The main results of this paper are:

\begin{enumerate}
  \item We construct a grid of self-consistent models spanning a broad physically motivated range of stellar mass ($M_\star = 0.1$--$2.0\,M_\odot$), gas disk mass ($M_{\rm gas} = 10^{-5}$--$10^{-1}\times M_\star$), dust-to-gas mass ratio ($\varepsilon = 0.003$--0.1), and characteristic radius ($R_{\rm c}=10$--$300$\,au).
  For each model, we compute synthetic observables commonly measured in ALMA surveys, including $L_{\rm mm}$, $L_{\coo}$, and $R_{\rm dust, 90}$.

  \item We test and include a supervised machine-learning regression model to map the observable vector $\mathbf{O}=(L_{\coo},\,L_{\rm mm},\,R_{90,\rm dust})$ to the physical parameter vector $\boldsymbol{\Theta}=(M_{\rm gas},\,\varepsilon,\,R_{\rm c})$ at fixed $M_\star$ based on the model grids.
  For $M_{\rm gas}$, measurement uncertainties are propagated via Monte Carlo sampling in observable space, typically contributing $\sim 5\%$ uncertainty and up to $\lesssim 25\%$.
  Uncertainty associated with the regression mapping is assessed using a reproducible 90/10 training--validation split and is $\lesssim 10\%$.
  Finally, to account for dominant systematic uncertainties from disk structure and chemistry (which are difficult to quantify and are not always included in published error budgets), we adopt a conservative factor-of-two uncertainty on the inferred gas masses.

  \item Applying \texttt{DiskMINT-GARDEN} to an archival sample of 34 disks with $>3\sigma$ detections in both ALMA continuum and C$^{18}$O line emission, we find that the inferred gas masses are broadly consistent with independent dynamical constraints and with thermochemical modeling with HD within the uncertainties of a factor of $\sim 2$.
  A small subset of disks shows larger discrepancies (e.g., due to disk geometric effects), and these cases highlight the importance of spatially resolved, disk-specific modeling such as using \texttt{DiskMINT}.

  \item Comparisons with $\mgas$ inferred from \texttt{DALI} with CO+$\ntwohp$ show that the previously inferred C or CO depletion factors are not necessary when including grain-surface chemistry. 
  Therefore, the current data are consistent with little change in disk chemistry due to disk evolutionary processes such as planetesimal formation.

  \item \texttt{DiskMINT-GARDEN} requires only widely available ALMA continuum and $\coo$ measurements -- without additional molecular tracers such as $\ntwohp$ that roughly doubles the required observing time -- making it directly applicable to existing and forthcoming large disk surveys.

\end{enumerate}

\texttt{DiskMINT-GARDEN} enables rapid and homogeneous gas-mass estimates for large ALMA surveys, making it well suited for population-level studies of disk evolution across ages and environments.
Furthermore, our \texttt{DiskMINT} framework can be extended to incorporate additional observables -- for example, multi-band continuum slopes and other molecular tracers (e.g., HD) -- thereby expanding the parameter space that can be constrained and reducing chemical degeneracies.

The \texttt{DiskMINT-GARDEN} model grids, including their inputs (e.g., $M_\star$, $\mgas$, $\mdust$, $R_{\rm c}$) and outputs (e.g., $L_{\coo\,(2-1)}$, $L_{\coo\,(3-2)}$, $L_{\rm mm}$, $R_{\rm 90,\,dust}$), together with the trained regression model and inference tools, are released as part of the open-source \texttt{DiskMINT} \texttt{v1.7.0} on \texttt{GitHub}.
For users who wish to go beyond the grid-level inference, the best-fit parameters
returned by \texttt{DiskMINT-GARDEN} can serve directly as inputs to a full
\texttt{DiskMINT} run, providing the complete disk thermochemical structure --
including the gas temperature $T_{\rm gas}$ and density $n_{\rm gas}$ profiles --
for detailed modeling of individual targets.

\section*{Acknowledgments}

The authors thank the anonymous referee and Dr. C. Longarini for helpful suggestions and comments.
D.D., U.G., and I.P. acknowledge support from the NASA/XRP research grant 80NSSC20K0273.
D.D. and I.P. also acknowledge support from the Collaborative NSF Astronomy \& Astrophysics Research grant (ID: 2205870). 
I.P. also acknowledges partial support by the National Aeronautics and Space Administration under agreement No. 80NSSC21K0593 for the program ``Alien Earths''.

This work made use of the High Performance Computing (HPC) resources, which are supported by the University of Arizona TRIF, UITS, and Research, Innovation, and Impact (RII), and maintained by the University of Arizona Research Technologies department.

This paper makes use of the following ALMA data: ADS/JAO.ALMA
\#2012.1.00158.S,
\#2012.1.00422.S,
\#2012.1.00870.S,
\#2015.1.00192.S,
\#2015.1.00678.S,
\#2015.1.00686.S,
\#2015.1.01017.S,
\#2016.1.00311.S,
\#2016.1.00484.L,
\#2016.1.00629.S,
\#2016.1.00724.S,
\#2017.1.00069.S,
\#2017.1.00940.S,
\#2017.1.01404.S,
\#2017.1.01419.S,
\#2018.1.00689.S,
\#2018.1.00945.S,
\#2018.1.01055.L,
\#2019.1.01683.S,
\#2021.1.00128.L,
\#2021.1.01123.L,
\#2022.1.00485.S,
\#2023.1.00334.S.
ALMA is a partnership of ESO (representing its member states), NSF (USA) and NINS (Japan), together with NRC (Canada), NSTC and ASIAA (Taiwan), and KASI (Republic of Korea), in cooperation with the Republic of Chile. The Joint ALMA Observatory is operated by ESO, AUI/NRAO and NAOJ. The National Radio Astronomy Observatory is a facility of the National Science Foundation operated under cooperative agreement by Associated Universities, Inc.

\textit{Facility:} SMA, ALMA.

\textit{Software:}
All figures were generated with the \texttt{Python}-based package \texttt{MATPLOTLIB} \citep{Hunter2007}.
This research made use of \texttt{RADMC-3D} \citep[][]{Dullemond_radmc-3d_2012}, \texttt{Optool} \citep[][]{dominik_optool_2021}, \texttt{GoFish} \citep[][]{teague_gofish_2019}, \texttt{Astropy} \citep{astropy:2018},  \texttt{Scipy} \citep{2020SciPy-NMeth}, \texttt{Scikit-learn} \citep[][]{scikit-learn}, and \texttt{XGBoost} \citep[][]{chen_xgboost_2026}.
The models in this work are created with \texttt{DiskMINT} \citep[][]{deng_diskmint_2023, deng_diskmint_2025} v1.7.0 (we save this frozen version on \texttt{zenodo} \citealt{Deng_2026_diskmint_zenodo}), and they can also be downloaded from our public \texttt{GitHub} repository: \url{https://github.com/DingshanDeng/DiskMINT}.

\newpage

\appendix
\restartappendixnumbering
\twocolumngrid 


\section{Optical Depth Calculation}
\label{appendix:optical_depth}

The reliability of millimeter continuum and $\coo$ emission as disk-mass tracers depends on whether the emission remains optically thin.
Once optical depth becomes significant, the luminosity can saturate and become less sensitive to the total mass.
In Section~\ref{subsubsec:optical_depth}, we argued that this effect is most relevant for compact, high-surface-density disks.

To quantify this behavior across the full \texttt{DiskMINT-GARDEN} grid, we define the optically thin mass fraction, $f_{\rm thin}$, for both the 1.3\,mm continuum and the $\coo$ line,
\begin{equation}
\label{equation:fthin}
    f_{\rm thin} = \frac{M_{\mathrm{thin}}}{M_{\rm total}}.
\end{equation}
For each model, we compute the radial optical-depth profile from the adopted surface-density distributions, determine the radius where $\tau=1$, and evaluate the fraction of the total mass beyond that radius that is optically thin ($M_{\mathrm{thin}}$).
This provides a simple, uniform diagnostic of when the corresponding luminosity begins to depart from optically thin scaling.
The detailed calculations for the continuum and line cases are described separately below.

\subsection{Optical depth of the dust continuum emission}

To assess whether the millimeter dust continuum emission is optically thin or thick in our disk models, we estimate the dust optical depth at 1.3\,mm using the modeled surface density profiles.
For the dust surface density, following Equation~\ref{equation:sigma},
\begin{equation}
  \Sigma_{\rm d}(r) = \Sigma_{\rm d,\,1}
  \left(\frac{r}{1\,{\rm au}}\right)^{-1}
  \exp\!\left(-\frac{r}{R_{\rm c}}\right),
\end{equation}
corresponding to a power-law index $\gamma=1$ and an exponential taper beyond the characteristic radius $R_{\rm c}$.

The normalization $\Sigma_1$ is determined by requiring that the surface density profile integrates to the total dust mass $M_{\rm dust}$,
\begin{equation}
  M_{\rm dust} = 2\pi \int_0^{\infty} \Sigma_{\rm d}(r)\, r\, dr.
\end{equation}

With the dust properties adopted in this work (Table~\ref{tab:grid_props}), the dust mass absorption opacity at 1.3\,mm of $\kappa_{1.3}\sim1.0~{\rm cm^2\,g^{-1}}$, the vertical dust continuum optical depth at radius $r$ is then
\begin{equation}
  \tau_{1.3}(r) = \kappa_{1.3}\, \Sigma_{\rm d}(r).
\end{equation}

Then, we quantify the fraction of the dust mass that are optically thin by
\begin{equation}
  f_{\rm dust,\,thin} \equiv \frac{M_{\rm dust}(r>r_{\tau_{1.3}=1})}{M_{\rm dust}},
\end{equation}
where $r_{\tau=1}$ is the outermost radius at which $\tau_{1.3} \geq 1$.

\subsection{Optical depth of the line emission}

We estimate the C$^{18}$O line optical depth for each model using the 2D molecular
abundance distribution produced by the DiskMINT chemistry network.
Rather than assuming a fixed abundance, we read the spatially resolved
C$^{18}$O number density $n_{\rm C^{18}O}(r,z)$ directly from the chemistry output
and integrate vertically at each radius,
\begin{equation}
  N_{\rm C^{18}O}(r) = 2\int_0^{z_{\rm max}} n_{\rm C^{18}O}(r,z)\, dz,
  \label{eq:Nc18o}
\end{equation}
where the factor of 2 accounts for the symmetry of both disk surfaces.
This approach naturally incorporates the effects of CO freeze-out,
photodissociation, and other chemistry that shapes the vertical abundance
structure.

We check if the LTE approximation is valid by comparing the gas density in the $\coo$-emitting regions with the critical densities of the modeled transitions. Over the relevant 20--50\,K temperature range, $n_{\rm crit}\sim 1\times10^4\,{\rm cm^{-3}}$ for $\coo$\,(2--1) and $n_{\rm crit}\sim 3\times10^4\,{\rm cm^{-3}}$ for $\coo$\,(3--2). 
In the stress-test models with lowest-surface-densities, including $M_{\rm gas}=10^{-6}\,M_\odot$ and $R_{\rm c}=300\,{\rm au}$, the $\coo$-emitting regions remain at densities comparable to or above these critical densities. 
We therefore consider LTE adequate for the integrated line luminosities used in this survey-scale grid.

When in LTE, the column density in the lower rotational level
$\ell$ of the $J_u \to \ell$ transition is
\begin{equation}
  N_{\ell}(r) = N_{\rm C^{18}O}(r)\,
  \frac{g_{\ell}\,\exp\!\left[-E_{\ell}/k T_{\rm ex}(r)\right]}{Q(T_{\rm ex}(r))},
\end{equation}
where $T_{\rm ex}(r)$ is the column-density-averaged gas temperature from the
model at radius $r$, $E_\ell = hBJ_\ell(J_\ell+1)$ is the energy of the lower
level, and $Q(T)$ is the rotational partition function.

The line-center optical depth is
\begin{equation}
  \tau_0(r) =
  \frac{A_{u\ell}\,c^3}{8\pi\nu^3}\,
  \frac{g_u}{g_\ell}\,
  \frac{N_{\ell}(r)}{\Delta v(r)}\,
  \left[1-\exp\!\left(-\frac{h\nu}{kT_{\rm ex}(r)}\right)\right],
  \label{eq:tau0}
\end{equation}
where $\nu = 219.560\,\rm GHz$, $A_{u\ell} = 6.011\times 10^{-7}\,\rm s^{-1}$,
$g_u = 5$, and $g_\ell = 3$ are the spectroscopic parameters for the
C$^{18}$O\,(2--1) transition, and the $\Delta v(r)$ is the thermal line broadening.

From $\tau_0(r)$, we define the optically thin gas mass fraction
\begin{equation}
  f_{\rm gas,\,thin} \equiv \frac{M_{\rm gas}(r>r_{\tau_0=1})}{M_{\rm gas}},
\end{equation}
where $r_{\tau=1}$ is the outermost radius at which $\tau_0 \geq 1$.

\subsection{Dust-to-gas mass ratio}
\label{appendix:dust-to-gas_mass_ratio}

In addition to disk size ($R_{\rm c}$), the dust-to-gas mass ratio ($\varepsilon$) influences the thermochemical structure of the disk and therefore the emergent continuum and CO isotopologue emission.
Varying $\varepsilon$ modifies the amount of the dust particles, further affecting their UV shielding and disk thermal structure, which in turn affects both the excitation and optical depth of molecular lines.
Here we illustrate the sensitivity of $L_{\rm C^{18}O}$ and millimeter continuum emission to changes in $\varepsilon$ across the \texttt{DiskMINT-GARDEN} grid.

\begin{figure*}
  \centering
  \includegraphics[width=1.0\linewidth]{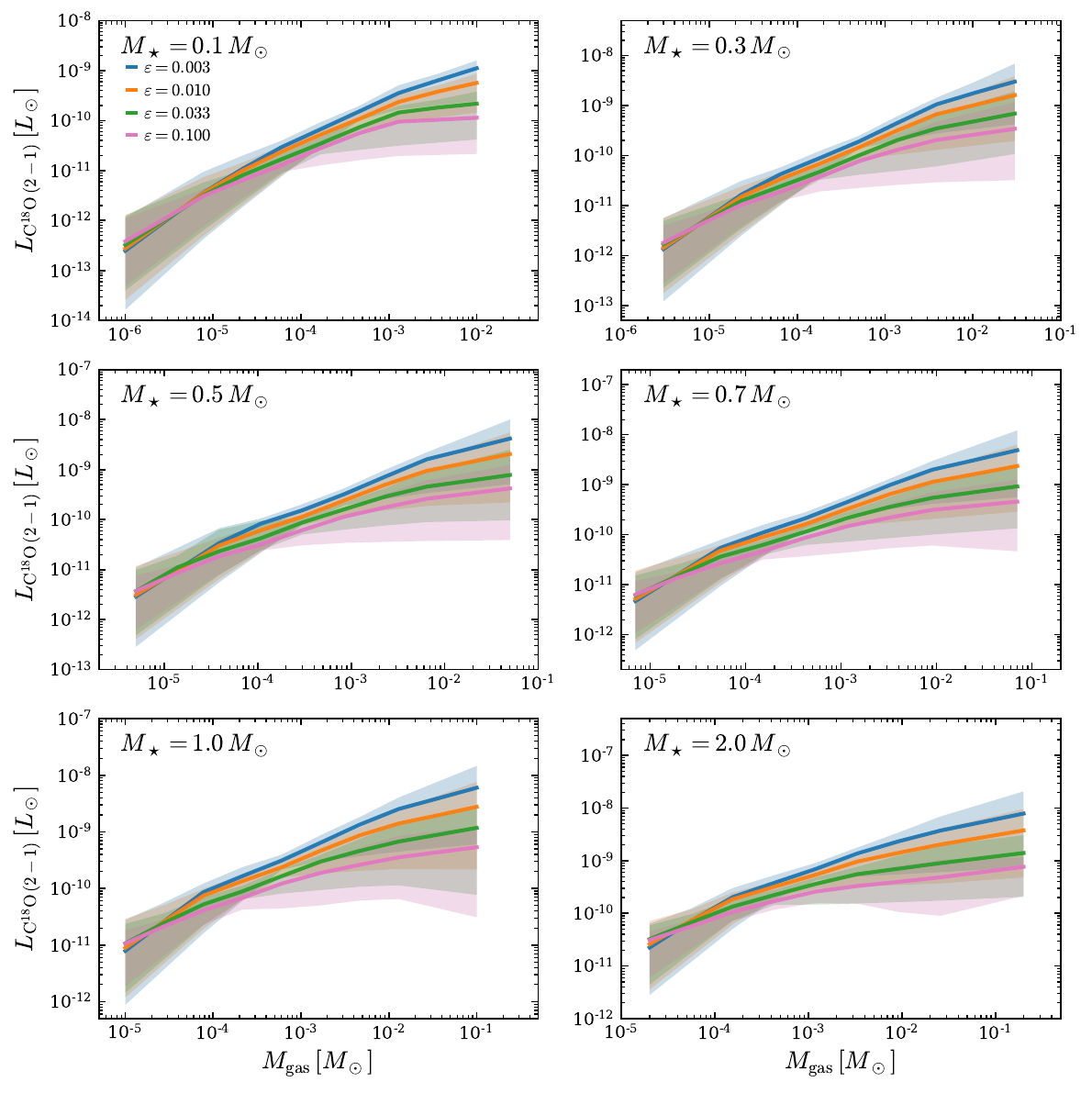}
  \caption{Grid-predicted relationship between the $\coo$\,(2-1) line luminosity and gas disk mass, shown in six panels for different stellar masses ($M_\star=0.1,\,0.3,\,0.5,\,0.7,\,1.0$ and $2.0\,M_\odot$). Colored curves indicate different dust-to-gas ratio $\varepsilon$ from 0.1 to 0.003. For each $\varepsilon$, the shaded band spans the range of models obtained by varying the characteristic radius $R_{\rm c}$, with the median value of each shade shown in the color solid lines.}
  \label{fig:grid_lc18o_mgas_dtgcolor}
\end{figure*}

Figure~\ref{fig:grid_lc18o_mgas_dtgcolor} shows the grid-predicted relationship between $L_{\rm C^{18}O}$ and $M_{\rm gas}$ for different values of $\varepsilon$.
While changes in $\varepsilon$ introduce measurable variations in line luminosity through altered temperature and shielding conditions, the overall scaling between $L_{\rm C^{18}O}$ and $M_{\rm gas}$ is primarily governed by disk size ($R_{\rm c}$), with $\varepsilon$ contributing a secondary modulation.

\begin{figure*}
  \centering
  \includegraphics[width=1.0\linewidth]{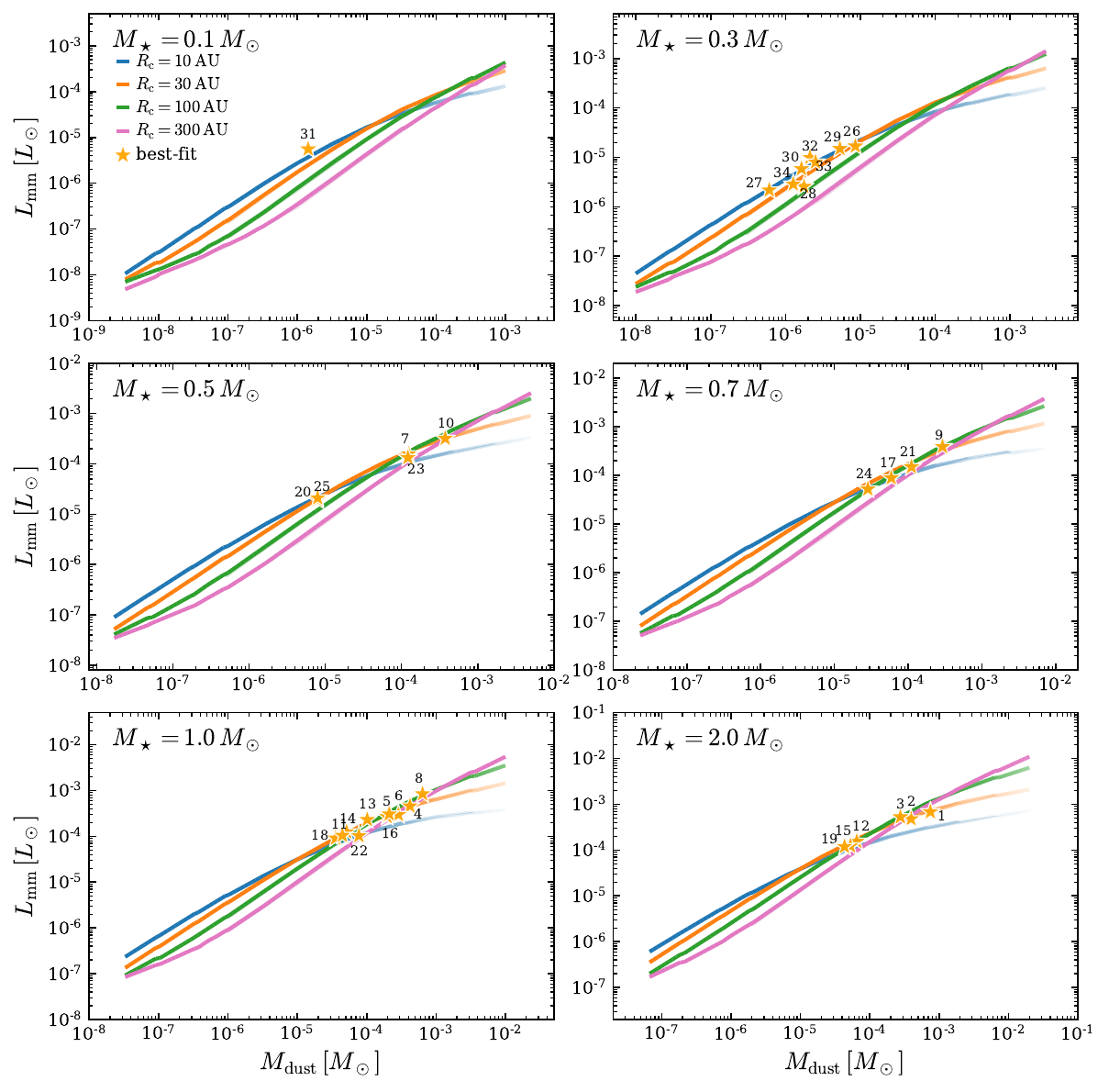}
  \caption{
    Relationship between the dust disk mass $M_{\rm dust}$ and the continuum luminosity, with the inferred continuum optical depth encoded by the curve transparency.
    Each panel corresponds to a different stellar mass ($M_\star=0.1$, $0.3$, $0.5$, $0.7$, $1.0$, and $2.0\,M_\odot$).
    Colored curves show grid models with different characteristic radii $R_{\rm c}$, while shaded bands indicate variations in the dust-to-gas ratio ($\varepsilon$).
    The transparency of each model segment is set by the estimated continuum optical depth $\tau_0$: more opaque segments correspond to more optically thin emission, while more transparent segments indicate higher optical depth and increasing saturation.
    Orange stars mark the locations of the observational sample, labeled in order from the brightest to the faintest $L_{\rm C^{18}O}$ (same as Figure~\ref{fig:compare_lit}). 
  }
  \label{fig:lmm_mdust_withdata}
\end{figure*}

Figure~\ref{fig:lmm_mdust_withdata} presents the corresponding relationship between dust mass $M_{\rm dust}$ and millimeter continuum luminosity.
As expected, continuum emission scales approximately linearly with $M_{\rm dust}$ in the optically thin regime, with deviations emerging only when the continuum becomes optically thick in compact, high-surface-density disks.

\section{XGBoost Implementation and Validation}
\label{appendix:ML_details}

\begin{figure*}
  \centering
  \includegraphics[width=1.0\linewidth]{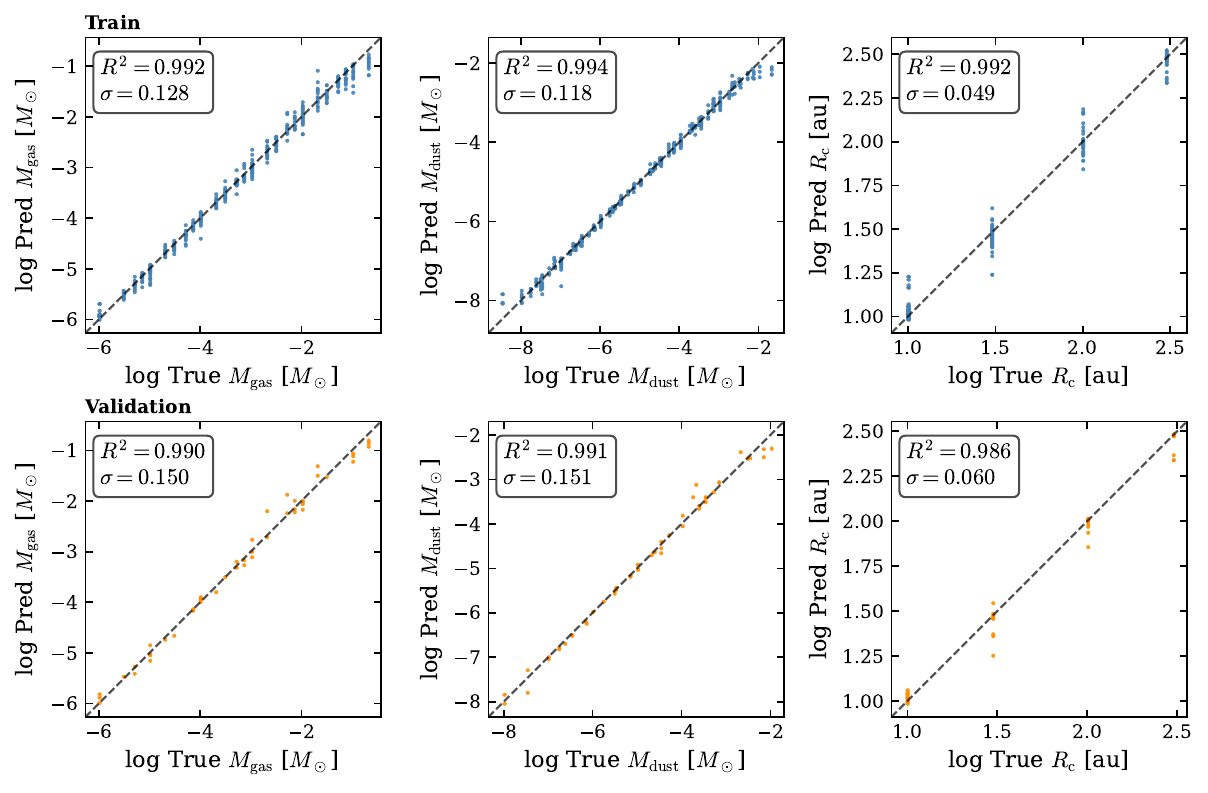}
  \caption{Validation of the regression model using a 90/10 train-validation split of the disk-model grid. Each column corresponds to one predicted physical parameter ($M_{\rm gas}$, $M_{\rm dust}$, and $R_{\rm c}$). The top row shows predictions for the training subset (90\% of the grid) and the bottom row for the held-out validation subset (10\% of the grid). In each panel, points compare the predicted values to the true grid values used for training; the dashed line indicates the one-to-one relation. We annotate the coefficient of determination ($R^2$) and the residual scatter ($\sigma$, computed in $\log_{10}$ space).}
  \label{fig:train_test_sample}
\end{figure*}

\subsection{Model Architecture and Hyperparameter Choices}

The machine-learning regression framework described in Section~\ref{subsec:infer_mass} is implemented using the gradient boosting library \texttt{XGBoost}\footnote{\url{https://github.com/dmlc/xgboost/?tab=readme-ov-file}} \citep{chen_xgboost_2026}.
\texttt{XGBoost} implements a parallelized gradient boosting algorithm that constructs the predictive mapping as an ensemble of decision trees.
Each physical parameter ($M_{\rm gas}$, $M_{\rm dust}$, and $R_{\rm c}$) is predicted using an independent regressor trained on the same observable inputs.
Training is performed in $\log_{10}$ space for both observables and target parameters to reflect the approximate power-law scalings present in the model grid.

We split the \texttt{DiskMINT-GARDEN} grid into a reproducible 90/10 train–validation split, using the 90\% subset for training and the 10\% subset for validation.
During training, early stopping is implemented using the validation subset to prevent overfitting.
The adopted hyperparameters of the \texttt{XGBoost} regressor are:

\begin{itemize}
    \item \texttt{n\_estimators} = 2000. This specifies the maximum number of sequential decision trees that can be added to the ensemble. 
    \item \texttt{learning\_rate} = 0.02. This controls the contribution of each individual tree to the final model. A small learning rate (0.02) ensures that the mapping is constructed gradually, improving stability and reducing the risk of overfitting.
    \item \texttt{max\_depth} = 5. This limits the depth of each tree, thereby controlling the complexity of the learned mapping. Restricting the depth prevents the model from fitting fine-scale structure associated with discrete grid sampling.
    \item $L_1$ regularization \texttt{reg\_alpha} = 1.0 and $L_2$ regularization \texttt{reg\_lambda} = 1.0. They correspond to regularization terms applied to the tree weights. These penalties discourage overly complex tree structures and improve generalization across the grid.
    \item \texttt{early\_stopping\_rounds} = 50. This halts training if the validation loss does not improve for 50 consecutive boosting iterations. This mechanism determines the effective model complexity and provides an additional safeguard against overfitting.
\end{itemize}

We adopt the default regression objective of \texttt{XGBoost} (\texttt{reg:squarederror}), which minimizes the regularized mean-squared error between predicted and true grid parameters.
These choices were selected to balance model flexibility and regularization.
Limiting tree depth and enforcing a minimum child weight reduces sensitivity to sparsely sampled regions of parameter space.
$L_{1}$ and $L_{2}$ penalties discourage overly complex tree structures, while early stopping halts training once validation performance ceases to improve.

\subsection{Validation Performance}

Figure~\ref{fig:train_test_sample} presents the validation performance of the model.
The top row shows predictions for the training subset (90\% of the grid), and the bottom row shows predictions for the validation subset (10\% of the grid).
In each panel, predicted parameters are compared directly to the true grid values.

The model reproduces the underlying grid behavior with high fidelity.
For all three inferred parameters, the coefficient of determination is $R^2 \approx 0.99$ on the validation subset, indicating that $\gtrsim 0.99$ of the variance in the grid parameters is captured by the regression model.
The residual scatter $\sigma \sim 0.15$, measured in $\log_{10}$ space, remains small across the full parameter range, indicating no strong systematic degradation in performance.
The similarity between training and validation performance demonstrates that the adopted regularization and early stopping strategy successfully prevents overfitting.
These results indicate that the learned mapping reliably interpolates across the model grid.

\subsection{Observational Error Propagation}

Statistical uncertainties from observational measurements are estimated by propagating measurement errors through the trained regression model.
For each target, we generate Monte Carlo realizations of the observable vector $\mathbf{O}$ using the reported observational uncertainties, assuming a Gaussian distribution centered on the reported value.
Each realization is passed through the regression model to obtain a distribution of inferred parameters $\mathbf{\Theta}$, from which confidence intervals are derived.
For most targets, the resulting statistical uncertainties are small ($\lesssim 5\%$), but can reach $\lesssim 25\%$ for lower signal-to-noise (SNR$\sim3$) observations.

\subsection{Domain Control and Extrapolation Handling}

To ensure reliable application to observed targets, we implement explicit domain checks in observable space.
For each object, its observable vector is compared against the feature-space envelope spanned by the training grid.
Targets whose observables fall outside the grid bounds are flagged (see Table~\ref{tab:disk_examples} in Appendix~\ref{appendix:obs_info}).
For those flagged targets, we conservatively clip the predicted parameters to the limited range of the grid.
This procedure prevents uncontrolled extrapolation beyond the domain sampled by the thermochemical models.

Objects requiring substantial clipping or lying near grid boundaries are treated with caution and flagged (Appendix~\ref{appendix:obs_info}).


\section{\texorpdfstring{Model Grids: Disk Sizes, Dust-to-gas mass ratio, and $\coo\,(3-2)$ line}{Model Tracks: Disk Sizes, Dust-to-gas mass ratio, and C18O (3-2) line}}
\label{appendix:additional_model_figure}

\subsection{Disk Sizes}
\label{appendix:disk_size}

Disk size represents another key structural parameter that can potentially affect the applicability of our grid-based inference.
In \texttt{DiskMINT-GARDEN}, the gas and dust are assumed to share the same radial surface-density profile with a constant dust-to-gas mass ratio ($\sigmadust = \varepsilon\,\sigmagas$), characterized by a common scale radius $R_{\rm c}$.
If the observed dust distribution deviates significantly from the range spanned by the grid, this assumption may no longer hold and could introduce systematic uncertainties in the inferred $M_{\rm gas}$.

\begin{figure}
    \centering
    \includegraphics[width=1.0\linewidth]{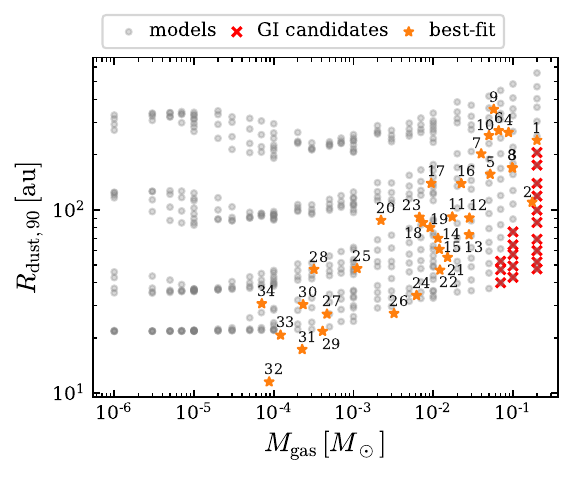}
    \caption{
    Distribution of the emulated disk-model grid in the $R_{\rm dust,90}$--$M_{\rm gas}$ plane. Gray points show all models in the grid. 
    Models flagged as gravitationally unstable (GI) are highlighted with red ``$\times$'' markers. 
    Star symbols mark the best-fit for each observed target, plotted at their predicted gas mass ($M_{\rm gas,pred}$) and observed 90\% dust radius ($R_{\rm dust,90\%}$), and they are labeled in order from the brightest to the faintest $L_{\rm C^{18}O}$ (same as Figure~\ref{fig:compare_lit}).
    Two disks (ID 1 \& 2) are approaching the GI boundary, and two (ID 31 \& 32) have observed $R_{\rm dust, 90}$ values that fall below the minimum $R_{\rm dust,90\%}$ in the grid.
    }
    \label{fig:Rdust_Mgas_withdata}
\end{figure}

Two systems (ID 31: J16054540-2023088 \& 32: Sz~72) have observed $R_{\rm dust,90}$ values that fall below the minimum radii represented in the model grid (Figure~\ref{fig:Rdust_Mgas_withdata}).
These sources may host intrinsically smaller disks than assumed in our models, or they may reflect radial drift and dust evolution that produce compact millimeter dust emission.
In the latter scenario, the dust and gas distributions may no longer be co-spatial, and the $\sigmadust$ should have a profile closer to the star compared to the $\sigmagas$.

We find that the disk sizes are important in determining the $L_{\rm C^{18}O}$–$M_{\rm gas}$ relation, and such a relation is illustrated in Figure~\ref{fig:lc18o_mgas_withdata}.
At fixed stellar mass and $M_{\rm gas}$, varying $R_{\rm c}$ produces up to two orders of magnitude variation in $L_{\coo}$ (with $\varepsilon$ producing comparatively smaller variations; see Appendix~\ref{appendix:additional_model_figure}).
We also use the line transparency to show the line optical depth of the model grids, with solid lines representing models with optically thin $\coo$ line emission and transparent points for optically thick ones (see Appendix~\ref{appendix:optical_depth} on how the optical depths are evaluated).
Compact disks reach higher surface densities and therefore become optically thick at lower gas masses, leading to saturation of the line luminosity.
In contrast, more extended disks remain optically thin over a wider range of $M_{\rm gas}$ and exhibit a steeper dependence of luminosity on gas mass.

Figure~\ref{fig:Rdust_Mgas_withdata} shows the full \texttt{DiskMINT-GARDEN} grid in the $R_{\rm dust,90\%}$–$M_{\rm gas}$ plane.
Gray points represent all emulated models, while red ``$\times$'' symbols highlight models flagged as GI candidates ($Q<1$).
As expected, gravitationally unstable models occupy the region of high gas mass and relatively compact disk radii, corresponding to enhanced surface densities.

The star symbols mark the best-fit gas masses predicted for the observed targets.
Most targets lie comfortably within the gravitationally stable region of the grid.
Two systems (ID 1: HD~34282 and 2: MWC~480) approach the GI boundary; for these disks, the steady and vertically smooth disk structure assumed in the grid may become less appropriate.
If a disk were truly close to the instability threshold, its mass distribution might deviate from the simplified prescriptions adopted here.
We note, however, that the inferred total $\mgas$ values for these systems (ID 1 and 2) are in good agreement with the dynamical masses within the uncertainties (Figures~\ref{fig:compare_dyn} \&~\ref{fig:compare_lit}).
Nevertheless, spatially resolved, disk-specific modeling would help better constrain the radial surface-density profile and assess the true dynamical state of these disks.

\subsection{\texorpdfstring{$\coo\,(3-2)$ line luminosity}{C18O (3-2) line luminosity}}
\label{appendix:lc18o_32_luminosity}

\begin{figure*}
  \centering
  \includegraphics[width=1.0\linewidth]{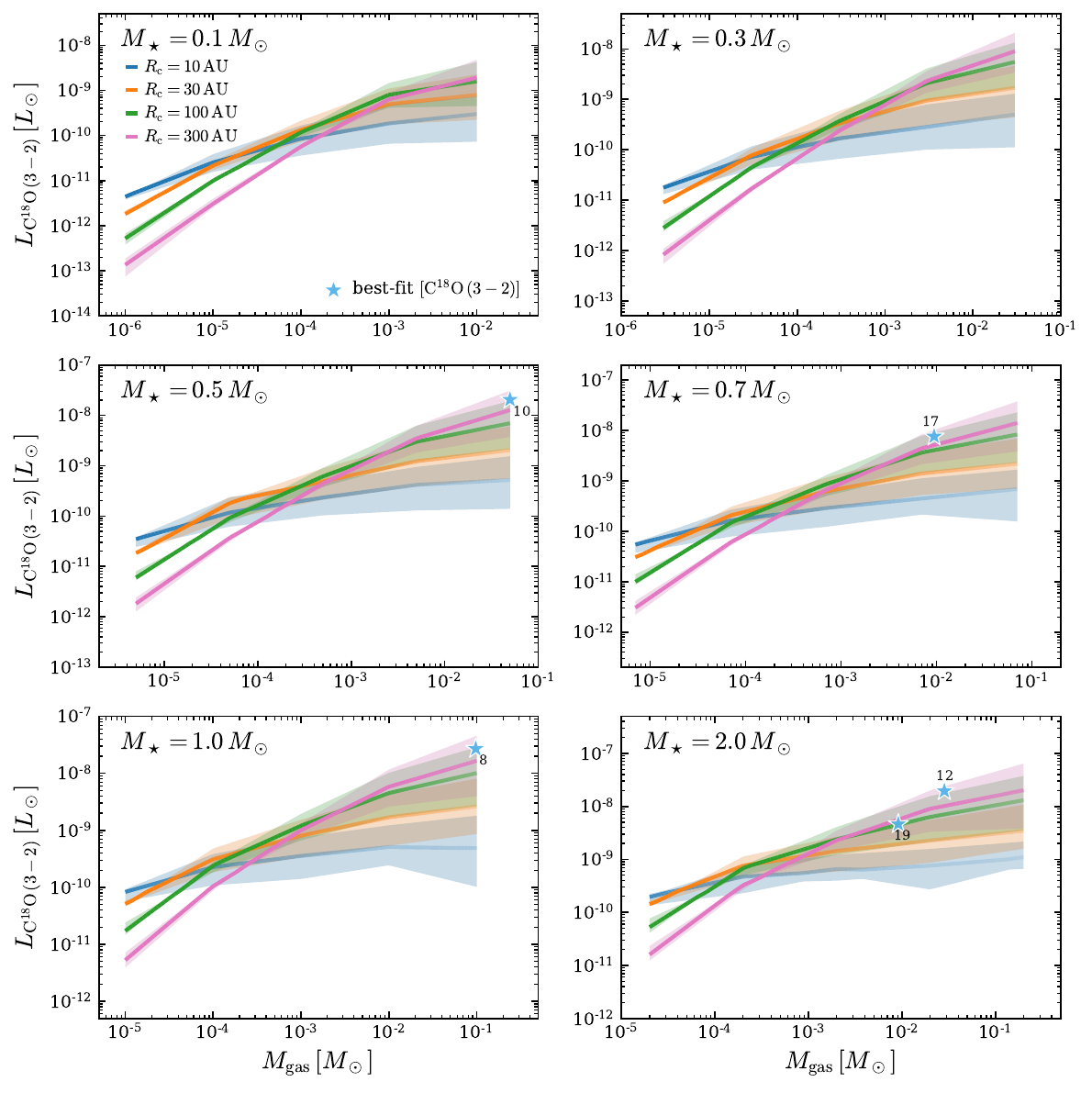}
  \caption{
    Relationship between the gas disk mass $M_{\rm gas}$ and the $\coo$\,(3-2) line luminosity, with the inferred C$^{18}$O line optical depth encoded by the curve transparency.
    The color notation follows the same as Figure~\ref{fig:lc18o_mgas_withdata}, but here we show the $\coo$\,(3-2) line luminosity instead of the (2-1) line luminosity, and only plot the five sources whose $\mgas$ is inferred from the (3-2) line (ID 8, 10, 12, 17, 19; see Table~\ref{tab:disk_examples} in Appendix~\ref{appendix:obs_info}).
  }
  \label{fig:lc18o_32_mgas_withdata}
\end{figure*}

In Section~\ref{sec:application}, we apply \texttt{DiskMINT-GARDEN} to a sample of disks with $>3\sigma$ detections in both ALMA continuum and $\coo$ line emission, using the (2-1) line luminosity as one of the key observables for inferring $M_{\rm gas}$ when the (2-1) line is detected.
However, there are five targets (ID 8, 10, 12, 17, 19; see Table~\ref{tab:disk_examples} in Appendix~\ref{appendix:obs_info}) for which the (2-1) line is not detected, and their $M_{\rm gas}$ are inferred from the (3-2) line luminosity instead.
To illustrate the relationship between $L_{\coo\,(3-2)}$ and $M_{\rm gas}$, we show in Figure~\ref{fig:lc18o_32_mgas_withdata} the model tracks of $L_{\coo\,(3-2)}$ as a function of $M_{\rm gas}$.

\section{Observational information and results for archival data}
\label{appendix:obs_info}

In this work, we applied \texttt{DiskMINT-GARDEN} to a sample of 34 disks that have both millimeter wavelength continuum and $\coo$ line measurements with ${\rm SNR} > 3$.
Here we summarize their observations, references, and inferred quantities in Table~\ref{tab:disk_examples}.

\begin{deluxetable*}{rllccccccccc}
\tablecaption{Observation and inferred values of disk parameters in the ALMA sample\label{tab:disk_examples}}
\tablewidth{0pt}
\tabletypesize{\scriptsize}
\tablehead{
\colhead{ID} & \colhead{Source} & \colhead{$M_{\star}$} & \colhead{$L_{\coo\,(2-1)}$} & \colhead{$L_{\coo\,(3-2)}$} & \colhead{$L_{\text{mm}}$} & \colhead{$R_{\text{dust,90}}$} & \colhead{Ref.\tablenotemark{a}} & \colhead{$M_{\text{gas,pred}}$} & \colhead{$M_{\text{dust,pred}}$} & \colhead{$R_{\text{c,pred}}$} & \colhead{Flag\tablenotemark{b}} \\
\colhead{} & \colhead{} & \colhead{($M_{\odot}$)} & \colhead{($L_{\odot}$)} & \colhead{($L_{\odot}$)} & \colhead{($L_{\odot}$)} & \colhead{(au)} & \colhead{} & \colhead{($M_{\odot}$)} & \colhead{($M_{\odot}$)} & \colhead{(au)} & \colhead{}
}
\startdata
1 & HD 34282 & 1.61 & $4.13 \times 10^{-8}$ & \nodata & $6.81 \times 10^{-4}$ & 239.40 & (10, 9, 13, 14) & $1.08 \times 10^{-1}$ & $7.48 \times 10^{-4}$ & 250.28 & b \\
2 & MWC 480 & 2.10 & $1.81 \times 10^{-8}$ & \nodata & $4.79 \times 10^{-4}$ & 110.00 & (10, 6, 13, 4) & $8.49 \times 10^{-2}$ & $3.94 \times 10^{-4}$ & 48.90 & b \\
3 & HD 163296 & 2.00 & $1.35 \times 10^{-8}$ & \nodata & $5.26 \times 10^{-4}$ & 169.00 & (10, 6, 13, 2) & $5.46 \times 10^{-2}$ & $2.77 \times 10^{-4}$ & 132.28 & - \\
4 & IM Lup & 1.10 & $9.10 \times 10^{-9}$ & \nodata & $4.55 \times 10^{-4}$ & 264.00 & (10, 6, 13, 2) & $8.79 \times 10^{-2}$ & $4.21 \times 10^{-4}$ & 258.39 & - \\
5 & LkCa 15 & 1.14 & $6.76 \times 10^{-9}$ & \nodata & $2.68 \times 10^{-4}$ & 156.30 & (10, 7, 13, 14) & $5.18 \times 10^{-2}$ & $1.97 \times 10^{-4}$ & 137.12 & - \\
6 & GM Aur & 1.10 & $6.32 \times 10^{-9}$ & \nodata & $2.95 \times 10^{-4}$ & 270.00 & (10, 6, 13, 4) & $6.61 \times 10^{-2}$ & $2.87 \times 10^{-4}$ & 295.74 & - \\
7 & DM Tau & 0.45 & $6.05 \times 10^{-9}$ & \nodata & $1.56 \times 10^{-4}$ & 201.90 & (10, 3, 13, 14) & $3.99 \times 10^{-2}$ & $1.24 \times 10^{-4}$ & 230.14 & - \\
8 & RXJ 1615.3-3255 & 1.14 & \nodata & $2.70 \times 10^{-8}$ & $8.42 \times 10^{-4}$ & 169.60 & (10, 13, 13, 14) & $8.05 \times 10^{-2}$ & $6.39 \times 10^{-4}$ & 91.19 & - \\
9 & V1094 Sco & 0.82 & $5.32 \times 10^{-9}$ & \nodata & $3.86 \times 10^{-4}$ & 353.01 & (12, 12, 12, 15) & $5.74 \times 10^{-2}$ & $2.92 \times 10^{-4}$ & 300.00 & - \\
10 & Elias 2-27 & 0.49 & \nodata & $2.06 \times 10^{-8}$ & $3.20 \times 10^{-4}$ & 254.00 & (10, 13, 13, 2) & $3.85 \times 10^{-2}$ & $3.75 \times 10^{-4}$ & 295.45 & - \\
11 & MWC 758 & 1.40 & $4.35 \times 10^{-9}$ & \nodata & $1.24 \times 10^{-4}$ & 91.40 & (10, 9, 13, 14) & $1.73 \times 10^{-2}$ & $5.13 \times 10^{-5}$ & 101.80 & - \\
12 & HD 135344B & 1.61 & \nodata & $1.95 \times 10^{-8}$ & $1.54 \times 10^{-4}$ & 90.20 & (10, 13, 13, 14) & $1.92 \times 10^{-2}$ & $6.50 \times 10^{-5}$ & 99.23 & - \\
13 & CQ Tau & 1.40 & $3.05 \times 10^{-9}$ & \nodata & $2.30 \times 10^{-4}$ & 73.10 & (10, 9, 13, 14) & $2.81 \times 10^{-2}$ & $1.02 \times 10^{-4}$ & 39.63 & - \\
14 & RXJ 1852.3-3700 & 1.03 & $2.50 \times 10^{-9}$ & \nodata & $9.07 \times 10^{-5}$ & 69.90 & (10, 13, 13, 14) & $1.15 \times 10^{-2}$ & $3.57 \times 10^{-5}$ & 36.59 & - \\
15 & V4046 Sgr & 1.73 & $2.30 \times 10^{-9}$ & \nodata & $1.22 \times 10^{-4}$ & 60.90 & (10, 3, 13, 14) & $1.19 \times 10^{-2}$ & $5.27 \times 10^{-5}$ & 29.48 & - \\
16 & AS 209 & 1.20 & $1.80 \times 10^{-9}$ & \nodata & $3.04 \times 10^{-4}$ & 139.00 & (10, 6, 13, 2) & $2.25 \times 10^{-2}$ & $2.12 \times 10^{-4}$ & 75.20 & - \\
17 & AA Tau & 0.79 & \nodata & $7.62 \times 10^{-9}$ & $8.80 \times 10^{-5}$ & 139.40 & (10, 8, 13, 14) & $5.83 \times 10^{-3}$ & $5.91 \times 10^{-5}$ & 104.78 & - \\
18 & RXJ 1842.9-3532 & 1.07 & $1.24 \times 10^{-9}$ & \nodata & $1.04 \times 10^{-4}$ & 85.20 & (10, 13, 13, 14) & $7.41 \times 10^{-3}$ & $4.46 \times 10^{-5}$ & 67.19 & - \\
19 & HD 143006 & 1.56 & \nodata & $4.63 \times 10^{-9}$ & $1.19 \times 10^{-4}$ & 79.90 & (10, 13, 13, 14) & $4.20 \times 10^{-3}$ & $4.33 \times 10^{-5}$ & 44.58 & - \\
20 & J16120668-3010270 & 0.51 & $6.03 \times 10^{-10}$ & \nodata & $1.92 \times 10^{-5}$ & 87.63 & (11, 11, 11, 15) & $2.21 \times 10^{-3}$ & $7.89 \times 10^{-6}$ & 100.08 & - \\
21 & TW Hya & 0.80 & $4.62 \times 10^{-10}$ & \nodata & $1.48 \times 10^{-4}$ & 55.00 & (1, 5, 1, 2) & $1.51 \times 10^{-2}$ & $1.11 \times 10^{-4}$ & 16.26 & - \\
22 & PDS 66 & 1.28 & $4.62 \times 10^{-10}$ & \nodata & $1.02 \times 10^{-4}$ & 46.90 & (10, 8, 13, 14) & $1.21 \times 10^{-2}$ & $7.73 \times 10^{-5}$ & 11.26 & - \\
23 & Sz 71 & 0.42 & $4.18 \times 10^{-10}$ & \nodata & $1.33 \times 10^{-4}$ & 91.07 & (12, 12, 12, 15) & $6.71 \times 10^{-3}$ & $1.22 \times 10^{-4}$ & 96.75 & - \\
24 & Sz 65 & 0.68 & $2.96 \times 10^{-10}$ & \nodata & $5.19 \times 10^{-5}$ & 34.10 & (12, 12, 12, 15) & $6.17 \times 10^{-3}$ & $2.84 \times 10^{-5}$ & 10.00 & - \\
25 & J16082324-1930009 & 0.56 & $2.94 \times 10^{-10}$ & \nodata & $2.10 \times 10^{-5}$ & 47.81 & (11, 11, 11, 15) & $1.11 \times 10^{-3}$ & $7.95 \times 10^{-6}$ & 29.84 & - \\
26 & J16221532-2511349 & 0.29 & $2.79 \times 10^{-10}$ & \nodata & $1.68 \times 10^{-5}$ & 27.20 & (11, 11, 11, 15) & $3.22 \times 10^{-3}$ & $8.56 \times 10^{-6}$ & 10.15 & s \\
27 & J16020757-2257467 & 0.37 & $1.74 \times 10^{-10}$ & \nodata & $2.17 \times 10^{-6}$ & 26.98 & (11, 11, 11, 15) & $4.65 \times 10^{-4}$ & $6.07 \times 10^{-7}$ & 10.31 & - \\
28 & J16202863-2442087 & 0.34 & $1.52 \times 10^{-10}$ & \nodata & $2.57 \times 10^{-6}$ & 47.34 & (11, 11, 11, 15) & $3.20 \times 10^{-4}$ & $1.76 \times 10^{-6}$ & 30.56 & - \\
29 & J16085324-3914401 & 0.31 & $9.75 \times 10^{-11}$ & \nodata & $1.49 \times 10^{-5}$ & 21.75 & (12, 12, 12, 15) & $4.10 \times 10^{-4}$ & $5.32 \times 10^{-6}$ & 10.15 & s \\
30 & Sz 131 & 0.31 & $8.79 \times 10^{-11}$ & \nodata & $5.86 \times 10^{-6}$ & 30.40 & (12, 12, 12, 15) & $2.33 \times 10^{-4}$ & $1.63 \times 10^{-6}$ & 11.05 & - \\
31 & J16054540-2023088 & 0.13 & $7.97 \times 10^{-11}$ & \nodata & $5.55 \times 10^{-6}$ & 17.31 & (11, 11, 11, 15) & $2.27 \times 10^{-4}$ & $1.43 \times 10^{-6}$ & 10.17 & s \\
32 & Sz 72 & 0.39 & $5.37 \times 10^{-11}$ & \nodata & $9.92 \times 10^{-6}$ & 11.52 & (12, 12, 12, 15) & $8.76 \times 10^{-5}$ & $2.12 \times 10^{-6}$ & 10.10 & s \\
33 & Sz 66 & 0.30 & $4.94 \times 10^{-11}$ & \nodata & $8.00 \times 10^{-6}$ & 20.72 & (12, 12, 12, 15) & $1.22 \times 10^{-4}$ & $2.52 \times 10^{-6}$ & 10.16 & s \\
34 & Sz 95 & 0.30 & $3.19 \times 10^{-11}$ & \nodata & $2.86 \times 10^{-6}$ & 30.76 & (12, 12, 12, 15) & $7.08 \times 10^{-5}$ & $1.27 \times 10^{-6}$ & 11.16 & - \\
\enddata
\tablenotetext{a}{The ``Ref.'' entry is a 4-tuple giving the literature sources for $(M_\star,\; L_{\coo\,(2-1)}\,{\rm or}\,L_{\coo\,(3-2)},\; L_{\rm mm},\; R_{\rm dust,90})$ in that order.}
\tablenotetext{b}{Flag: (s) small-disk sample; (b) bright-disk sample.}
\tablecomments{References: (1) \citet{andrews_resolved_2011}, (2) \citet{huang_disk_2018}, (3) \citet{flaherty_measuring_2020}, (4) \citet{law_molecules_2021}, (5) \citet{calahan_tw_2021}, (6) \citet{oberg_disk_2011}, (7) \citet{sturm_disentangling_2023}, (8) \citet{ribas_alma_2023}, (9) \citet{stapper_constraining_2024}, (10) \citet{izquierdo_exoalma_2025}, (11) \citet{agurto-gangas_alma_2025}, (12) \citet{Deng_2025_AGEPRO_III_Lupus}, (13) \citet{trapman_exoalma_2025}, (14) \citet{curone_exoalma_2025}, (15) \citet{vioque_alma_2025}. }
\end{deluxetable*}

\section{Grain-surface chemistry and its impact on CO column densities}
\label{appendix:compare_column_density}

\begin{figure}
    \centering
    \includegraphics[width=1.0\linewidth]{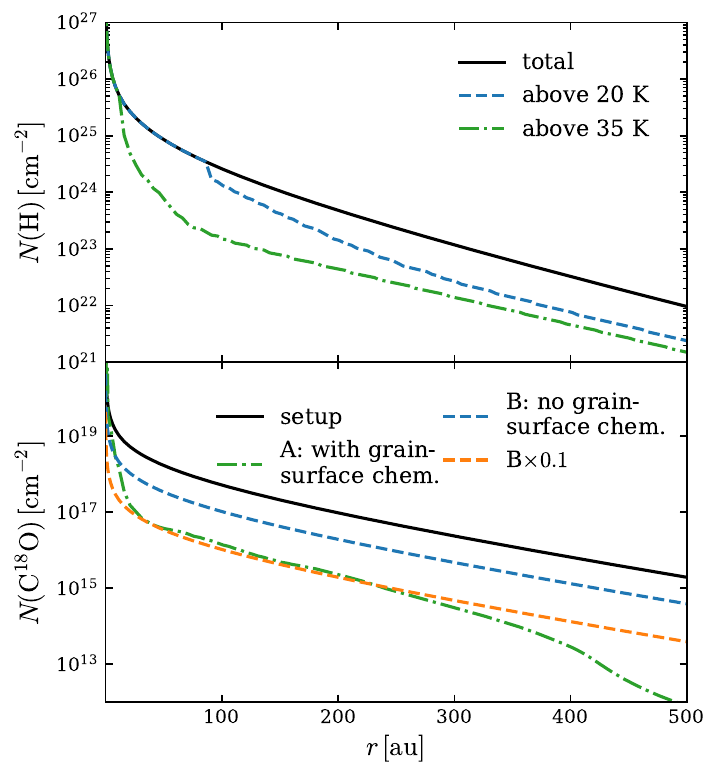}
    \caption{
    Radial column densities of hydrogen (top) and C$^{18}$O (bottom) predicted by the \texttt{DiskMINT} models.
    The top panel shows the total hydrogen column density as well as the fraction of gas above gas temperature of 20\,K ($\sim\co$ snowline) and 35\,K ($\sim \co_{2}$ snowline).
    The bottom panel compares the resulting C$^{18}$O column densities for models with and without grain-surface chemistry, where the $\co \rightarrow \co_2$ conversion is the main reaction.
    Models with the grain-surface chemistry result in a higher CO snowline, and show that the previously inferred C or CO depletion factors (typically a factor of $\sim 10$ as needed in models without grain-surface chemistry) are not necessary.
    }
    \label{fig:compare_density}
\end{figure}

Figure~\ref{fig:compare_density} compares radial column densities predicted by \texttt{DiskMINT} with or without grain-surface chemistry, and illustrates the effects on gas phase CO abundance and the resulting C$^{18}$O  column density in our models.

The upper panel shows the total hydrogen column density together with the column densities of gas above temperature thresholds of 20\,K and 35\,K, approximately corresponding to the vertical locations of the CO and CO$_2$ snowlines, respectively.

The lower panel shows the resulting C$^{18}$O column densities predicted by the \texttt{DiskMINT} models.
In our framework (green dash-dotted line, model A), grain-surface chemistry converts CO into the more stable CO$_2$ ice \citep[][for details]{Ruaud2022}, raising the CO snowline toward the CO$_2$ snowline and reducing the gas-phase CO abundance in the emitting layers of the disk.
As a result, the predicted $\coo$ column densities are naturally reduced without invoking an explicit CO depletion factor.
In contrast, if this process is not included in the thermochemical model and a 20K CO freeze-out temperature is assumed (model~B), then the column density of $\coo$ is overestimated by a factor $\sim $ 10, and it needs to be scaled down by a factor of $\sim 0.1$ to match the model~A (see orange line labeled `B $\times 0.1$' in Fig.~\ref{fig:compare_density}), which corresponds to a typical factor of $\sim 10$ CO or C depletion from empirical estimates to match with observations \citep[e.g.,][]{trapman_exoalma_2025, Trapman_AGEPRO_V_gas_masses}. 



\bibliography{ref}{}

@ARTICLE{Molyarova2017ApJ...849..130M,
       author = {{Molyarova}, Tamara and {Akimkin}, Vitaly and {Semenov}, Dmitry and {Henning}, Thomas and {Vasyunin}, Anton and {Wiebe}, Dmitri},
        title = "{Gas Mass Tracers in Protoplanetary Disks: CO is Still the Best}",
      journal = {\apj},
         year = 2017,
        month = nov,
       volume = {849},
       number = {2},
          eid = {130},
        pages = {130},
          doi = {10.3847/1538-4357/aa9227},
archivePrefix = {arXiv},
       eprint = {1710.02993},
 primaryClass = {astro-ph.EP},
       adsurl = {https://ui.adsabs.harvard.edu/abs/2017ApJ...849..130M}
}

@article{deng_diskmint_2023,
    title = {{DiskMINT}: {A} {Tool} to {Estimate} {Disk} {Masses} with {CO} {Isotopologues}},
    volume = {954},
    issn = {0004-637X, 1538-4357},
    shorttitle = {{DiskMINT}},
    url = {https://iopscience.iop.org/article/10.3847/1538-4357/acdfcc},
    doi = {10.3847/1538-4357/acdfcc},
    language = {en},
    number = {2},
    urldate = {2024-02-15},
    journal = {The Astrophysical Journal},
    author = {Deng, Dingshan and Ruaud, Maxime and Gorti, Uma and Pascucci, Ilaria},
    month = sep,
    year = {2023},
    pages = {165},
}

@article{deng_diskmint_2025,
	title = {{DiskMINT}: {Self}-consistent {Thermochemical} {Disk} {Models} with {Radially} {Varying} {Gas} and {Dust}—{Application} to the {Massive}, {CO}-{Rich} {Disk} of {IM} {Lup}},
	volume = {995},
	url = {https://doi.org/10.3847/1538-4357/ae0e66},
	doi = {10.3847/1538-4357/ae0e66},
	number = {1},
	journal = {The Astrophysical Journal},
	publisher = {The American Astronomical Society},
	author = {Deng, Dingshan and Gorti, Uma and Pascucci, Ilaria and Ruaud, Maxime},
	month = dec,
	year = {2025},
	pages = {98},
}

@article{min_multiwavelength_2016,
    title = {Multiwavelength optical properties of compact dust aggregates in protoplanetary disks},
    volume = {585},
    issn = {0004-6361},
    url = {https://ui.adsabs.harvard.edu/abs/2016A&A...585A..13M},
    doi = {10.1051/0004-6361/201526048},
    urldate = {2025-02-20},
    journal = {Astronomy and Astrophysics},
    author = {Min, M. and Rab, Ch. and Woitke, P. and Dominik, C. and Ménard, F.},
    month = jan,
    year = {2016},
    note = {ADS Bibcode: 2016A\&A...585A..13M},
    pages = {A13},
}

@article{lynden-bell_evolution_1974,
    title = {The evolution of viscous discs and the origin of the nebular variables.},
    volume = {168},
    issn = {0035-8711},
    url = {https://ui.adsabs.harvard.edu/abs/1974MNRAS.168..603L},
    doi = {10.1093/mnras/168.3.603},
    urldate = {2025-02-21},
    journal = {Monthly Notices of the Royal Astronomical Society},
    author = {Lynden-Bell, D. and Pringle, J. E.},
    month = sep,
    year = {1974},
    note = {Publisher: OUP
ADS Bibcode: 1974MNRAS.168..603L},
    pages = {603--637},
}

@misc{Dullemond_radmc-3d_2012,
    title = {\{{RADMC}\}-\{{3D}\}: \{{A}\} multi-purpose radiative transfer tool - \{{NASA}\}/\{{ADS}\}},
    url = {https://ui.adsabs.harvard.edu/abs/2012ascl.soft02015D/abstract},
    urldate = {2020-09-03},
    author = {Dullemond, C. P.; and Juhasz, A. ; and Pohl, A. ; and Sereshti, F. ; and Shetty, R. ; and Peters, T. ; and Commercon, B. ; and Flock, M.},
    year = {2012},
    note = {Publication Title: Astrophysics Source Code Library, record ascl:1202.015},
}

@article{gorti_photoevaporation_2009,
    title = {Photoevaporation of {Circumstellar} {Disks} {By} {Far}-{Ultraviolet}, {Extreme}-{Ultraviolet} and {X}-{Ray} {Radiation} from the {Central} {Star}},
    volume = {690},
    issn = {0004-637X},
    url = {https://ui.adsabs.harvard.edu/abs/2009ApJ...690.1539G},
    doi = {10.1088/0004-637X/690/2/1539},
    urldate = {2024-04-19},
    journal = {The Astrophysical Journal},
    author = {Gorti, U. and Hollenbach, D.},
    month = jan,
    year = {2009},
    note = {Publisher: IOP
ADS Bibcode: 2009ApJ...690.1539G},
    pages = {1539--1552},
}

@article{gorti_time_2009,
    title = {Time {Evolution} of {Viscous} {Circumstellar} {Disks} due to {Photoevaporation} by {Far}-{Ultraviolet}, {Extreme}-{Ultraviolet}, and {X}-ray {Radiation} from the {Central} {Star}},
    volume = {705},
    issn = {0004-637X},
    url = {https://ui.adsabs.harvard.edu/abs/2009ApJ...705.1237G},
    doi = {10.1088/0004-637X/705/2/1237},
    urldate = {2025-08-08},
    journal = {The Astrophysical Journal},
    publisher = {IOP},
    author = {Gorti, U. and Dullemond, C. P. and Hollenbach, D.},
    month = nov,
    year = {2009},
    note = {ADS Bibcode: 2009ApJ...705.1237G},
    pages = {1237--1251},
}

@article{alcala_x-shooter_2017,
    title = {X-shooter spectroscopy of young stellar objects in {Lupus}: {Accretion} properties of class {II} and transitional objects⋆},
    volume = {600},
    issn = {0004-6361, 1432-0746},
    shorttitle = {X-shooter spectroscopy of young stellar objects in {Lupus}},
    url = {http://www.aanda.org/10.1051/0004-6361/201629929},
    doi = {10.1051/0004-6361/201629929},
    language = {en},
    urldate = {2024-03-14},
    journal = {Astronomy \& Astrophysics},
    author = {Alcalá, J. M. and Manara, C. F. and Natta, A. and Frasca, A. and Testi, L. and Nisini, B. and Stelzer, B. and Williams, J. P. and Antoniucci, S. and Biazzo, K. and Covino, E. and Esposito, M. and Getman, F. and Rigliaco, E.},
    month = apr,
    year = {2017},
    pages = {A20},
}

@ARTICLE{Valenti_IUE_2003,
       author = {{Valenti}, Jeff A. and {Fallon}, Adam A. and {Johns-Krull}, Christopher M.},
        title = "{An IUE Atlas of Pre-Main-Sequence Stars. III. Co-added Final Archive Spectra from the Long-Wavelength Cameras}",
      journal = {\apjs},
         year = 2003,
        month = aug,
       volume = {147},
       number = {2},
        pages = {305-336},
          doi = {10.1086/375445},
       adsurl = {https://ui.adsabs.harvard.edu/abs/2003ApJS..147..305V}
}

@article{baraffe_new_2015,
    title = {New evolutionary models for pre-main sequence and main sequence low-mass stars down to the hydrogen-burning limit},
    volume = {577},
    issn = {0004-6361},
    url = {https://ui.adsabs.harvard.edu/abs/2015A&A...577A..42B},
    doi = {10.1051/0004-6361/201425481},
    urldate = {2024-09-13},
    journal = {Astronomy and Astrophysics},
    author = {Baraffe, Isabelle and Homeier, Derek and Allard, France and Chabrier, Gilles},
    month = may,
    year = {2015},
    note = {ADS Bibcode: 2015A\&A...577A..42B},
    pages = {A42},
}

@ARTICLE{Deng_2025_AGEPRO_III_Lupus,
       author = {Deng, Dingshan and {Vioque}, Miguel and {Pascucci}, Ilaria and {P{\'e}rez}, Laura M. and {Zhang}, Ke and {Kurtovic}, Nicol{\'a}s and {Trapman}, Leon and {TorresVillanueva}, Estephani E. and {Agurto-Gangas}, Carolina and {Carpenter}, John and {Pinilla}, Paola and {Gorti}, Uma and {Tabone}, Beno{\^\i}t and {Sierra}, Anibal and {Rosotti}, Giovanni P. and {Cieza}, Lucas A. and {Anania}, Rossella and {Gonz{\'a}lez-Ruilova}, Camilo and {Hogerheijde}, Michiel R. and {Miley}, James and {Ruiz-Rodriguez}, Dary A. and {Ruaud}, Maxime and {Schwarz}, Kamber},
        title = "{The ALMA Survey of Gas Evolution of PROtoplanetary Disks (AGE-PRO). III. Dust and Gas Disk Properties in the Lupus Star-forming Region}",
      journal = {\apj},
         year = 2025,
        month = aug,
       volume = {989},
       number = {1},
          eid = {3},
        pages = {3},
          doi = {10.3847/1538-4357/add43a},
archivePrefix = {arXiv},
       eprint = {2506.10734},
 primaryClass = {astro-ph.EP},
       adsurl = {https://ui.adsabs.harvard.edu/abs/2025ApJ...989....3D}
}

@software{Deng_2026_diskmint_zenodo,
  author       = {{Deng}, Dingshan and
                  {Ruaud}, Maxime and
                  {Gorti}, Uma and
                  {Pascucci}, Ilaria},
  title        = {DingshanDeng/DiskMINT: DiskMINT v1.7.0 Release},
  month        = TBA,
  year         = 2026,
  publisher    = {Zenodo},
  version      = {v1.7.0},
  doi          = {10.5281/zenodo.8115820},
  url          = {https://doi.org/10.5281/zenodo.8115820},
}

@ARTICLE{Trapman_AGEPRO_V_gas_masses,
       author = {{Trapman}, Leon and {Zhang}, Ke and {Rosotti}, Giovanni P. and {Pinilla}, Paola and {Tabone}, Beno{\^\i}t and {Pascucci}, Ilaria and {Agurto-Gangas}, Carolina and {Anania}, Rossella and {Carpenter}, John and {Cieza}, Lucas A. and {Deng}, Dingshan and {Gonz{\'a}lez-Ruilova}, Camilo and {Hogerheijde}, Michiel R. and {Kurtovic}, Nicol{\'a}s T. and {Kuznetsova}, Aleksandra and {Miley}, James and {P{\'e}rez}, Laura M. and {Ruiz-Rodriguez}, Dary A. and {Schwarz}, Kamber and {Sierra}, Anibal and {TorresVillanueva}, Estephani and {Vioque}, Miguel},
        title = "{The ALMA Survey of Gas Evolution of PROtoplanetary Disks (AGE-PRO). V. Protoplanetary Gas Disk Masses}",
      journal = {\apj},
         year = 2025,
        month = aug,
       volume = {989},
       number = {1},
          eid = {5},
        pages = {5},
          doi = {10.3847/1538-4357/adcd6e},
archivePrefix = {arXiv},
       eprint = {2506.10738},
 primaryClass = {astro-ph.EP},
       adsurl = {https://ui.adsabs.harvard.edu/abs/2025ApJ...989....5T}
}

@INPROCEEDINGS{Miotello_PPVII_2023,
       author = {{Miotello}, A. and {Kamp}, I. and {Birnstiel}, T. and {Cleeves}, L.~C. and {Kataoka}, A.},
        title = "{Setting the Stage for Planet Formation: Measurements and Implications of the Fundamental Disk Properties}",
    booktitle = {Protostars and Planets VII},
         year = 2023,
       editor = {{Inutsuka}, S. and {Aikawa}, Y. and {Muto}, T. and {Tomida}, K. and {Tamura}, M.},
       series = {Astronomical Society of the Pacific Conference Series},
       volume = {534},
        month = jul,
        pages = {501},
          doi = {10.48550/arXiv.2203.09818},
archivePrefix = {arXiv},
       eprint = {2203.09818},
 primaryClass = {astro-ph.EP},
       adsurl = {https://ui.adsabs.harvard.edu/abs/2023ASPC..534..501M}
}

@ARTICLE{lodato_dynamical_mass_2023,
       author = {{Lodato}, G. and {Rampinelli}, L. and {Viscardi}, E. and {Longarini}, C. and {Izquierdo}, A. and {Paneque-Carre{\~n}o}, T. and {Testi}, L. and {Facchini}, S. and {Miotello}, A. and {Veronesi}, B. and {Hall}, C.},
        title = "{Dynamical mass measurements of two protoplanetary discs}",
      journal = {\mnras},
         year = 2023,
        month = jan,
       volume = {518},
       number = {3},
        pages = {4481-4493},
          doi = {10.1093/mnras/stac3223},
archivePrefix = {arXiv},
       eprint = {2211.03712},
 primaryClass = {astro-ph.EP},
       adsurl = {https://ui.adsabs.harvard.edu/abs/2023MNRAS.518.4481L}
}

@article{Ansdell_lupus_2016,
   author = {M. Ansdell and J. P. Williams and N. van der Marel and J. M. Carpenter and G. Guidi and M. Hogerheijde and G. S. Mathews and C. F. Manara and A. Miotello and A. Natta and I. Oliveira and M. Tazzari and L. Testi and E. F. van Dishoeck and S. E. van Terwisga},
   doi = {10.3847/0004-637X/828/1/46},
   issn = {1538-4357},
   issue = {1},
   journal = {The Astrophysical Journal},
   month = {8},
   pages = {46},
   publisher = {American Astronomical Society},
   title = {ALMA SURVEY OF LUPUS PROTOPLANETARY DISKS. I. DUST AND GAS MASSES},
   volume = {828},
   url = {https://iopscience.iop.org/article/10.3847/0004-637X/828/1/46},
   year = {2016},
}

@ARTICLE{Pascucci_mass_2016,
       author = {{Pascucci}, I. and {Testi}, L. and {Herczeg}, G.~J. and {Long}, F. and {Manara}, C.~F. and {Hendler}, N. and {Mulders}, G.~D. and {Krijt}, S. and {Ciesla}, F. and {Henning}, Th. and {Mohanty}, S. and {Drabek-Maunder}, E. and {Apai}, D. and {Sz{\H{u}}cs}, L. and {Sacco}, G. and {Olofsson}, J.},
        title = "{A Steeper than Linear Disk Mass-Stellar Mass Scaling Relation}",
      journal = {\apj},
         year = 2016,
        month = nov,
       volume = {831},
       number = {2},
          eid = {125},
        pages = {125},
          doi = {10.3847/0004-637X/831/2/125},
archivePrefix = {arXiv},
       eprint = {1608.03621},
 primaryClass = {astro-ph.EP},
       adsurl = {https://ui.adsabs.harvard.edu/abs/2016ApJ...831..125P}
}

@article{woitke_diana_2019,
   author = {P Woitke and I Kamp and S Antonellini and F Anthonioz and C Baldovin-Saveedra and A Carmona and O Dionatos and C Dominik and J Greaves and M Güdel and J D Ilee and A Liebhardt and F Menard and M Min and C Pinte and C Rab and L Rigon and W F Thi and N Thureau and L B F M Waters},
   doi = {10.1088/1538-3873/aaf4e5},
   issn = {0004-6280},
   journal = {Publications of the Astronomical Society of the Pacific},
   pages = {64301},
   title = {Consistent \{Dust\} and \{Gas\} \{Models\} for \{Protoplanetary\} \{Disks\}. \{III\}. \{Models\} for \{Selected\} \{Objects\} from the \{FP7\} \{DIANA\} \{Project\}},
   volume = {131},
   url = {http://adsabs.harvard.edu/abs/2019PASP..131f4301W},
   year = {2019},
}

@article{Trapman_HD_2017,
   author = {L. Trapman and A. Miotello and M. Kama and E. F. Van Dishoeck and S. Bruderer},
   doi = {10.1051/0004-6361/201630308},
   issn = {14320746},
   journal = {Astronomy and Astrophysics},
   month = {9},
   pages = {A69},
   publisher = {EDP Sciences},
   title = {Far-infrared HD emission as a measure of protoplanetary disk mass},
   volume = {605},
   url = {https://ui.adsabs.harvard.edu/abs/2017A%26A...605A..69T/abstract},
   year = {2017},
}

@article{Ruaud2022,
   author = {Maxime Ruaud and Uma Gorti and David J. Hollenbach},
   doi = {10.3847/1538-4357/ac3826},
   issn = {0004-637X},
   issue = {1},
   journal = {The Astrophysical Journal},
   month = {1},
   pages = {49},
   title = {C <sup>18</sup> O Emission as an Effective Measure of Gas Masses of Protoplanetary Disks},
   volume = {925},
   url = {https://iopscience.iop.org/article/10.3847/1538-4357/ac3826},
   year = {2022},
}

@article{Ruaud2019,
   author = {Maxime Ruaud and Uma Gorti},
   doi = {10.3847/1538-4357/ab4996},
   issn = {23318422},
   issue = {2},
   journal = {The Astrophysical Journal},
   pages = {146},
   publisher = {IOP Publishing},
   title = {A three-phase approach to grain surface chemistry in protoplanetary disks: Gas, ice surfaces and ice mantles of dust grains},
   volume = {885},
   url = {http://dx.doi.org/10.3847/1538-4357/ab4996},
   year = {2019},
}

@ARTICLE{McClure_HD_2016,
       author = {{McClure}, M.~K. and {Bergin}, E.~A. and {Cleeves}, L.~I. and {van Dishoeck}, E.~F. and {Blake}, G.~A. and {Evans}, N.~J., II and {Green}, J.~D. and {Henning}, Th. and {{\"O}berg}, K.~I. and {Pontoppidan}, K.~M. and {Salyk}, C.},
        title = "{Mass Measurements in Protoplanetary Disks from Hydrogen Deuteride}",
      journal = {\apj},
         year = 2016,
        month = nov,
       volume = {831},
       number = {2},
          eid = {167},
        pages = {167},
          doi = {10.3847/0004-637X/831/2/167},
archivePrefix = {arXiv},
       eprint = {1608.07817},
 primaryClass = {astro-ph.SR},
       adsurl = {https://ui.adsabs.harvard.edu/abs/2016ApJ...831..167M}
}

@ARTICLE{Bergin_HD_2013,
       author = {{Bergin}, Edwin A. and {Cleeves}, L. Ilsedore and {Gorti}, Uma and {Zhang}, Ke and {Blake}, Geoffrey A. and {Green}, Joel D. and {Andrews}, Sean M. and {Evans}, Neal J., II and {Henning}, Thomas and {{\"O}berg}, Karin and {Pontoppidan}, Klaus and {Qi}, Chunhua and {Salyk}, Colette and {van Dishoeck}, Ewine F.},
        title = "{An old disk still capable of forming a planetary system}",
      journal = {\nat},
         year = 2013,
        month = jan,
       volume = {493},
       number = {7434},
        pages = {644-646},
          doi = {10.1038/nature11805},
archivePrefix = {arXiv},
       eprint = {1303.1107},
 primaryClass = {astro-ph.SR},
       adsurl = {https://ui.adsabs.harvard.edu/abs/2013Natur.493..644B}
}

@Book{Bergin_n_Williams_mass_2017,
author="Bergin, Edwin A.
and Williams, Jonathan P.",
title="The Determination of Protoplanetary Disk Masses",
bookTitle="Formation, Evolution, and Dynamics of Young Solar Systems",
year="2017",
publisher="Springer International Publishing",
address="Cham",
isbn="978-3-319-60609-5",
doi="10.1007/978-3-319-60609-5_1",
url="https://doi.org/10.1007/978-3-319-60609-5_1"
}

@ARTICLE{Bosman_n_Banzatti_twhya_2019,
       author = {{Bosman}, Arthur D. and {Banzatti}, Andrea},
        title = "{The dry and carbon-poor inner disk of TW Hydrae: evidence for a massive icy dust trap}",
      journal = {\aap},
         year = 2019,
        month = dec,
       volume = {632},
          eid = {L10},
        pages = {L10},
          doi = {10.1051/0004-6361/201936638},
archivePrefix = {arXiv},
       eprint = {1911.11152},
 primaryClass = {astro-ph.SR},
       adsurl = {https://ui.adsabs.harvard.edu/abs/2019A&A...632L..10B}
}

@ARTICLE{Krijt_COdep_2020,
       author = {{Krijt}, Sebastiaan and {Bosman}, Arthur D. and {Zhang}, Ke and {Schwarz}, Kamber R. and {Ciesla}, Fred J. and {Bergin}, Edwin A.},
        title = "{CO Depletion in Protoplanetary Disks: A Unified Picture Combining Physical Sequestration and Chemical Processing}",
      journal = {\apj},
         year = 2020,
        month = aug,
       volume = {899},
       number = {2},
          eid = {134},
        pages = {134},
          doi = {10.3847/1538-4357/aba75d},
archivePrefix = {arXiv},
       eprint = {2007.09517},
 primaryClass = {astro-ph.SR},
       adsurl = {https://ui.adsabs.harvard.edu/abs/2020ApJ...899..134K}
}

@ARTICLE{Powell_COdep_2022,
       author = {{Powell}, Diana and {Gao}, Peter and {Murray-Clay}, Ruth and {Zhang}, Xi},
        title = "{Depletion of gaseous CO in protoplanetary disks by surface-energy-regulated ice formation}",
      journal = {Nature Astronomy},
         year = 2022,
        month = aug,
       volume = {6},
        pages = {1147-1155},
          doi = {10.1038/s41550-022-01741-9},
archivePrefix = {arXiv},
       eprint = {2208.13806},
 primaryClass = {astro-ph.EP},
       adsurl = {https://ui.adsabs.harvard.edu/abs/2022NatAs...6.1147P}
}

@article{pascucci_noCOdep_2023,
	title = {Large {Myr}-old {Disks} {Are} {Not} {Severely} {Depleted} of {Gas}-phase {CO} or {Carbon}},
	volume = {953},
	issn = {0004-637X, 1538-4357},
	url = {https://iopscience.iop.org/article/10.3847/1538-4357/ace4bf},
	doi = {10.3847/1538-4357/ace4bf},
	language = {en},
	number = {2},
	urldate = {2024-02-15},
	journal = {The Astrophysical Journal},
	author = {Pascucci, Ilaria and Skinner, Bennett N. and Deng, Dingshan and Ruaud, Maxime and Gorti, Uma and Schwarz, Kamber R. and Chapillon, Edwige and Vioque, Miguel and Miley, James},
	month = aug,
	year = {2023},
	pages = {183},
}

@article{Miotello2014,
   author = {A. Miotello and S. Bruderer and E. F. Van Dishoeck},
   doi = {10.1051/0004-6361/201424712},
   issn = {14320746},
   journal = {Astronomy and Astrophysics},
   month = {12},
   publisher = {EDP Sciences},
   title = {Protoplanetary disk masses from CO isotopologue line emission},
   volume = {572},
   year = {2014},
}

@article{Miotello2016,
   author = {A. Miotello and E. F. van Dishoeck and M. Kama and S. Bruderer},
   doi = {10.1051/0004-6361/201628159},
   issn = {0004-6361},
   journal = {Astronomy \& Astrophysics},
   month = {10},
   pages = {A85},
   publisher = {EDP Sciences},
   title = {Determining protoplanetary disk gas masses from CO isotopologues line observations},
   volume = {594},
   url = {http://www.aanda.org/10.1051/0004-6361/201628159},
   year = {2016},
}

@article{miotello_lupus_2017,
   author = {A. Miotello and E. F. van Dishoeck and J. P. Williams and M. Ansdell and G. Guidi and M. Hogerheijde and C. F. Manara and M. Tazzari and L. Testi and N. van der Marel and S. van Terwisga},
   doi = {10.1051/0004-6361/201629556},
   issn = {0004-6361},
   journal = {Astronomy \& Astrophysics},
   month = {3},
   pages = {A113},
   title = {Lupus disks with faint CO isotopologues: low gas/dust or high carbon depletion?},
   volume = {599},
   url = {http://www.aanda.org/10.1051/0004-6361/201629556},
   year = {2017},
}

@ARTICLE{Long_Cham_2017,
       author = {{Long}, Feng and {Herczeg}, Gregory J. and {Pascucci}, Ilaria and {Drabek-Maunder}, Emily and {Mohanty}, Subhanjoy and {Testi}, Leonardo and {Apai}, Daniel and {Hendler}, Nathan and {Henning}, Thomas and {Manara}, Carlo F. and {Mulders}, Gijs D.},
        title = "{An ALMA Survey of CO Isotopologue Emission from Protoplanetary Disks in Chamaeleon I}",
      journal = {\apj},
         year = 2017,
        month = aug,
       volume = {844},
       number = {2},
          eid = {99},
        pages = {99},
          doi = {10.3847/1538-4357/aa78fc},
archivePrefix = {arXiv},
       eprint = {1706.03320},
 primaryClass = {astro-ph.SR},
       adsurl = {https://ui.adsabs.harvard.edu/abs/2017ApJ...844...99L}
}

@article{calahan_tw_2021,
    title = {The {TW} {Hya} {Rosetta} {Stone} {Project}. {III}. {Resolving} the {Gaseous} {Thermal} {Profile} of the {Disk}},
    volume = {908},
    issn = {0004-637X, 1538-4357},
    url = {https://iopscience.iop.org/article/10.3847/1538-4357/abd255},
    doi = {10.3847/1538-4357/abd255},
    language = {en},
    number = {1},
    urldate = {2024-04-03},
    journal = {The Astrophysical Journal},
    author = {Calahan, Jenny K. and Bergin, Edwin and Zhang, Ke and Teague, Richard and Cleeves, Ilsedore and Bergner, Jennifer and Blake, Geoffrey A. and Cazzoletti, Paolo and Guzmán, Viviana and R. Hogerheijde, Michiel and Huang, Jane and Kama, Mihkel and Loomis, Ryan and Öberg, Karin and Qi, Charlie and Van Dishoeck, Ewine F. and Van Scheltinga, Jeroen Terwisscha and Walsh, Catherine and Wilner, David},
    month = feb,
    year = {2021},
    pages = {8},
}

@article{andrews_disk_2018,
    title = {The \{{Disk}\} \{{Substructures}\} at \{{High}\} \{{Angular}\} \{{Resolution}\} \{{Project}\} (\{{DSHARP}\}). \{{I}\}. \{{Motivation}\}, \{{Sample}\}, \{{Calibration}\}, and \{{Overview}\}},
    volume = {869},
    url = {http://adsabs.harvard.edu/abs/2018ApJ...869L..41A},
    doi = {10.3847/2041-8213/aaf741},
    urldate = {2020-08-21},
    journal = {The Astrophysical Journal Letters},
    author = {Andrews, Sean M and Huang, Jane and Pérez, Laura M and Isella, Andrea and Dullemond, Cornelis P and Kurtovic, Nicolás T and Guzmán, Viviana V and Carpenter, John M and Wilner, David J and Zhang, Shangjia and Zhu, Zhaohuan and Birnstiel, Tilman and Bai, Xue-Ning and Benisty, Myriam and Hughes, A Meredith and Öberg, Karin I and Ricci, Luca},
    year = {2018},
    pages = {L41},
}

@article{oberg_molecules_2021,
    title = {Molecules with {ALMA} at {Planet}-forming {Scales} ({MAPS}). {I}. {Program} {Overview} and {Highlights}},
    volume = {257},
    issn = {0067-0049},
    doi = {10.3847/1538-4365/ac1432},
    number = {1},
    journal = {The Astrophysical Journal Supplement Series},
    author = {Öberg, Karin I. and Guzmán, Viviana V. and Walsh, Catherine and Aikawa, Yuri and Bergin, Edwin A. and Law, Charles J. and Loomis, Ryan A. and Alarcón, Felipe and Andrews, Sean M. and Bae, Jaehan and Bergner, Jennifer B. and Boehler, Yann and Booth, Alice S. and Bosman, Arthur D. and Calahan, Jenny K. and Cataldi, Gianni and Cleeves, L. Ilsedore and Czekala, Ian and Furuya, Kenji and Huang, Jane and Ilee, John D. and Kurtovic, Nicolas T. and Le Gal, Romane and Liu, Yao and Long, Feng and Ménard, François and Nomura, Hideko and Pérez, Laura M. and Qi, Chunhua and Schwarz, Kamber R. and Sierra, Anibal and Teague, Richard and Tsukagoshi, Takashi and Yamato, Yoshihide and van ’t Hoff, Merel L. R. and Waggoner, Abygail R. and Wilner, David J. and Zhang, Ke},
    month = nov,
    year = {2021},
    note = {arXiv: 2109.06268
Publisher: American Astronomical Society},
    pages = {1},
}

@article{huang_disk_2018,
    title = {The \{{Disk}\} \{{Substructures}\} at \{{High}\} \{{Angular}\} \{{Resolution}\} \{{Project}\} (\{{DSHARP}\}). \{{II}\}. \{{Characteristics}\} of \{{Annular}\} \{{Substructures}\}},
    volume = {869},
    issn = {0004-637X},
    url = {https://ui.adsabs.harvard.edu/abs/2018ApJ...869L..42H/abstract},
    doi = {10.3847/2041-8213/aaf740},
    number = {2},
    urldate = {2020-08-21},
    journal = {The Astrophysical Journal},
    author = {Huang, Jane and Andrews, Sean M and Dullemond, Cornelis P and Isella, Andrea and Pérez, Laura M and Guzmán, Viviana V and Öberg, Karin I and Zhu, Zhaohuan and Zhang, Shangjia and Bai, Xue-Ning and Benisty, Myriam and Birnstiel, Tilman and Carpenter, John M and Hughes, A Meredith and Ricci, Luca and Weaver, Erik and Wilner, David J},
    year = {2018},
    pages = {L42},
}

@article{martire_rotation_2024,
    title = {Rotation curves in protoplanetary disks with thermal stratification: {Physical} model and observational evidence in {MAPS} disks},
    volume = {686},
    copyright = {https://creativecommons.org/licenses/by/4.0},
    issn = {0004-6361, 1432-0746},
    shorttitle = {Rotation curves in protoplanetary disks with thermal stratification},
    url = {https://www.aanda.org/10.1051/0004-6361/202348546},
    doi = {10.1051/0004-6361/202348546},
    language = {en},
    urldate = {2024-12-03},
    journal = {Astronomy \& Astrophysics},
    author = {Martire, P. and Longarini, C. and Lodato, G. and Rosotti, G. P. and Winter, A. and Facchini, S. and Hardiman, C. and Benisty, M. and Stadler, J. and Izquierdo, A. F. and Testi, Leonardo},
    month = jun,
    year = {2024},
    pages = {A9},
}

@article{dominik_optool_2021,
    title = {{OpTool}: {Command}-line driven tool for creating complex dust opacities},
    shorttitle = {{OpTool}},
    url = {https://ui.adsabs.harvard.edu/abs/2021ascl.soft04010D},
    urldate = {2025-02-20},
    journal = {Astrophysics Source Code Library},
    author = {Dominik, Carsten and Min, Michiel and Tazaki, Ryo},
    month = apr,
    year = {2021},
    note = {ADS Bibcode: 2021ascl.soft04010D},
    pages = {ascl:2104.010},
}

@INPROCEEDINGS{Manara2022DemographicsFormation,
       author = {{Manara}, C.~F. and {Ansdell}, M. and {Rosotti}, G.~P. and {Hughes}, A.~M. and {Armitage}, P.~J. and {Lodato}, G. and {Williams}, J.~P.},
        title = "{Demographics of Young Stars and their Protoplanetary Disks: Lessons Learned on Disk Evolution and its Connection to Planet Formation}",
    booktitle = {Protostars and Planets VII},
         year = 2023,
       editor = {{Inutsuka}, S. and {Aikawa}, Y. and {Muto}, T. and {Tomida}, K. and {Tamura}, M.},
       series = {Astronomical Society of the Pacific Conference Series},
       volume = {534},
        month = jul,
        pages = {539},
          doi = {10.48550/arXiv.2203.09930},
archivePrefix = {arXiv},
       eprint = {2203.09930},
 primaryClass = {astro-ph.SR},
       adsurl = {https://ui.adsabs.harvard.edu/abs/2023ASPC..534..539M}
}

@article{birnstiel_disk_2018,
    title = {The {Disk} {Substructures} at {High} {Angular} {Resolution} {Project} ({DSHARP}). {V}. {Interpreting} {ALMA} {Maps} of {Protoplanetary} {Disks} in {Terms} of a {Dust} {Model}},
    volume = {869},
    issn = {2041-8213},
    url = {http://dx.doi.org/10.3847/2041-8213/aaf743},
    doi = {10.3847/2041-8213/aaf743},
    number = {2},
    journal = {The Astrophysical Journal},
    author = {Birnstiel, Tilman and Dullemond, Cornelis P. and Zhu, Zhaohuan and Andrews, Sean M. and Bai, Xue-Ning and Wilner, David J. and Carpenter, John M. and Huang, Jane and Isella, Andrea and Benisty, Myriam and Pérez, Laura M. and Zhang, Shangjia},
    year = {2018},
    note = {arXiv: 1812.04043
Publisher: IOP Publishing},
    pages = {L45},
}

@article{visser_photodissociation_2009,
    title = {{The photodissociation and chemistry of {\{}CO{\}} isotopologues: applications to interstellar clouds and circumstellar disks}},
    shorttitle = {The photodissociation and chemistry of {\{}CO{\}} isotop},
    year = {2009},
    journal = {Astronomy {\&} Astrophysics},
    author = {Visser, R and van Dishoeck, E F and Black, J H},
    number = {2},
    pages = {323--343},
    volume = {503},
    url = {http://www.aanda.org/10.1051/0004-6361/200912129},
    doi = {10.1051/0004-6361/200912129},
    issn = {0004-6361, 1432-0746}
}

@misc{paneque-carreno_vertical_2025,
    title = {Vertical {CO} surfaces as a probe for protoplanetary disk mass and carbon depletion},
    url = {https://ui.adsabs.harvard.edu/abs/2025arXiv250108294P},
    urldate = {2025-01-15},
    author = {Paneque-Carreño, T. and Miotello, A. and van Dishoeck, E. F. and Rosotti, G. and Tabone, B.},
    month = jan,
    year = {2025},
    note = {Publication Title: arXiv e-prints
ADS Bibcode: 2025arXiv250108294P},
}

@article{birnstiel_dust_2024,
    title = {Dust {Growth} and {Evolution} in {Protoplanetary} {Disks}},
    volume = {62},
    issn = {0066-4146},
    url = {https://ui.adsabs.harvard.edu/abs/2024ARA&A..62..157B},
    doi = {10.1146/annurev-astro-071221-052705},
    urldate = {2025-01-28},
    journal = {Annual Review of Astronomy and Astrophysics},
    author = {Birnstiel, Tilman},
    month = sep,
    year = {2024},
    note = {ADS Bibcode: 2024ARA\&A..62..157B},
    pages = {157--202},
}

@article{trapman_exoalma_2025,
	title = {{exoALMA}. {XIII}. {Gas} {Masses} from {N2H}+ and {C18O}: {A} {Comparison} of {Measurement} {Techniques} for {Protoplanetary} {Gas} {Disk} {Masses}},
	volume = {984},
	issn = {2041-8205},
	shorttitle = {{exoALMA}. {XIII}. {Gas} {Masses} from {N2H}+ and {C18O}},
	url = {https://dx.doi.org/10.3847/2041-8213/adc430},
	doi = {10.3847/2041-8213/adc430},
	language = {en},
	number = {1},
	urldate = {2025-04-29},
	journal = {The Astrophysical Journal Letters},
	author = {{Trapman}, Leon and Longarini, Cristiano and Rosotti, Giovanni P. and Andrews, Sean M. and Bae, Jaehan and Barraza-Alfaro, Marcelo and Benisty, Myriam and Cataldi, Gianni and Curone, Pietro and Czekala, Ian and Facchini, Stefano and Fasano, Daniele and Flock, Mario and Fukagawa, Misato and Galloway-Sprietsma, Maria and Garg, Himanshi and Hall, Cassandra and Huang, Jane and Ilee, John D. and Izquierdo, Andres F. and Kanagawa, Kazuhiro and Lesur, Geoffroy and Lodato, Giuseppe and Loomis, Ryan A. and Orihara, Ryuta and Paneque-Carreno, Teresa and Pinte, Christophe and Price, Daniel and Stadler, Jochen and Teague, Richard and van Terwisga, Sierk and Testi, Leonardo and Yen, Hsi-Wei and Wafflard-Fernandez, Gaylor and Wilner, David J. and Winter, Andrew J. and Wölfer, Lisa and Yoshida, Tomohiro C. and Zawadzki, Brianna and Zhang, Ke},
	month = apr,
	year = {2025},
	note = {Publisher: The American Astronomical Society},
	pages = {L18},
}

@article{youdin_streaming_2005,
    title = {Streaming {Instabilities} in {Protoplanetary} {Disks}},
    volume = {620},
    issn = {0004-637X},
    url = {https://ui.adsabs.harvard.edu/abs/2005ApJ...620..459Y},
    doi = {10.1086/426895},
    urldate = {2025-05-01},
    journal = {The Astrophysical Journal},
    author = {Youdin, Andrew N. and Goodman, Jeremy},
    month = feb,
    year = {2005},
    note = {Publisher: IOP
ADS Bibcode: 2005ApJ...620..459Y},
    pages = {459--469},
}

@article{li_thresholds_2021,
    title = {Thresholds for {Particle} {Clumping} by the {Streaming} {Instability}},
    volume = {919},
    issn = {0004-637X},
    url = {https://ui.adsabs.harvard.edu/abs/2021ApJ...919..107L},
    doi = {10.3847/1538-4357/ac0e9f},
    urldate = {2025-05-01},
    journal = {The Astrophysical Journal},
    author = {Li, Rixin and Youdin, Andrew N.},
    month = oct,
    year = {2021},
    note = {Publisher: IOP
ADS Bibcode: 2021ApJ...919..107L},
    pages = {107},
}

@article{bruderer_warm_2012,
    title = {The warm gas atmosphere of the {HD} 100546 disk seen by {Herschel}: {Evidence} of a gas-rich, carbon-poor atmosphere?},
    volume = {541},
    issn = {00046361},
    doi = {10.1051/0004-6361/201118218},
    journal = {Astronomy and Astrophysics},
    author = {Bruderer, S. and Van Dishoeck, E. F. and Doty, S. D. and Herczeg, G. J.},
    year = {2012},
    note = {arXiv: 1201.4860},
}

@article{bruderer_survival_2013,
    title = {Survival of molecular gas in cavities of transition disks: {I}. {CO}},
    volume = {559},
    issn = {14320746},
    doi = {10.1051/0004-6361/201321171},
    journal = {Astronomy and Astrophysics},
    author = {Bruderer, Simon},
    year = {2013},
    note = {arXiv: 1308.2966
Publisher: EDP Sciences},
}

@article{oberg_disk_2011,
    title = {Disk {Imaging} {Survey} of {Chemistry} with {SMA}. {II}. {Southern} {Sky} {Protoplanetary} {Disk} {Data} and {Full} {Sample} {Statistics}},
    volume = {734},
    issn = {0004-637X},
    url = {https://ui.adsabs.harvard.edu/abs/2011ApJ...734...98O},
    doi = {10.1088/0004-637X/734/2/98},
    urldate = {2025-05-21},
    journal = {The Astrophysical Journal},
    author = {Öberg, Karin I. and Qi, Chunhua and Fogel, Jeffrey K. J. and Bergin, Edwin A. and Andrews, Sean M. and Espaillat, Catherine and Wilner, David J. and Pascucci, Ilaria and Kastner, Joel H.},
    month = jun,
    year = {2011},
    note = {Publisher: IOP
ADS Bibcode: 2011ApJ...734...98O},
    pages = {98},
}

@article{sturm_edge-protoplanetary_2023,
    title = {The edge-on protoplanetary disk {HH} 48 {NE}. {I}. {Modeling} the geometry and stellar parameters},
    volume = {677},
    issn = {0004-6361},
    url = {https://ui.adsabs.harvard.edu/abs/2023A&A...677A..17S},
    doi = {10.1051/0004-6361/202346052},
    urldate = {2025-02-03},
    journal = {Astronomy and Astrophysics},
    author = {Sturm, J. A. and McClure, M. K. and Law, C. J. and Harsono, D. and Bergner, J. B. and Dartois, E. and Drozdovskaya, M. N. and Ioppolo, S. and Öberg, K. I. and Palumbo, M. E. and Pendleton, Y. J. and Rocha, W. R. M. and Terada, H. and Urso, R. G.},
    month = sep,
    year = {2023},
    note = {Publisher: EDP
ADS Bibcode: 2023A\&A...677A..17S},
    pages = {A17},
}

@article{bergner_jwst_2024,
    title = {{JWST} {Ice} {Band} {Profiles} {Reveal} {Mixed} {Ice} {Compositions} in the {HH} 48 {NE} {Disk}},
    volume = {975},
    issn = {0004-637X},
    url = {https://ui.adsabs.harvard.edu/abs/2024ApJ...975..166B},
    doi = {10.3847/1538-4357/ad79fc},
    urldate = {2025-02-06},
    journal = {The Astrophysical Journal},
    author = {Bergner, Jennifer B. and Sturm, J. A. and Piacentino, Elettra L. and McClure, M. K. and Öberg, Karin I. and Boogert, A. C. A. and Dartois, E. and Drozdovskaya, M. N. and Fraser, H. J. and Harsono, Daniel and Ioppolo, Sergio and Law, Charles J. and Lis, Dariusz C. and McGuire, Brett A. and Melnick, Gary J. and Noble, Jennifer A. and Palumbo, M. E. and Pendleton, Yvonne J. and Perotti, Giulia and Qasim, Danna and Rocha, W. R. M. and van Dishoeck, E. F.},
    month = nov,
    year = {2024},
    note = {Publisher: IOP
ADS Bibcode: 2024ApJ...975..166B},
    pages = {166},
}

@misc{ruaud_cold_2024,
    title = {Cold water emission cannot be used to infer depletion of bulk elemental oxygen [{O}/{H}] in disks},
    url = {https://ui.adsabs.harvard.edu/abs/2024arXiv240604457R},
    urldate = {2024-06-10},
    author = {Ruaud, Maxime and Gorti, Uma},
    month = jun,
    year = {2024},
    note = {Publication Title: arXiv e-prints
ADS Bibcode: 2024arXiv240604457R},
}

@article{Hunter2007,
  title={Matplotlib: A 2D graphics environment},
  author={Hunter, John D},
  journal={Computing in science \& engineering},
  volume={9},
  number={3},
  pages={90--95},
  year={2007},
  publisher={IEEE}
}

@ARTICLE{astropy:2018,
       author = {{Astropy Collaboration} and {Price-Whelan}, A.~M. and
         {Sip{\H{o}}cz}, B.~M. and {G{\"u}nther}, H.~M. and {Lim}, P.~L. and
         {Crawford}, S.~M. and {Conseil}, S. and {Shupe}, D.~L. and
         {Craig}, M.~W. and {Dencheva}, N. and {Ginsburg}, A. and {Vand
        erPlas}, J.~T. and {Bradley}, L.~D. and {P{\'e}rez-Su{\'a}rez}, D. and
         {de Val-Borro}, M. and {Aldcroft}, T.~L. and {Cruz}, K.~L. and
         {Robitaille}, T.~P. and {Tollerud}, E.~J. and {Ardelean}, C. and
         {Babej}, T. and {Bach}, Y.~P. and {Bachetti}, M. and {Bakanov}, A.~V. and
         {Bamford}, S.~P. and {Barentsen}, G. and {Barmby}, P. and
         {Baumbach}, A. and {Berry}, K.~L. and {Biscani}, F. and {Boquien}, M. and
         {Bostroem}, K.~A. and {Bouma}, L.~G. and {Brammer}, G.~B. and
         {Bray}, E.~M. and {Breytenbach}, H. and {Buddelmeijer}, H. and
         {Burke}, D.~J. and {Calderone}, G. and {Cano Rodr{\'\i}guez}, J.~L. and
         {Cara}, M. and {Cardoso}, J.~V.~M. and {Cheedella}, S. and {Copin}, Y. and
         {Corrales}, L. and {Crichton}, D. and {D'Avella}, D. and {Deil}, C. and
         {Depagne}, {\'E}. and {Dietrich}, J.~P. and {Donath}, A. and
         {Droettboom}, M. and {Earl}, N. and {Erben}, T. and {Fabbro}, S. and
         {Ferreira}, L.~A. and {Finethy}, T. and {Fox}, R.~T. and
         {Garrison}, L.~H. and {Gibbons}, S.~L.~J. and {Goldstein}, D.~A. and
         {Gommers}, R. and {Greco}, J.~P. and {Greenfield}, P. and
         {Groener}, A.~M. and {Grollier}, F. and {Hagen}, A. and {Hirst}, P. and
         {Homeier}, D. and {Horton}, A.~J. and {Hosseinzadeh}, G. and {Hu}, L. and
         {Hunkeler}, J.~S. and {Ivezi{\'c}}, {\v{Z}}. and {Jain}, A. and
         {Jenness}, T. and {Kanarek}, G. and {Kendrew}, S. and {Kern}, N.~S. and
         {Kerzendorf}, W.~E. and {Khvalko}, A. and {King}, J. and {Kirkby}, D. and
         {Kulkarni}, A.~M. and {Kumar}, A. and {Lee}, A. and {Lenz}, D. and
         {Littlefair}, S.~P. and {Ma}, Z. and {Macleod}, D.~M. and
         {Mastropietro}, M. and {McCully}, C. and {Montagnac}, S. and
         {Morris}, B.~M. and {Mueller}, M. and {Mumford}, S.~J. and {Muna}, D. and
         {Murphy}, N.~A. and {Nelson}, S. and {Nguyen}, G.~H. and
         {Ninan}, J.~P. and {N{\"o}the}, M. and {Ogaz}, S. and {Oh}, S. and
         {Parejko}, J.~K. and {Parley}, N. and {Pascual}, S. and {Patil}, R. and
         {Patil}, A.~A. and {Plunkett}, A.~L. and {Prochaska}, J.~X. and
         {Rastogi}, T. and {Reddy Janga}, V. and {Sabater}, J. and
         {Sakurikar}, P. and {Seifert}, M. and {Sherbert}, L.~E. and
         {Sherwood-Taylor}, H. and {Shih}, A.~Y. and {Sick}, J. and
         {Silbiger}, M.~T. and {Singanamalla}, S. and {Singer}, L.~P. and
         {Sladen}, P.~H. and {Sooley}, K.~A. and {Sornarajah}, S. and
         {Streicher}, O. and {Teuben}, P. and {Thomas}, S.~W. and
         {Tremblay}, G.~R. and {Turner}, J.~E.~H. and {Terr{\'o}n}, V. and
         {van Kerkwijk}, M.~H. and {de la Vega}, A. and {Watkins}, L.~L. and
         {Weaver}, B.~A. and {Whitmore}, J.~B. and {Woillez}, J. and
         {Zabalza}, V. and {Astropy Contributors}},
        title = "{The Astropy Project: Building an Open-science Project and Status of the v2.0 Core Package}",
      journal = {\aj},
         year = "2018",
        month = "Sep",
       volume = {156},
       number = {3},
          eid = {123},
        pages = {123},
          doi = {10.3847/1538-3881/aabc4f},
archivePrefix = {arXiv},
       eprint = {1801.02634},
 primaryClass = {astro-ph.IM},
       adsurl = {https://ui.adsabs.harvard.edu/abs/2018AJ....156..123A}
}

@ARTICLE{2020SciPy-NMeth,
  author  = {Virtanen, Pauli and Gommers, Ralf and Oliphant, Travis E. and
            Haberland, Matt and Reddy, Tyler and Cournapeau, David and
            Burovski, Evgeni and Peterson, Pearu and Weckesser, Warren and
            Bright, Jonathan and {van der Walt}, St{\'e}fan J. and
            Brett, Matthew and Wilson, Joshua and Millman, K. Jarrod and
            Mayorov, Nikolay and Nelson, Andrew R. J. and Jones, Eric and
            Kern, Robert and Larson, Eric and Carey, C J and
            Polat, {\.I}lhan and Feng, Yu and Moore, Eric W. and
            {VanderPlas}, Jake and Laxalde, Denis and Perktold, Josef and
            Cimrman, Robert and Henriksen, Ian and Quintero, E. A. and
            Harris, Charles R. and Archibald, Anne M. and
            Ribeiro, Ant{\^o}nio H. and Pedregosa, Fabian and
            {van Mulbregt}, Paul and {SciPy 1.0 Contributors}},
  title   = {{{SciPy} 1.0: Fundamental Algorithms for Scientific
            Computing in Python}},
  journal = {Nature Methods},
  year    = {2020},
  volume  = {17},
  pages   = {261--272},
  adsurl  = {https://rdcu.be/b08Wh},
  doi     = {10.1038/s41592-019-0686-2},
}

@article{teague_gofish_2019,
    title = {{GoFish}: {Fishing} for {Line} {Observations} in {Protoplanetary} {Disks}},
    volume = {4},
    doi = {10.21105/joss.01632},
    number = {41},
    urldate = {2022-07-19},
    journal = {Journal of Open Source Software},
    author = {Teague, Richard},
    month = sep,
    year = {2019},
    note = {Publisher: The Open Journal},
    pages = {1632},
}

@misc{zwicky_dancing_2025,
    title = {Dancing on the {Grain}: {Variety} of {CO} and its isotopologue fluxes as a result of surface chemistry and {T} {Tauri} disk properties},
    shorttitle = {Dancing on the {Grain}},
    url = {https://ui.adsabs.harvard.edu/abs/2025arXiv250615508Z},
    urldate = {2025-06-19},
    publisher = {arXiv},
    author = {Zwicky, L. and Molyarova, T. and K\'{o}sp\'{a}l, \'{A}. and \'{A}brah\'{a}m, P.},
    month = jun,
    year = {2025},
    note = {ADS Bibcode: 2025arXiv250615508Z},
}

@article{longarini_exoalma_2025,
    title = {{exoALMA}. {XII}. {Weighing} and {Sizing} {exoALMA} {Disks} with {Rotation} {Curve} {Modelling}},
    volume = {984},
    issn = {2041-8205},
    url = {https://dx.doi.org/10.3847/2041-8213/adc431},
    doi = {10.3847/2041-8213/adc431},
    language = {en},
    number = {1},
    urldate = {2025-04-29},
    journal = {The Astrophysical Journal Letters},
    publisher = {The American Astronomical Society},
    author = {Longarini, Cristiano and Lodato, Giuseppe and Rosotti, Giovanni and Andrews, Sean and Winter, Andrew and Stadler, Jochen and Izquierdo, Andrés and Galloway-Sprietsma, Maria and Facchini, Stefano and Curone, Pietro and Benisty, Myriam and Teague, Richard and Bae, Jaehan and Barraza-Alfaro, Marcelo and Cataldi, Gianni and Czekala, Ian and Cuello, Nicolás and Fasano, Daniele and Flock, Mario and Fukagawa, Misato and Garg, Himanshi and Hall, Cassandra and Hammond, Iain and Hardiman, Caitlyn and Hilder, Thomas and Huang, Jane and Ilee, John D. and Isella, Andrea and Kanagawa, Kazuhiro and Lesur, Geoffroy and Loomis, Ryan A. and Ménard, Francois and Orihara, Ryuta and Pinte, Christophe and Price, Daniel and Testi, Leonardo and Fernandez, Gaylor Wafflard- and Wölfer, Lisa and Yen, Hsi-Wei and Yoshida, Tomohiro C. and Zawadzki, Brianna},
    month = apr,
    year = {2025},
    pages = {L17},
}

@article{anderson_new_2022,
    title = {New {Constraints} on {Protoplanetary} {Disk} {Gas} {Masses} in {Lupus}},
    volume = {927},
    issn = {0004-637X, 1538-4357},
    url = {https://iopscience.iop.org/article/10.3847/1538-4357/ac517e},
    doi = {10.3847/1538-4357/ac517e},
    language = {en},
    number = {2},
    urldate = {2024-02-14},
    journal = {The Astrophysical Journal},
    author = {Anderson, Dana E. and Cleeves, L. Ilsedore and Blake, Geoffrey A. and Bergin, Edwin A. and Zhang, Ke and Carpenter, John M. and Schwarz, Kamber R.},
    month = mar,
    year = {2022},
    pages = {229},
}

@article{trapman_novel_2022,
    title = {A {Novel} {Way} of {Measuring} the {Gas} {Disk} {Mass} of {Protoplanetary} {Disks} {Using} {N} 2 {H} + and {C} 18 {O}},
    volume = {926},
    issn = {2041-8205},
    doi = {10.3847/2041-8213/ac4f47},
    number = {1},
    journal = {The Astrophysical Journal Letters},
    publisher = {American Astronomical Society},
    author = {Trapman, Leon and Zhang, Ke and van ’t Hoff, Merel L. R. and Hogerheijde, Michiel R. and Bergin, Edwin A.},
    month = feb,
    year = {2022},
    note = {arXiv: 2201.09900},
    pages = {L2},
}

@article{zhang_alma_2025,
    title = {The {ALMA} {Survey} of {Gas} {Evolution} of {PROtoplanetary} {Disks} ({AGE}-{PRO}). {I}. {Program} {Overview} and {Summary} of {First} {Results}},
    volume = {989},
    issn = {0004-637X},
    url = {https://ui.adsabs.harvard.edu/abs/2025ApJ...989....1Z},
    doi = {10.3847/1538-4357/addebe},
    urldate = {2025-09-08},
    journal = {The Astrophysical Journal},
    author = {Zhang, Ke and Pérez, Laura M. and Pascucci, Ilaria and Pinilla, Paola and Cieza, Lucas A. and Carpenter, John and Trapman, Leon and Deng, Dingshan and Agurto-Gangas, Carolina and Sierra, Anibal and Kurtovic, Nicolás T. and Ruiz-Rodriguez, Dary A. and Vioque, Miguel and Miley, James and Tabone, Benoît and González-Ruilova, Camilo and Anania, Rossella and Rosotti, Giovanni P. and TorresVillanueva, Estephani and Hogerheijde, Michiel R. and Schwarz, Kamber and Kuznetsova, Aleksandra},
    month = aug,
    year = {2025},
    note = {ADS Bibcode: 2025ApJ...989....1Z},
    pages = {1},
}

@article{teague_exoalma_2025,
    title = {{exoALMA}. {I}. {Science} {Goals}, {Project} {Design}, and {Data} {Products}},
    volume = {984},
    issn = {2041-8205},
    url = {https://dx.doi.org/10.3847/2041-8213/adc43b},
    doi = {10.3847/2041-8213/adc43b},
    language = {en},
    number = {1},
    urldate = {2025-04-29},
    journal = {The Astrophysical Journal Letters},
    publisher = {The American Astronomical Society},
    author = {Teague, Richard and Benisty, Myriam and Facchini, Stefano and Fukagawa, Misato and Pinte, Christophe and Andrews, Sean M. and Bae, Jaehan and Barraza-Alfaro, Marcelo and Cataldi, Gianni and Cuello, Nicolás and Curone, Pietro and Czekala, Ian and Fasano, Daniele and Flock, Mario and Galloway-Sprietsma, Maria and Garg, Himanshi and Hall, Cassandra and Hammond, Iain and Hilder, Thomas and Huang, Jane and Ilee, John D. and Izquierdo, Andrés F. and Kanagawa, Kazuhiro and Lesur, Geoffroy and Lodato, Giuseppe and Longarini, Cristiano and Loomis, Ryan A. and Masset, Frédéric and Menard, Francois and Orihara, Ryuta and Price, Daniel J. and Rosotti, Giovanni and Stadler, Jochen and Testi, Leonardo and Yen, Hsi-Wei and Wafflard-Fernandez, Gaylor and Wilner, David J. and Winter, Andrew J. and Wölfer, Lisa and Yoshida, Tomohiro C. and Zawadzki, Brianna},
    month = apr,
    year = {2025},
    pages = {L6},
}

@article{agurto-gangas_alma_2025,
    title = {The {ALMA} {Survey} of {Gas} {Evolution} of {PROtoplanetary} {Disks} ({AGE}-{PRO}). {IV}. {Dust} and {Gas} {Disk} {Properties} in the {Upper} {Scorpius} {Star}-forming {Region}},
    volume = {989},
    issn = {0004-637X},
    url = {https://ui.adsabs.harvard.edu/abs/2025ApJ...989....4A},
    doi = {10.3847/1538-4357/adc7ab},
    urldate = {2025-09-08},
    journal = {The Astrophysical Journal},
    author = {Agurto-Gangas, Carolina and Pérez, Laura M. and Sierra, Anibal and Miley, James and Zhang, Ke and Pascucci, Ilaria and Pinilla, Paola and Deng, Dingshan and Carpenter, John and Trapman, Leon and Vioque, Miguel and Rosotti, Giovanni P. and Kurtovic, Nicolas and Cieza, Lucas A. and Anania, Rossella and Tabone, Benoît and Schwarz, Kamber and Hogerheijde, Michiel R. and TorresVillanueva, Estephani E. and Ruiz-Rodriguez, Dary A. and González-Ruilova, Camilo},
    month = aug,
    year = {2025},
    note = {ADS Bibcode: 2025ApJ...989....4A},
    pages = {4},
}

@article{vioque_alma_2025,
    title = {The {ALMA} {Survey} of {Gas} {Evolution} of {PROtoplanetary} {Disks} ({AGE}-{PRO}). {X}. {Dust} {Substructures}, {Disk} {Geometries}, and {Dust}-disk {Radii}},
    volume = {989},
    issn = {0004-637X},
    url = {https://ui.adsabs.harvard.edu/abs/2025ApJ...989....9V},
    doi = {10.3847/1538-4357/adc7b0},
    urldate = {2025-09-08},
    journal = {The Astrophysical Journal},
    author = {Vioque, Miguel and Kurtovic, Nicolás T. and Trapman, Leon and Sierra, Anibal and Pérez, Laura M. and Zhang, Ke and Curone, Pietro and Rosotti, Giovanni P. and Carpenter, John and Tabone, Benoît and Pinilla, Paola and Deng, Dingshan and Pascucci, Ilaria and Miley, James and Agurto-Gangas, Carolina and Cieza, Lucas A. and Anania, Rossella and Ruiz-Rodriguez, Dary A. and González-Ruilova, Camilo and TorresVillanueva, Estephani E. and Kuznetsova, Aleksandra},
    month = aug,
    year = {2025},
    note = {ADS Bibcode: 2025ApJ...989....9V},
    pages = {9},
}

@article{galloway-sprietsma_exoalma_2025,
    title = {{exoALMA}. {V}. {Gaseous} {Emission} {Surfaces} and {Temperature} {Structures}},
    volume = {984},
    issn = {2041-8205},
    url = {https://dx.doi.org/10.3847/2041-8213/adc437},
    doi = {10.3847/2041-8213/adc437},
    language = {en},
    number = {1},
    urldate = {2025-04-29},
    journal = {The Astrophysical Journal Letters},
    publisher = {The American Astronomical Society},
    author = {Galloway-Sprietsma, Maria and Bae, Jaehan and Izquierdo, Andrés F. and Stadler, Jochen and Longarini, Cristiano and Teague, Richard and Andrews, Sean M. and Winter, Andrew J. and Benisty, Myriam and Facchini, Stefano and Rosotti, Giovanni and Zawadzki, Brianna and Pinte, Christophe and Fasano, Daniele and Barraza-Alfaro, Marcelo and Cataldi, Gianni and Cuello, Nicolás and Curone, Pietro and Czekala, Ian and Flock, Mario and Fukagawa, Misato and Gardner, Charles H. and Garg, Himanshi and Hall, Cassandra and Huang, Jane and Ilee, John D. and Kanagawa, Kazuhiro and Lesur, Geoffroy and Lodato, Giuseppe and Loomis, Ryan A. and Menard, Francois and Orihara, Ryuta and Price, Daniel J. and Wafflard-Fernandez, Gaylor and Wilner, David J. and Wölfer, Lisa and Yen, Hsi-Wei and Yoshida, Tomohiro C.},
    month = apr,
    year = {2025},
    pages = {L10},
}

@article{feiden_magnetic_2016,
    title = {Magnetic inhibition of convection and the fundamental properties of low-mass stars. {III}. {A} consistent 10 {Myr} age for the {Upper} {Scorpius} {OB} association},
    volume = {593},
    issn = {0004-6361},
    url = {https://ui.adsabs.harvard.edu/abs/2016A&A...593A..99F},
    doi = {10.1051/0004-6361/201527613},
    urldate = {2024-09-13},
    journal = {Astronomy and Astrophysics},
    author = {Feiden, Gregory A.},
    month = sep,
    year = {2016},
    note = {ADS Bibcode: 2016A\&A...593A..99F},
    pages = {A99},
}

@article{toomre_gravitational_1964,
    title = {On the gravitational stability of a disk of stars.},
    volume = {139},
    issn = {0004-637X},
    url = {https://ui.adsabs.harvard.edu/abs/1964ApJ...139.1217T},
    doi = {10.1086/147861},
    urldate = {2026-01-26},
    journal = {The Astrophysical Journal},
    publisher = {IOP},
    author = {Toomre, A.},
    month = may,
    year = {1964},
    note = {ADS Bibcode: 1964ApJ...139.1217T},
    pages = {1217--1238},
}

@article{qi_probing_2019,
	title = {Probing {CO} and {N2} {Snow} {Surfaces} in {Protoplanetary} {Disks} with {N2H}+ {Emission}},
	volume = {882},
	issn = {1538-4357},
	url = {http://dx.doi.org/10.3847/1538-4357/ab35d3},
	doi = {10.3847/1538-4357/ab35d3},
	number = {2},
	journal = {arXiv},
	publisher = {IOP Publishing},
	author = {Qi, Chunhua and Öberg, Karin I. and Espaillat, Catherine C. and Robinson, Connor E. and Andrews, Sean M. and Wilner, David J. and Blake, Geoffrey A. and Bergin, Edwin A. and Cleeves, L. Ilsedore},
	year = {2019},
	note = {arXiv: 1907.10647},
	pages = {160},
}

@article{flaherty_measuring_2020,
	title = {Measuring {Turbulent} {Motion} in {Planet}-forming {Disks} with {ALMA}: {A} {Detection} around {DM} {Tau} and {Nondetections} around {MWC} 480 and {V4046} {Sgr}},
	volume = {895},
	issn = {15384357},
	url = {https://ui.adsabs.harvard.edu/abs/2020ApJ...895..109F/abstract},
	doi = {10.3847/1538-4357/ab8cc5},
	number = {2},
	urldate = {2022-07-29},
	journal = {The Astrophysical Journal},
	publisher = {American Astronomical Society},
	author = {Flaherty, Kevin and Hughes, A. Meredith and Simon, Jacob B. and Qi, Chunhua and Bai, Xue-Ning and Bulatek, Alyssa and Andrews, Sean M. and Wilner, David J. and Kóspál, Ágnes},
	month = jun,
	year = {2020},
	note = {arXiv: 2004.12176},
	pages = {109},
}

@article{stapper_constraining_2024,
	title = {Constraining the gas mass of {Herbig} disks using {CO} isotopologues},
	volume = {682},
	issn = {0004-6361},
	url = {https://ui.adsabs.harvard.edu/abs/2024A&A...682A.149S},
	doi = {10.1051/0004-6361/202347271},
	urldate = {2025-08-06},
	journal = {Astronomy and Astrophysics},
	publisher = {EDP},
	author = {Stapper, L. M. and Hogerheijde, M. R. and van Dishoeck, E. F. and Lin, L. and Ahmadi, A. and Booth, A. S. and Grant, S. L. and Immer, K. and Leemker, M. and Pérez-Sánchez, A. F.},
	month = feb,
	year = {2024},
	note = {ADS Bibcode: 2024A\&A...682A.149S},
	pages = {A149},
}

@article{sturm_disentangling_2023,
	title = {Disentangling the protoplanetary disk gas mass and carbon depletion through combined atomic and molecular tracers},
	volume = {670},
	issn = {0004-6361},
	url = {https://ui.adsabs.harvard.edu/abs/2023A&A...670A..12S},
	doi = {10.1051/0004-6361/202244227},
	urldate = {2026-01-26},
	journal = {Astronomy and Astrophysics},
	publisher = {EDP},
	author = {Sturm, J. A. and Booth, A. S. and McClure, M. K. and Leemker, M. and van Dishoeck, E. F.},
	month = feb,
	year = {2023},
	note = {ADS Bibcode: 2023A\&A...670A..12S},
	pages = {A12},
}

@article{loomis_unbiased_2020,
	title = {An {Unbiased} {ALMA} {Spectral} {Survey} of the {LkCa} 15 and {MWC} 480 {Protoplanetary} {Disks}},
	volume = {893},
	issn = {0004-637X},
	url = {https://ui.adsabs.harvard.edu/abs/2020ApJ...893..101L},
	doi = {10.3847/1538-4357/ab7cc8},
	urldate = {2026-01-26},
	journal = {The Astrophysical Journal},
	publisher = {IOP},
	author = {Loomis, Ryan A. and Öberg, Karin I. and Andrews, Sean M. and Bergin, Edwin and Bergner, Jennifer and Blake, Geoffrey A. and Cleeves, L. Ilsedore and Czekala, Ian and Huang, Jane and Le Gal, Romane and Ménard, Francois and Pegues, Jamila and Qi, Chunhua and Walsh, Catherine and Williams, Jonathan P. and Wilner, David J.},
	month = apr,
	year = {2020},
	note = {ADS Bibcode: 2020ApJ...893..101L},
	pages = {101},
}

@article{qi_chemical_2015,
	title = {Chemical {Imaging} of the {CO} {Snow} {Line} in the {HD} 163296 {Disk}},
	volume = {813},
	issn = {0004-637X},
	url = {https://ui.adsabs.harvard.edu/abs/2015ApJ...813..128Q},
	doi = {10.1088/0004-637X/813/2/128},
	urldate = {2026-01-26},
	journal = {The Astrophysical Journal},
	publisher = {IOP},
	author = {Qi, Chunhua and Öberg, Karin I. and Andrews, Sean M. and Wilner, David J. and Bergin, Edwin A. and Hughes, A. Meredith and Hogherheijde, Michiel and D'Alessio, Paola},
	month = nov,
	year = {2015},
	note = {ADS Bibcode: 2015ApJ...813..128Q},
	pages = {128},
}

@article{ribas_alma_2023,
  title = {The {ALMA} view of {MP} {Mus} ({PDS} 66): {A} protoplanetary disk with no visible gaps down to 4 au scales},
  volume = {673},
  issn = {0004-6361},
  url = {https://ui.adsabs.harvard.edu/abs/2023A&A...673A..77R},
  doi = {10.1051/0004-6361/202245637},
  journal = {Astronomy and Astrophysics},
  publisher = {EDP},
  author = {Ribas, {\'A}. and Mac{\'i}as, E. and Weber, P. and P{\'e}rez, S. and Cuello, N. and Dong, R. and Aguayo, A. and C{\'a}ceres, C. and Carpenter, J. and Dent, W. R. F. and de Gregorio-Monsalvo, I. and Duch{\^e}ne, G. and Espaillat, C. C. and Riviere-Marichalar, P. and Villenave, M.},
  month = may,
  year = {2023},
  note = {ADS Bibcode: 2023A\&A...673A..77R},
  pages = {A77},
}

@article{paneque-carreno_directly_2023,
	title = {Directly tracing the vertical stratification of molecules in protoplanetary disks},
	volume = {669},
	issn = {0004-6361},
	url = {https://ui.adsabs.harvard.edu/abs/2023A&A...669A.126P},
	doi = {10.1051/0004-6361/202244428},
	urldate = {2026-01-26},
	journal = {Astronomy and Astrophysics},
	publisher = {EDP},
	author = {Paneque-Carreño, T. and Miotello, A. and van Dishoeck, E. F. and Tabone, B. and Izquierdo, A. F. and Facchini, S.},
	month = jan,
	year = {2023},
	note = {ADS Bibcode: 2023A\&A...669A.126P},
	pages = {A126},
}

@article{van_der_marel_resolved_2016,
	title = {Resolved gas cavities in transitional disks inferred from {CO} isotopologs with {ALMA}},
	volume = {585},
	issn = {0004-6361},
	url = {https://ui.adsabs.harvard.edu/abs/2016A&A...585A..58V},
	doi = {10.1051/0004-6361/201526988},
	urldate = {2026-01-26},
	journal = {Astronomy and Astrophysics},
	publisher = {EDP},
	author = {van der Marel, N. and van Dishoeck, E. F. and Bruderer, S. and Andrews, S. M. and Pontoppidan, K. M. and Herczeg, G. J. and van Kempen, T. and Miotello, A.},
	month = jan,
	year = {2016},
	note = {ADS Bibcode: 2016A\&A...585A..58V},
	pages = {A58},
}

@article{friedman_greedy_2001,
	title = {Greedy function approximation: {A} gradient boosting machine.},
	volume = {29},
	issn = {0090-5364, 2168-8966},
	shorttitle = {Greedy function approximation},
	url = {https://projecteuclid.org/journals/annals-of-statistics/volume-29/issue-5/Greedy-function-approximation-A-gradient-boosting-machine/10.1214/aos/1013203451.full},
	doi = {10.1214/aos/1013203451},
	language = {en},
	number = {5},
	urldate = {2026-02-18},
	journal = {The Annals of Statistics},
	publisher = {Institute of Mathematical Statistics},
	author = {Friedman, Jerome H.},
	month = oct,
	year = {2001},
	pages = {1189--1232},
}

@Manual{chen_xgboost_2026,
  title = {xgboost: Extreme Gradient Boosting},
  author = {Tianqi Chen and Tong He and Michael Benesty and Vadim Khotilovich and Yuan Tang and Hyunsu Cho and Kailong Chen and Rory Mitchell and Ignacio Cano and Tianyi Zhou and Mu Li and Junyuan Xie and Min Lin and Yifeng Geng and Yutian Li and Jiaming Yuan and David Cortes},
  year = {2026},
  note = {R package version 3.2.0.1},
  url = {https://github.com/dmlc/xgboost},
}

@article{veronesi_dynamical_2021,
    title = {A {Dynamical} {Measurement} of the {Disk} {Mass} in {Elias} 2–27},
    volume = {914},
    issn = {2041-8205},
    url = {https://doi.org/10.3847/2041-8213/abfe6a},
    doi = {10.3847/2041-8213/abfe6a},
    language = {en},
    number = {2},
    urldate = {2025-11-25},
    journal = {The Astrophysical Journal Letters},
    publisher = {The American Astronomical Society},
    author = {Veronesi, Benedetta and Paneque-Carreño, Teresa and Lodato, Giuseppe and Testi, Leonardo and Pérez, Laura M. and Bertin, Giuseppe and Hall, Cassandra},
    month = jun,
    year = {2021},
    pages = {L27},
}

@article{andrews_resolved_2011,
	title = {Resolved images of large cavities in protoplanetary transition disks},
	volume = {732},
	issn = {15384357},
	doi = {10.1088/0004-637X/732/1/42},
	number = {1},
	journal = {Astrophysical Journal},
	publisher = {Institute of Physics Publishing},
	author = {Andrews, Sean M. and Wilner, David J. and Espaillat, Catherine and Hughes, A. M. and Dullemond, C. P. and McClure, M. K. and Qi, Chunhua and Brown, J. M.},
	month = may,
	year = {2011},
	note = {arXiv: 1103.0284},
}

@article{law_molecules_2021,
	title = {Molecules with {ALMA} at {Planet}-forming {Scales} ({MAPS}). {III}. {Characteristics} of {Radial} {Chemical} {Substructures}},
	volume = {257},
	issn = {0067-0049, 1538-4365},
	url = {https://iopscience.iop.org/article/10.3847/1538-4365/ac1434},
	doi = {10.3847/1538-4365/ac1434},
	language = {en},
	number = {1},
	urldate = {2024-02-14},
	journal = {The Astrophysical Journal Supplement Series},
	author = {Law, Charles J. and Loomis, Ryan A. and Teague, Richard and Öberg, Karin I. and Czekala, Ian and Andrews, Sean M. and Huang, Jane and Aikawa, Yuri and Alarcón, Felipe and Bae, Jaehan and Bergin, Edwin A. and Bergner, Jennifer B. and Boehler, Yann and Booth, Alice S. and Bosman, Arthur D. and Calahan, Jenny K. and Cataldi, Gianni and Cleeves, L. Ilsedore and Furuya, Kenji and Guzmán, Viviana V. and Ilee, John D. and Le Gal, Romane and Liu, Yao and Long, Feng and Ménard, François and Nomura, Hideko and Qi, Chunhua and Schwarz, Kamber R. and Sierra, Anibal and Tsukagoshi, Takashi and Yamato, Yoshihide and Van ’T Hoff, Merel L. R. and Walsh, Catherine and Wilner, David J. and Zhang, Ke},
	month = nov,
	year = {2021},
	pages = {3},
}

@article{curone_exoalma_2025,
	title = {{exoALMA}. {IV}. {Substructures}, {Asymmetries}, and the {Faint} {Outer} {Disk} in {Continuum} {Emission}},
	volume = {984},
	issn = {2041-8205},
	url = {https://dx.doi.org/10.3847/2041-8213/adc438},
	doi = {10.3847/2041-8213/adc438},
	language = {en},
	number = {1},
	urldate = {2025-04-29},
	journal = {The Astrophysical Journal Letters},
	publisher = {The American Astronomical Society},
	author = {Curone, Pietro and Facchini, Stefano and Andrews, Sean M. and Testi, Leonardo and Benisty, Myriam and Czekala, Ian and Huang, Jane and Ilee, John D. and Isella, Andrea and Lodato, Giuseppe and Loomis, Ryan A. and Stadler, Jochen and Winter, Andrew J. and Bae, Jaehan and Barraza-Alfaro, Marcelo and Cataldi, Gianni and Cuello, Nicolás and Fasano, Daniele and Flock, Mario and Fukagawa, Misato and Galloway-Sprietsma, Maria and Garg, Himanshi and Hall, Cassandra and Izquierdo, Andrés F. and Kanagawa, Kazuhiro and Lesur, Geoffroy and Longarini, Cristiano and Menard, Francois and Orihara, Ryuta and Pinte, Christophe and Price, Daniel J. and Rosotti, Giovanni and Teague, Richard and Wafflard-Fernandez, Gaylor and Wilner, David J. and Wölfer, Lisa and Yen, Hsi-Wei and Yoshida, Tomohiro C. and Zawadzki, Brianna},
	month = apr,
	year = {2025},
	pages = {L9},
}

@article{izquierdo_exoalma_2025,
	title = {{exoALMA}. {III}. {Line}-intensity {Modeling} and {System} {Property} {Extraction} from {Protoplanetary} {Disks}},
	volume = {984},
	issn = {2041-8205},
	url = {https://dx.doi.org/10.3847/2041-8213/adc439},
	doi = {10.3847/2041-8213/adc439},
	language = {en},
	number = {1},
	urldate = {2025-05-13},
	journal = {The Astrophysical Journal Letters},
	publisher = {The American Astronomical Society},
	author = {Izquierdo, Andrés F. and Stadler, Jochen and Galloway-Sprietsma, Maria and Benisty, Myriam and Pinte, Christophe and Bae, Jaehan and Teague, Richard and Facchini, Stefano and Wölfer, Lisa and Longarini, Cristiano and Curone, Pietro and Andrews, Sean M. and Barraza-Alfaro, Marcelo and Cataldi, Gianni and Cuello, Nicolás and Czekala, Ian and Fasano, Daniele and Flock, Mario and Fukagawa, Misato and Garg, Himanshi and Hall, Cassandra and Hammond, Iain and Hilder, Thomas and Huang, Jane and Ilee, John D. and Isella, Andrea and Kanagawa, Kazuhiro and Lesur, Geoffroy and Lodato, Giuseppe and Loomis, Ryan A. and Orihara, Ryuta and Price, Daniel J. and Rosotti, Giovanni and Testi, Leonardo and Yen, Hsi-Wei and Wafflard-Fernandez, Gaylor and Wilner, David J. and Winter, Andrew J. and Yoshida, Tomohiro C. and Zawadzki, Brianna},
	month = apr,
	year = {2025},
	pages = {L8},
}

@ARTICLE{Smith_ice_NatAs_2025,
       author = {{Smith}, Z.~L. and {Dickinson}, H.~J. and {Fraser}, H.~J. and {McClure}, M.~K. and {Noble}, J.~A. and {Boogert}, A.~C.~A. and {Sun}, F. and {Egami}, E. and {Dartois}, E. and {Erkal}, J. and et al.},
        title = "{Cospatial ice mapping of H$_{2}$O with CO$_{2}$ and CO across a molecular cloud with JWST/NIRCam}",
      journal = {Nature Astronomy},
         year = 2025,
        month = jun,
       volume = {9},
        pages = {883-894},
          doi = {10.1038/s41550-025-02511-z},
       adsurl = {https://ui.adsabs.harvard.edu/abs/2025NatAs...9..883S}
}

@ARTICLE{Williams_n_Best_GasMass_2014,
       author = {{Williams}, Jonathan P. and {Best}, William M.~J.},
        title = "{A Parametric Modeling Approach to Measuring the Gas Masses of Circumstellar Disks}",
      journal = {\apj},
         year = 2014,
        month = jun,
       volume = {788},
       number = {1},
          eid = {59},
        pages = {59},
          doi = {10.1088/0004-637X/788/1/59},
archivePrefix = {arXiv},
       eprint = {1312.0151},
 primaryClass = {astro-ph.EP},
       adsurl = {https://ui.adsabs.harvard.edu/abs/2014ApJ...788...59W}
}

@article{qi_resolving_2011,
	title = {Resolving the {CO} {Snow} {Line} in the {Disk} around {HD} 163296},
	volume = {740},
	issn = {0004-637X},
	url = {https://ui.adsabs.harvard.edu/abs/2011ApJ...740...84Q},
	doi = {10.1088/0004-637X/740/2/84},
	urldate = {2025-11-11},
	journal = {The Astrophysical Journal},
	publisher = {IOP},
	author = {Qi, Chunhua and D'Alessio, Paola and Öberg, Karin I. and Wilner, David J. and Hughes, A. Meredith and Andrews, Sean M. and Ayala, Sandra},
	month = oct,
	year = {2011},
	note = {ADS Bibcode: 2011ApJ...740...84Q},
	pages = {84},
}

@article{caselli_co_1999,
	title = {{CO} {Depletion} in the {Starless} {Cloud} {Core} {L1544}},
	volume = {523},
	issn = {0004-637X},
	url = {https://ui.adsabs.harvard.edu/abs/1999ApJ...523L.165C},
	doi = {10.1086/312280},
	urldate = {2026-03-09},
	journal = {The Astrophysical Journal},
	publisher = {IOP},
	author = {Caselli, P. and Walmsley, C. M. and Tafalla, M. and Dore, L. and Myers, P. C.},
	month = oct,
	year = {1999},
	note = {ADS Bibcode: 1999ApJ...523L.165C},
	pages = {L165--L169},
}

@article{tafalla_systematic_2002,
	title = {Systematic {Molecular} {Differentiation} in {Starless} {Cores}},
	volume = {569},
	issn = {0004-637X},
	url = {https://ui.adsabs.harvard.edu/abs/2002ApJ...569..815T},
	doi = {10.1086/339321},
	urldate = {2026-03-09},
	journal = {The Astrophysical Journal},
	publisher = {IOP},
	author = {Tafalla, M. and Myers, P. C. and Caselli, P. and Walmsley, C. M. and Comito, C.},
	month = apr,
	year = {2002},
	note = {ADS Bibcode: 2002ApJ...569..815T},
	pages = {815--835},
}

@article{jorgensen_physical_2002,
	title = {Physical structure and {CO} abundance of low-mass protostellar envelopes},
	volume = {389},
	issn = {0004-6361},
	url = {https://ui.adsabs.harvard.edu/abs/2002A&A...389..908J},
	doi = {10.1051/0004-6361:20020681},
	urldate = {2026-03-09},
	journal = {Astronomy and Astrophysics},
	publisher = {EDP},
	author = {Jørgensen, J. K. and Schöier, F. L. and van Dishoeck, E. F.},
	month = jul,
	year = {2002},
	note = {ADS Bibcode: 2002A\&A...389..908J},
	pages = {908--930},
}

@article{qi_resolving_2008,
	title = {Resolving the {Chemistry} in the {Disk} of {TW} {Hydrae}. {I}. {Deuterated} {Species}},
	volume = {681},
	issn = {0004-637X},
	url = {https://ui.adsabs.harvard.edu/abs/2008ApJ...681.1396Q},
	doi = {10.1086/588516},
	urldate = {2026-03-09},
	journal = {The Astrophysical Journal},
	publisher = {IOP},
	author = {Qi, Chunhua and Wilner, David J. and Aikawa, Yuri and Blake, Geoffrey A. and Hogerheijde, Michiel R.},
	month = jul,
	year = {2008},
	note = {ADS Bibcode: 2008ApJ...681.1396Q},
	pages = {1396--1407},
}

@article{sturm_jwst_2023,
    title = {A {JWST} inventory of protoplanetary disk ices. {The} edge-on protoplanetary disk {HH} 48 {NE}, seen with the {Ice} {Age} {ERS} program},
    volume = {679},
    issn = {0004-6361},
    url = {https://ui.adsabs.harvard.edu/abs/2023A&A...679A.138S/abstract},
    doi = {10.1051/0004-6361/202347512},
    language = {en},
    urldate = {2026-02-12},
    journal = {Astronomy and Astrophysics},
    author = {Sturm, J. A. and McClure, M. K. and Beck, T. L. and Harsono, D. and Bergner, J. B. and Dartois, E. and Boogert, A. C. A. and Chiar, J. E. and Cordiner, M. A. and Drozdovskaya, M. N. and Ioppolo, S. and Law, C. J. and Linnartz, H. and Lis, D. C. and Melnick, G. J. and McGuire, B. A. and Noble, J. A. and Öberg, K. I. and Palumbo, M. E. and Pendleton, Y. J. and Perotti, G. and Pontoppidan, K. M. and Qasim, D. and Rocha, W. R. M. and Terada, H. and Urso, R. G. and van Dishoeck, E. F.},
    month = nov,
    year = {2023},
    pages = {A138},
}

@ARTICLE{Glenn_PRIMA_2025,
       author = {{Glenn}, Jason and {Meixner}, Margaret and {Bradford}, Charles M. and {Pontoppidan}, Klaus and {Pope}, Alexandra and {Kataria}, Tiffany and {Rocca}, Jennifer and {Luthman}, Elizabeth and {Armus}, Lee and {Baselmans}, Jochem and et al.},
        title = "{PRIMA mission concept}",
      journal = {Journal of Astronomical Telescopes, Instruments, and Systems},
         year = 2025,
        month = jul,
       volume = {11},
          eid = {031628},
        pages = {031628},
          doi = {10.1117/1.JATIS.11.3.031628},
       adsurl = {https://ui.adsabs.harvard.edu/abs/2025JATIS..11c1628G}
}

@article{oberg_astrochemistry_2021,
    title = {Astrochemistry and compositions of planetary systems},
    volume = {893},
    issn = {0370-1573},
    url = {https://ui.adsabs.harvard.edu/abs/2021PhR...893....1O},
    doi = {10.1016/j.physrep.2020.09.004},
    urldate = {2026-03-09},
    journal = {Physics Reports},
    publisher = {Elsevier},
    author = {Öberg, Karin I. and Bergin, Edwin A.},
    month = jan,
    year = {2021},
    note = {ADS Bibcode: 2021PhR...893....1O},
    pages = {1--48},
}

@article{woitke_diana_2016,
	title = {Consistent dust and gas models for protoplanetary disks: {I}. {Disk} shape, dust settling, opacities, and {PAHs}},
	volume = {586},
	issn = {0004-6361, 1432-0746},
	shorttitle = {Consistent dust and gas models for protoplanetary disks},
	url = {http://www.aanda.org/10.1051/0004-6361/201526538},
	doi = {10.1051/0004-6361/201526538},
	language = {en},
	urldate = {2024-02-15},
	journal = {Astronomy \& Astrophysics},
	author = {Woitke, P. and Min, M. and Pinte, C. and Thi, W.-F. and Kamp, I. and Rab, C. and Anthonioz, F. and Antonellini, S. and Baldovin-Saavedra, C. and Carmona, A. and Dominik, C. and Dionatos, O. and Greaves, J. and Güdel, M. and Ilee, J. D. and Liebhart, A. and Ménard, F. and Rigon, L. and Waters, L. B. F. M. and Aresu, G. and Meijerink, R. and Spaans, M.},
	month = feb,
	year = {2016},
	pages = {A103},
}

@ARTICLE{Aikawa_COsnowline_n2hp_2015,
       author = {{Aikawa}, Yuri and {Furuya}, Kenji and {Nomura}, Hideko and {Qi}, Chunhua},
        title = "{Analytical Formulae of Molecular Ion Abundances and the N$_{2}$H$^{+}$ Ring in Protoplanetary Disks}",
      journal = {\apj},
         year = 2015,
        month = jul,
       volume = {807},
       number = {2},
          eid = {120},
        pages = {120},
          doi = {10.1088/0004-637X/807/2/120},
archivePrefix = {arXiv},
       eprint = {1505.07550},
 primaryClass = {astro-ph.SR},
       adsurl = {https://ui.adsabs.harvard.edu/abs/2015ApJ...807..120A}
}

@article{booth_planet-forming_2019,
	title = {Planet-forming material in a protoplanetary disc: the interplay between chemical evolution and pebble drift},
	volume = {487},
	issn = {0035-8711},
	shorttitle = {Planet-forming material in a protoplanetary disc},
	url = {https://ui.adsabs.harvard.edu/abs/2019MNRAS.487.3998B},
	doi = {10.1093/mnras/stz1488},
	urldate = {2026-03-18},
	journal = {Monthly Notices of the Royal Astronomical Society},
	publisher = {OUP},
	author = {Booth, R. A. and Ilee, J. D.},
	month = aug,
	year = {2019},
	note = {ADS Bibcode: 2019MNRAS.487.3998B},
	pages = {3998--4011},
}

@article{furuya_different_2022,
	title = {Different {Degrees} of {Nitrogen} and {Carbon} {Depletion} in the {Warm} {Molecular} {Layers} of {Protoplanetary} {Disks}},
	volume = {938},
	issn = {0004-637X},
	url = {https://iopscience.iop.org/article/10.3847/1538-4357/ac9233},
	doi = {10.3847/1538-4357/ac9233},
	number = {1},
	journal = {The Astrophysical Journal},
	publisher = {American Astronomical Society},
	author = {Furuya, Kenji and Lee, Seokho and Nomura, Hideko},
	month = oct,
	year = {2022},
	pages = {29},
}

@article{arabhavi_minds_2025,
	title = {{MINDS}: {The} {Very} {Low}-mass {Star} and {Brown} {Dwarf} {Sample} {Hidden} {Water} in {Carbon}-dominated {Protoplanetary} {Disks}},
	volume = {984},
	issn = {0004-637X},
	shorttitle = {{MINDS}},
	url = {https://ui.adsabs.harvard.edu/abs/2025ApJ...984L..62A},
	doi = {10.3847/2041-8213/adc692},
	urldate = {2026-03-19},
	journal = {The Astrophysical Journal},
	publisher = {IOP},
	author = {Arabhavi, Aditya M. and Kamp, Inga and van Dishoeck, Ewine F. and Henning, Thomas and Jang, Hyerin and Christiaens, Valentin and Gasman, Danny and Pascucci, Ilaria and Perotti, Giulia and Grant, Sierra L. and Barrado, David and Güdel, Manuel and Lagage, Pierre-Olivier and Caratti o Garatti, Alessio and Lahuis, Fred and Waters, L. B. F. M. and Kaeufer, Till and Kanwar, Jayatee and Morales-Calderón, Maria and Schwarz, Kamber and Sellek, Andrew D. and Tabone, Benoît and Temmink, Milou and Vlasblom, Marissa},
	month = may,
	year = {2025},
	note = {ADS Bibcode: 2025ApJ...984L..62A},
	pages = {L62},
}

@article{van_t_hoff_robustness_2017,
	title = {Robustness of {N2H}+ as tracer of the {CO} snowline},
	volume = {599},
	issn = {0004-6361},
	url = {https://ui.adsabs.harvard.edu/abs/2017A&A...599A.101V},
	doi = {10.1051/0004-6361/201629452},
	urldate = {2026-03-19},
	journal = {Astronomy and Astrophysics},
	publisher = {EDP},
	author = {van 't Hoff, M. L. R. and Walsh, C. and Kama, M. and Facchini, S. and van Dishoeck, E. F.},
	month = mar,
	year = {2017},
	note = {ADS Bibcode: 2017A\&A...599A.101V},
	pages = {A101},
}

@article{scikit-learn,
  title={Scikit-learn: Machine Learning in {P}ython},
  author={Pedregosa, F. and Varoquaux, G. and Gramfort, A. and Michel, V.
          and Thirion, B. and Grisel, O. and Blondel, M. and Prettenhofer, P.
          and Weiss, R. and Dubourg, V. and Vanderplas, J. and Passos, A. and
          Cournapeau, D. and Brucher, M. and Perrot, M. and Duchesnay, E.},
  journal={Journal of Machine Learning Research},
  volume={12},
  pages={2825--2830},
  year={2011}
}

@misc{ricciardi_compact_2026,
    title = {Compact {CO} emission and no evidence of radial drift. {ALMA} observations of the faintest planet-forming disks in {Lupus}},
    url = {https://ui.adsabs.harvard.edu/abs/2026arXiv260411292R},
    doi = {10.48550/arXiv.2604.11292},
    urldate = {2026-05-22},
    publisher = {arXiv},
    author = {Ricciardi, Giulia and Zagaria, Francesco and Miotello, Anna and Manara, Carlo F. and Rosotti, Giovanni and Zallio, Luigi and Andrews, Sean and Booth, Richard and Carpenter, John and Cleeves, Ilse and Facchini, Stefano and Guzmán, Viviana V. and Toci, Claudia and Vioque, Miguel and Wilner, David and Williams, Jonathan P.},
    month = apr,
    year = {2026},
    note = {ADS Bibcode: 2026arXiv260411292R},
}
\bibliographystyle{aasjournalv7}



\end{document}